%% file: FEMOSAA.tex
\documentclass[table,format=acmsmall, review=false, screen=true]{acmart}

\DeclareFontFamily{OT1}{pzc}{}
\DeclareFontShape{OT1}{pzc}{m}{it}{<-> s * [1.10] pzcmi7t}{}
\DeclareMathAlphabet{\mathpzc}{OT1}{pzc}{m}{it}
\usepackage{xcolor}
\usepackage{tabularx}
\usepackage{color}
\usepackage{multirow} 
\usepackage{enumitem}
\usepackage{standalone}
\usepackage{algorithm}
\usepackage{algpseudocode}
\usepackage{dblfloatfix}
\usepackage{tikz}
\usepackage{pgfplots, filecontents}
\usepackage{pgfplotstable}
\pgfplotsset{
        /pgfplots/ybar legend/.style={
        /pgfplots/legend image code/.code={%
        \draw[##1,/tikz/.cd,bar width=3pt,yshift=-0.2em,bar shift=0pt]
                plot coordinates {(0cm,0.8em)};},
},
}
\usepackage{subcaption}
\usepackage{alphalph}

\usepackage[flushleft]{threeparttable}
\usepackage{balance}
\usepackage{tcolorbox}
\usetikzlibrary{patterns}

\input{tikz/detail-data}

\newcolumntype{P}[1]{>{\centering\arraybackslash}m{#1}}
\newcolumntype{Y}{>{\centering\arraybackslash}X}

\makeatletter
    \newcommand*{\algrule}[1][\algorithmicindent]{\makebox[#1][l]{\hspace*{.5em}\thealgruleextra\vrule height \thealgruleheight depth \thealgruledepth}}%
\newcommand*{\thealgruleextra}{}
\newcommand*{\thealgruleheight}{.75\baselineskip}
\newcommand*{\thealgruledepth}{.25\baselineskip}

\newcount\ALG@printindent@tempcnta
\def\ALG@printindent{%
    \ifnum \theALG@nested>0
        \ifx\ALG@text\ALG@x@notext
        \else
            \unskip
            \addvspace{-1pt}
            \ALG@printindent@tempcnta=1
            \loop
                \algrule[\csname ALG@ind@\the\ALG@printindent@tempcnta\endcsname]%
                \advance \ALG@printindent@tempcnta 1
            \ifnum \ALG@printindent@tempcnta<\numexpr\theALG@nested+1\relax
            \repeat
        \fi
    \fi
    }%
\usepackage{etoolbox}
\patchcmd{\ALG@doentity}{\noindent\hskip\ALG@tlm}{\ALG@printindent}{}{\errmessage{failed to patch}}
\makeatother

\newbox\statebox
\newcommand{\myState}[1]{%
    \setbox\statebox=\vbox{#1}%
    \edef\thealgruleheight{\dimexpr \the\ht\statebox+1pt\relax}%
    \edef\thealgruledepth{\dimexpr \the\dp\statebox+1pt\relax}%
    \ifdim\thealgruleheight<.75\baselineskip
        \def\thealgruleheight{\dimexpr .75\baselineskip+1pt\relax}%
    \fi
    \ifdim\thealgruledepth<.25\baselineskip
        \def\thealgruledepth{\dimexpr .25\baselineskip+1pt\relax}%
    \fi
    \State #1%
    \def\thealgruleheight{\dimexpr .75\baselineskip+1pt\relax}%
    \def\thealgruledepth{\dimexpr .25\baselineskip+1pt\relax}%
}

\begin{document}

\title{FEMOSAA: Feature Guided and Knee Driven Multi-Objective Optimization for Self-Adaptive Software} 
\author{Tao Chen}
\affiliation{%
  \institution{Department of Computing and Technology, Nottingham Trent University, UK, and CERCIA, School of Computer Science, University of Birmingham, UK}
  \streetaddress{Edgbaston}
  \city{Birmingham}
  \postcode{B15 2TT}
  \country{UK}}
\author{Ke Li}  
  \affiliation{%
  \institution{School of Computer Science and Engineering, University of Electronic Science and Technology of China, China, and Department of Computer Science, University of Exeter, UK}
  \city{Exeter}
  \postcode{EX4}
  \country{UK}}
\author{Rami Bahsoon}  
  \affiliation{%
  \institution{CERCIA, School of Computer Science, University of Birmingham, UK}
  \streetaddress{Edgbaston}
  \city{Birmingham}
  \postcode{B15 2TT}
  \country{UK}}
  \author{Xin Yao}  
  \affiliation{%
  \institution{Department of Computer Science and Engineering, Southern University of Science and Technology, Shenzhen 518055, China, and CERCIA, School of Computer Science, University of Birmingham, UK.}
  \streetaddress{Edgbaston}
  \city{Birmingham}
  \postcode{B15 2TT}
  \country{UK}}

\begin{abstract}
Self-adaptive software (SAS) can reconfigure itself to adapt to the changing environment at runtime, aiming for continually optimizing conflicted non-functional objectives, e.g., response time, energy consumption, throughput and cost \emph{etc}. In this paper, we present Feature guided and knEe driven Multi-Objective optimization for Self-Adaptive softwAre (FEMOSAA), a novel framework that automatically synergizes the feature model and Multi-Objective Evolutionary Algorithm (MOEA), to optimize SAS at runtime. FEMOSAA operates in two phases: at design time, FEMOSAA automatically transposes the engineers' design of SAS, expressed as a feature model, to fit the MOEA, creating new chromosome representation and reproduction operators. At runtime, FEMOSAA utilizes the feature model as domain knowledge to guide the search and further extend the MOEA, providing a larger chance for finding better solutions. In addition, we have designed a new method to search for the knee solutions, which can achieve a balanced trade-off. We comprehensively evaluated FEMOSAA on two running SAS: one is a highly complex SAS with various adaptable real-world software under the realistic workload trace; another is a service-oriented SAS that can be dynamically composed from services. In particular, we compared the effectiveness and overhead of FEMOSAA against four of its variants and three other search-based frameworks for SAS under various scenarios, including three commonly applied MOEAs, two workload patterns and diverse conflicting quality objectives. The results reveal the effectiveness of FEMOSAA and its superiority over the others with high statistical significance and non-trivial effect sizes.

\end{abstract}

\setcopyright{rightsretained} 
\acmJournal{TOSEM}
\acmYear{2018} \acmVolume{1} \acmNumber{1} \acmArticle{1} \acmMonth{1} \acmPrice{}\acmDOI{10.1145/3204459}

%
%

\ccsdesc[500]{Software and its engineering~Software performance}
\ccsdesc[500]{Software and its engineering~Search-based software engineering}

%
%


\keywords{feature model, search-based software engineering, multi-objective evolutionary algorithm, multi-objective optimization, self-adaptive system, performance engineering}


\thanks{This work is supported by the Ministry of Science and Technology of China (Grant No. 2017YFC0804003), Science and Technology Innovation Committee Foundation of Shenzhen (Grant No. ZDSYS201703031748284), and EPSRC (Grant Nos. EP/J017515/01 and EP/K001523). The co-corresponding author: Tao Chen (t.chen@cs.bham.ac.uk), Ke Li (keli.genius@gmail.com) and Xin Yao (xiny@sustc.edu.cn).}


\maketitle


\section{Introduction}
\label{sec:introduction}

Self-Adaptive Software (SAS) is a special type of software that is capable of adapting and reconfiguring itself at runtime, through a set of known features (e.g., CPU cap, thread pool size and cache size, \emph{etc}), according to the changing environment~\cite{deLemos2013}. One major goal of SAS is to continually optimize multiple and often conflicting non-functional objectives, e.g., response time versus energy consumption, throughput versus cost, \emph{etc}. However, given the dynamic and uncertain nature of running software, it is difficult to fully specify all possible conditions and their adaptation solutions at design time. Thus, designing an efficient and effective runtime optimization approach is necessary, yet challenging. Depending on the complexity of SAS, software engineers have exploited various search algorithms, e.g., exact or stochastic search, for continually finding the optimal (or near-optimal) adaptation solution for SAS at runtime~\cite{fusion}\cite{nsgaii-cloud2}\cite{tsc-chen-2015}\cite{plato}\cite{Pascual2015392}\cite{chen-gecco}\cite{chen-icpe}.




To optimize SAS at runtime using the search algorithms, there are two crucial challenges: (i) firstly, it is difficult to effectively and systematically convert the SAS design to the context of search algorithm while considering the right encoding of features in the representation of optimization, e.g., using only the features that contribute to different aspects of the variability of SAS. Here, the features might be categorical or numeric, where the former refers to those with distinct characteristics, e.g., the \emph{Cache} feature is \lq on\rq~or \lq off\rq; the latter denotes those that can be quantified, measured and sorted, e.g., the size of~\emph{maxThreads}. Furthermore, it is difficult to effectively and systematically handle the features' dependencies, e.g., one can change \emph{Cache Mode} only if the \emph{Cache} feature is \lq turned on\rq. Dependency can become even more complex in the presence of numeric features, e.g., in Tomcat~\cite{tomcat}, the size of~\emph{maxThreads} should not be less than the size of \emph{minSpareThreads}. Those conversion tasks are non-trivial as the design of SAS can be complex and most search algorithms cannot handle dependency constraints in nature. (ii) Secondly, optimizing multiple conflicting objectives and managing their trade-offs are complex and challenging, especially for SAS runtime. This is attributed to the huge number of alternative adaptation solutions and the required efficiency for the found solution to be effective. Moreover, the dynamic and uncertain nature of SAS further complicates the conflicting relations between objectives, rendering the trade-off surface difficult to be explored. Those challenges, when not appropriately addressed, can result in compromised quality, unacceptable running overhead and imbalanced trade-off in SAS runtime optimization.

Most existing work fails to handle the first challenge as they have relied on a manual and/or incomplete conversion of the SAS design into the search algorithm's context~\cite{plato}\cite{duse}\cite{nsgaii-cloud1}\cite{nsgaii-cloud2}, which renders the process expensive, non-systematic and error-prone. Moreover, the feature dependencies are often ignored, wasting the valuable function evaluations on invalid solutions at SAS runtime while providing no guarantee on finding the valid ones. Inspired by the applications of search algorithms to Software Product Line problems~\cite{sayyad2013scalable}, researchers~\cite{fusion}\cite{Pascual2015392} have combined the feature model~\cite{fm1990} with search algorithms to optimize SAS at runtime, considering categorical dependencies. However, numeric features are ignored and a solution often encodes all the features using a simple binary representation. This might lead to the \emph{curse of dimensionality}, and thereby entailing unnecessary complexity at SAS runtime. Further, existing approaches cannot prevent wasteful exploration of invalid solutions and difficult to handle the dependencies related to numeric features.


For the second challenge above, exact search~\cite{fusion}~\cite{moses}, with the helps of objective aggregation (e.g., a weighted sum), has been exploited for SAS runtime optimization. However, modern SAS often exhibits high variability, leading to an explosion of the search space of all possible solutions and rendering the problem intractable. Henceforth, exact search may fail to scale at runtime. In contrast, stochastic search, particularly Evolutionary Algorithms that are widely applied in Search-Based Software Engineering (SBSE), tends to be naturally robust in solving problems with extremely high number of alternatives and thus appealing for SAS optimization~\cite{6475391}. Those algorithms, when properly tailored, can lead to approximate and near-optimal solutions for complex software engineering problems with reasonable running time as of minutes, if not seconds~\cite{sip}. Furthermore, stochastic search has proven to be effective for many real-time systems~\cite{nsgaii-cloud1}\cite{valkyrie}\cite{nsgaii-cloud2}\cite{tsc-chen-2015}. Often, existing approaches rely on single-objective evolutionary algorithm to optimize SAS by simply transforming a multi-objective problem into an aggregated single-objective one~\cite{plato}\cite{valkyrie}. While objective aggregation might be preferable for some contexts, it has been shown that there are cases where assigning weights to different objectives is a non-trivial task for software engineers and the aggregation can hardly maintain a good diversity of the solutions~\cite{6475391}. To alleviate this issue, studies~\cite{duse}\cite{nsgaii-cloud1}~\cite{nsgaii-cloud2} have used NSGA-II~\cite{nsgaii}, a popular Multi-Objective Evolutionary Algorithm (MOEA), to optimize SAS without using the weighted aggregation; they have shown that MOEA can find more convergent and diverse solutions in the trade-off surface than optimizing via objective aggregation. However, NSGA-II has a coarse diversity preservation mechanism that is unable to provide well distributed solutions in certain cases~\cite{ZhangL07}. Therefore, it is desirable to have a general framework that can easily work with different MOEAs for optimizing SAS without suffering the limitation from one specific algorithm. In addition, given the fact that MOEAs produce a set of non-dominated solutions, there is no established method for the SAS to choose an appropriate one for adaptation at runtime, entailing the risk of imbalanced trade-offs.


To address the aforementioned challenges and limitations, this paper presents Feature guided and knEe driven Multi-Objective optimization for Self-Adaptive softwAre (FEMOSAA), a novel framework that automatically synergizes the feature model and a given MOEA, to optimize SAS at runtime. Specifically, our contributions include:


\begin{itemize}[leftmargin=0.5cm]

\item[---] We rely on the feature model to represent the design of a given SAS with explicit considerations of numeric features and their dependencies. In FEMOSAA, we provide an automatic and systematic approach to transpose a given design of SAS, expressed as a feature model, into the MOEA's context at design time. Further, such transposition extends the internal structure of MOEAs in order to improve their ability to search for better adaptation solutions at SAS runtime. Notably, we contribute to the following in the transposition approach:
\begin{enumerate}
\item To tailor the problem to be more suitable for SAS runtime, we discard the lengthy binary encoding. Instead, our approach identifies the \emph{\textbf{elitist features}} from the feature model to encode an elegant and polyadic chromosome representation in the MOEA. By elitist features, we refer to those that cannot be removed in the optimization without damaging the original variability of SAS while minimizing the length of encoding. The benefit of such encoding is that (i) it is intuitive, simpler and enable direct dependency extraction and (ii) reducing the number of genes helps to greatly shrink the search space and simplify the dependency constraints, which also improves the quality of the solutions found while shortening the running time of MOEA.


\item To better guide the search and avoid exploring invalid solutions, our approach extracts the feature dependencies with respect to the elitist features. Then, these dependencies are injected into the basic mutation and crossover operators of the MOEA to create new dependency aware operators. These operators can systematically steer the MOEA to focus on exploring the valid solutions of SAS, creating a larger chance to find better ones.
\end{enumerate}
\item[---] Without loss of generality, we design FEMOSAA in such a way that it can be seamlessly integrated with different MOEAs\footnote{In addition to MOEAs, FEMOSAA also works with single-objective evolutionary algorithms in which case the knee selection method would be deactivated.} to optimize SAS at runtime. The elitist features and extracted dependencies, as processed by the transposition approach at design time, are used to guide the running behaviors of a given MOEA for SAS runtime optimization. In this work, we run FEMOSAA with three fundamentally distinct yet widely-used MOEAs in the literature, i.e., MOEA based Decomposition with STable-Matching model (MOEA/D-STM)~\cite{LiZKLW14}, Non-dominated Sort Genetic Algorithm-II (NSGA-II)~\cite{nsgaii} and Indicator Based Evolutionary Algorithm (IBEA)~\cite{ZitzlerK04}.




\item[---] To achieve a balanced trade-off in SAS optimization, FEMOSAA identifies knee solutions automatically from the final non-dominated set. The knee solutions often imply well balanced trade-offs, such that any improvement on one objective of a knee will cause relatively severe degradations on others.

\item[---] We conduct comprehensive experiments on two running SAS: one is a highly complex SAS that consists of the \emph{eBay}-like RUBiS benchmark~\cite{rubis} and a set of real-world adaptable software (i.e., Apache Tomcat~\cite{tomcat}, MySQL~\cite{mysql}, Ehcache~\cite{ehcache} and Xen~\cite{xen}) under the realistic FIFA98 workload trace~\cite{fifa98}; another is a service-oriented SAS that can be dynamically composed by various services. We compare FEMOSAA with four of its variants (e.g., without dependency aware operators) and three other state-of-the-art frameworks (i.e., DUSE~\cite{duse}, PLATO~\cite{plato} and FUSION~\cite{fusion}) under various scenarios, including three commonly applied MOEAs (i.e., MOEA/D-STM, NSGA-II and IBEA) and two different workload patterns\footnote{Different workload patterns will create diverse behaviors of the SAS.} (i.e., read-write and read-only) along with diverse conflicting quality objectives. The experiments reveal the effectiveness of FEMOSAA and its superiority over the others when optimizing conflicting objectives for SAS, with statistically significant results and non-trivial effect sizes.


\end{itemize}

The contributions have clear impact on the synergy between software engineering for SAS and evolutionary computation as FEMOSAA combines the strengths from both fields. Unlike many SBSE work that simply formulates the software engineering problem as a classic optimization problem for some MOEAs, our deeper synergy takes one step further by automatically and dynamically extracting the domain information of SAS to extend the internal structure of MOEA, improving its search ability. As a result, to control and exploit the power of MOEAs, the software engineers of SAS only need to provide the feature model when using FEMOSAA, without being an expert on MOEA. In addition, FEMOSAA improves MOEA and provides insights for MOEA researchers to design better algorithms for SAS, since the identified elitist features and their dependencies serve as the engineers' systematic domain knowledge by which we can reduce the search space and better guide the search, providing a larger chance for finding better solutions.

The reminder of this paper is organized as follows: Section~\ref{sec:mot} illustrates a detailed motivating example of SAS. Section~\ref{sec:bg} presents the background and the extended notions of numeric features in the feature model. Section~\ref{sec:overview} gives an overview of FEMOSAA. Section~\ref{sec:fm-to-ea-moea} illustrates our approach that transposes a feature model to MOEA. Section~\ref{sec:moead-stm} presents how the internal structure of existing MOEAs can be extended to combine with our dependency aware operators and knee selection. Experimental results, verifiability and threats to validity are discussed in Section~\ref{sec:exp}. Finally, Sections~\ref{sec:rw} and~\ref{sec:con} present related work and conclusion respectively.

\section{A Detailed Motivating Scenario of Self-Adaptive Software}
  \label{sec:mot}

\begin{figure}[t!]
\centering
  \includegraphics[width=0.7\textwidth]{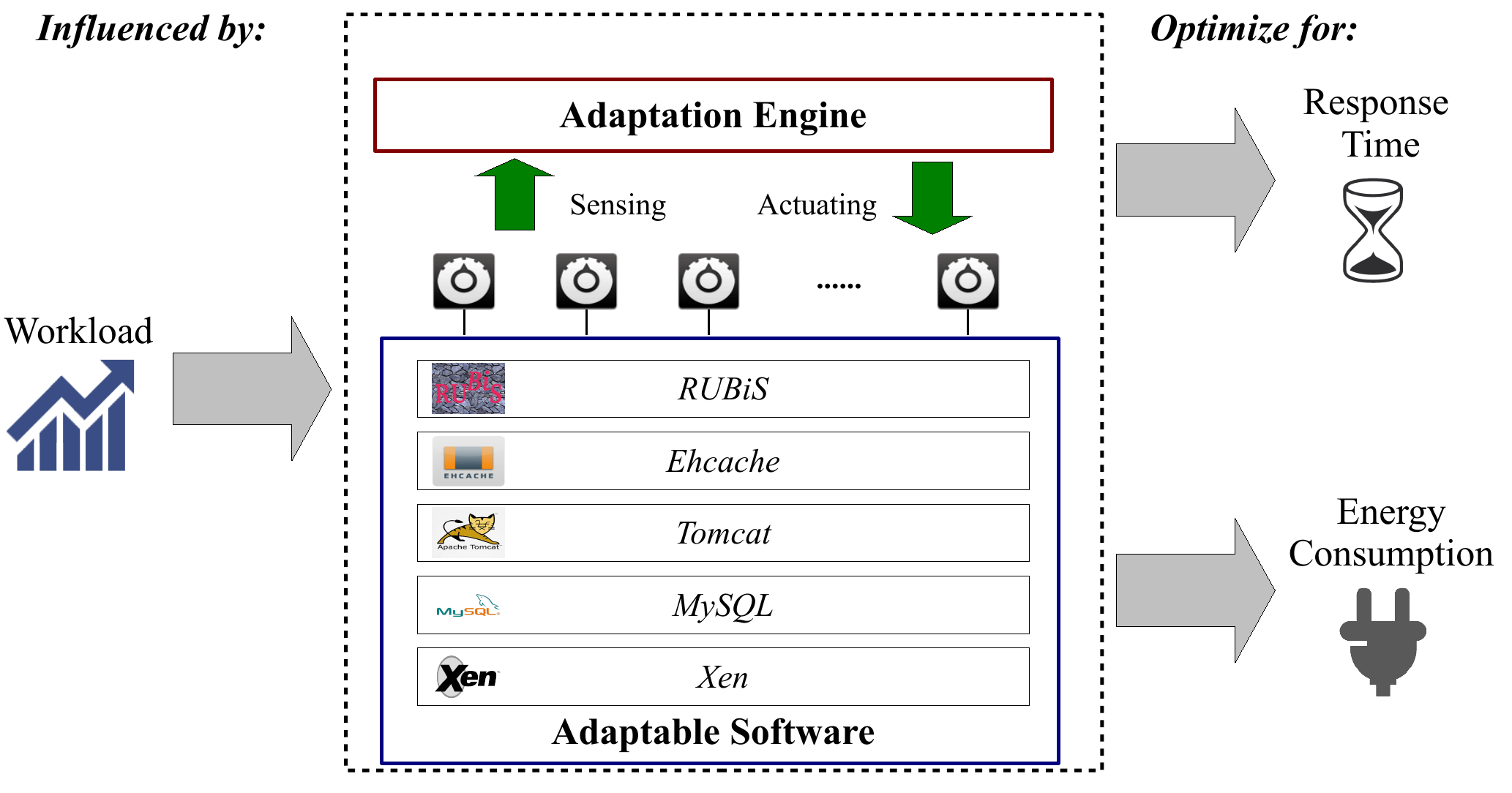}
  \caption{An example of SAS.}
  \label{fig:example}
  \vspace{-0.5cm}
  \end{figure}

While our work can be applied to different contexts that demand runtime adaptation, we draw on a representative and realistic SAS to motivate and illustrate the need. As shown in Figure~\ref{fig:example}, like many SAS, the SAS example consists of two parts: an adaptable software that is being managed at runtime, and an engine that controls the adaptation. Additionally, the SAS contains a complex software stack consisting of RUBiS\footnote{An eBay like software application with 26 services.}~\cite{rubis}, Apache Tomcat~\cite{tomcat}, Ehcache~\cite{ehcache} and MySQL~\cite{mysql}, running on the virtualization hypervisor Xen~\cite{xen}. The RUBiS benchmark serves as a representative of many real-world software applications that offers diverse functionalities and services to many end-users concurrently. We can see from Figure~\ref{fig:example}, as the case of most practical software applications, the SAS's software stack contains many off-the-shelf real-world software as described above. Each of the software supports various control features, which, together with those from the other software in the stack, can be changed dynamically on-the-fly to influence the runtime behaviors of the software system. Example of the control features includes the number of threads, the memory allocation and enabling/disabling cache mechanism, \emph{etc}. By design, all the possible configurations of control features form the search space, or variability of the SAS. As the workload changes, the SAS is capable of adapting the features at runtime to optimize for various non-functional quality attributes, e.g., response time. To achieve such goal, thanks to the rapid development of search algorithms, through which the SAS is often designed as to continually search for the combination of feature configurations that lead to the optimal (or near optimal) quality at runtime. However, to effectively and efficiently engineer the SAS in this way is challenging for the following reasons:

\textbf{Encoding the Features from the SAS Design.} Consider a complex SAS which contains many features and configurations, systematically and generically choosing the right features and encode them into the representation of search algorithm is difficult. To optimize the SAS at runtime, such representation defines the fundamental search space of the problem to be explored, therefore the encoding of features could have positive or negative impact on the search ability of a potential search algorithm. Given that some features in the SAS design do not contribute to the SAS's variability or they can represent the same aspect of variability~\cite{Benavides2010615}, existing work~\cite{fusion}~\cite{Pascual2015392} that simply encodes all features in a binary format is unnecessary. Suppose a feature model with 100 features, binary representation can easily create a search space of $2^{100}$ and this, as we will show in Section~\ref{eval-chro}, can negatively affect the adaptation quality and overhead.

\textbf{Handling the Dependencies in the SAS Design.} Many widely-used exact and stochastic search algorithms, e.g., MOEA, are not designed to handle dependencies constraints. This makes the treatment of dependencies difficult especially when the dependencies in SAS come in a mixture of categorical dependencies, e.g., \emph{Cache Mode} $\mathpzc{require}$ \emph{Cache}, and numeric ones, e.g., \emph{maxThreads} $\geq$ \emph{minSpareThreads}. As we will show in Section~\ref{eval-dep}, those dependencies, when ignored~\cite{plato}~\cite{duse}~\cite{valkyrie} or incorrectly handled~\cite{fusion}~\cite{Pascual2015392} (as in existing work), can degrade the adaptation quality.

\textbf{Explosion of the Search Space.} Modern SAS often has high variability leading to an explosion of the search space. For example, the original design of the SAS shown in Figure~\ref{fig:example} has a search space of more than a billion, which we will elaborate in details at Section~\ref{sec:fm}.

\textbf{Trade-off on the Conflicting Objectives.} SAS often exhibits multiple conflicting quality objectives, e.g., response time versus energy consumption and throughput versus cost, which need to be optimized simultaneously and trade-off needs to be made to comprise some of them. In general, many existing approaches~\cite{plato} have assumed that the relative importance of objectives can be correctly quantified as numeric weights, which has been found to be difficult in some cases~\cite{6475391}. Those weights, when inappropriately specified and expressed, would inevitably create negative impact on the search process and result in unwanted bad adaptation quality. It is even more difficult to achieve balanced trade-off.

These difficulties motivate our work, which automatically synergizes the feature model of SAS and a given MOEA, creating feature guided MOEA with knee selection, to optimize SAS at runtime.

\begin{algorithm}[htbp]
\scriptsize
\caption{General algorithmic process of MOEA}
\label{alg:moea}
\begin{algorithmic}[1]
\Require given mutation rate $r_m$, crossover rate $r_c$ and the maximum number of evaluation $eval_{max}$, which is often equivalent to the size of population $\times$ the maximum number of generations
\Ensure a set of optimized non-dominated solutions

\State \textbf{start evolution}
\State $P=\emptyset$
\State $eval=0$
\For{$i=1$ to $P_{size}$}
\State $S=$ \textproc{getRandomSolution()}
\State \textproc{evaluateFitness($S$)}
\State $eval=eval+1$
\State $P=P+S$
\EndFor
\While{$eval<eval_{max}$}
\State $P_0:=\emptyset$
\While{$|P_0|\leq P_{size}$}
\State $parents:=$ \textproc{doMatingSelection($P$)}
\State $offspring:=$ \textproc{doCrossover($parents$, $r_c$)}
\For{\textbf{each} solution $S$ in $offspring$}
\State \textproc{doMutation($S$, $r_m$)}
\EndFor
\State \textproc{evaluateFitness($offspring$)}
\State $eval:=eval+|offspring|$
\State $P_0:=P_0\cup offspring$
\EndWhile
\State $P:=P\cup P_0$
\State \textproc{doSurvivalSelection($P$, $P_{size}$)}
\EndWhile
\State\Return \textproc{getNonDominatedSolutions($P$)}
\State \textbf{end evolution}
\end{algorithmic}
\end{algorithm}

\section{Background and Preliminaries}
  \label{sec:bg}

\subsection{Multi-Objective Evolutionary Algorithm (MOEA)}
\label{sec:MOEA}

\begin{figure}[t!]
\centering
    \minipage{0.62\textwidth}
    \centering
   \includegraphics[width=1.1\textwidth]{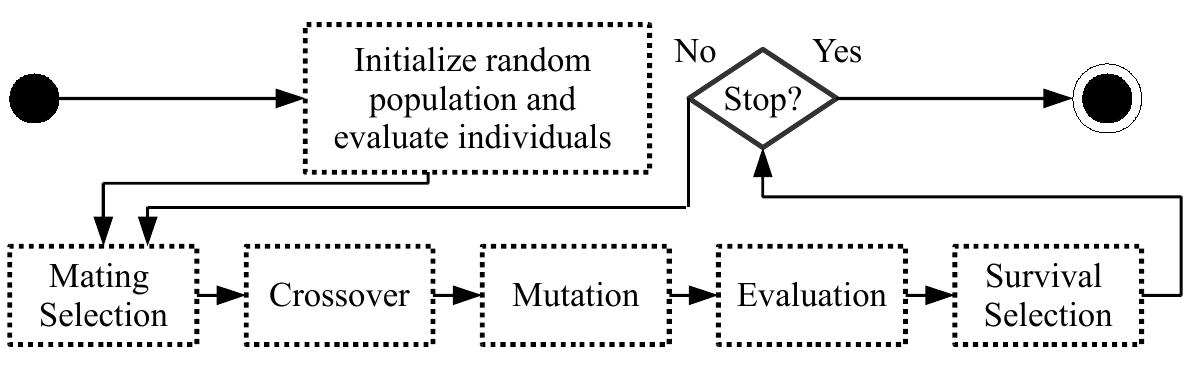}
    \caption{The general workflow of MOEA.}
     \label{fig:moea-flow}
\endminipage\hfill
  \minipage{0.38\textwidth}
  \centering
  \includestandalone[width=0.6\textwidth]{tikz/knee-example}
       \caption{Pareto optimal and knee solutions.}
 \label{fig:knee-example}
\endminipage\hfill
 \end{figure}

Evolutionary algorithm, a stochastic search-based meta-heuristic, has been widely accepted as a major approach for solving multi-objective optimization problems~\cite{Deb:2001}, in which case it is also known as MOEA. In MOEA, the population contains a set of solutions (individuals), each of which is represented by a fixed-length thread-like chromosome carrying different values at each gene. As shown in Figure~\ref{fig:moea-flow} and Algorithm~\ref{alg:moea}, the evolutionary search of MOEA starts after the initialization of the population (line 2 to 9). During the search process, the elite information can propagate from the parents to the offspring via some random and probabilistic reproduction operations (i.e., crossover and mutation) upon the mating parents chosen from the mating selection procedure. Inspired by the \textit{survival of the fittest} rule from the evolutionism, the survival selection preserves the high quality individuals, having superior fitness values, to the next iteration (generation), as shown from line 10 to 24. The evolution process repeats until a stopping criteria, e.g., a predefined function evaluation threshold, is satisfied. The major difference between MOEA and the classic single-objective evolutionary algorithm lies in the mating and survival selection mechanisms. In particular, instead of finding a single optimal (or near optimal) solution as in the single-objective evolutionary algorithm, MOEA aims to find a set of non-dominated solutions\footnote{A solution dominates another if it has at least one objective better than another while all other objectives are not worse than another. Non-dominated solutions denote those solutions that are not dominated by any other solutions in the set.} that approximates the \emph{Pareto Front} with a good convergence and uniform distribution (line 25). Notably, for every solution in the non-dominated set, any improvement of an objective will result in a degradation for at least one other objective. 

Generally, the existing MOEAs can be divided into the following three categories according to the survival selection mechanisms:

\begin{itemize}[leftmargin=0.5cm]
    \item \emph{Decomposition-based method}: The MOEA decomposes the original multi-objective optimization problem into several single-objective optimization subproblems by linear or non-linear aggregation methods~\cite{nonlinear}. Then, it uses a population-based technique to solve these subproblems in a collaborative manner. MOEA/D~\cite{ZhangL07}, MOEA/D-STM~\cite{LiZKLW14} and NSGA-III~\cite{DebJ14} are the representative algorithms of this sort.

    \item\emph{Pareto-based method}: The MOEA uses Pareto dominance relation as the primary selection criterion to push the solutions towards the Pareto front as close as possible. In the meanwhile, it employs some density estimation techniques, e.g., the crowding distance~\cite{nsgaii} and the clustering analysis~\cite{SPEA2}, to maintain the population diversity. The representative algorithms are NSGA-II~\cite{nsgaii}, SPEA2~\cite{SPEA2} and PAES~\cite{KnowlesC00} \textit{etc}.

    \item\emph{Indicator-based method}: Here, sophisticated performance indicators are designed to measure the overall quality of a solution set. The representative algorithm is IBEA~\cite{ZitzlerK04}, which transfers the multi-objective optimization problem into a new single-objective one that aims to find the optimal set of solutions with respect to a given indicator.
\end{itemize}




\subsection{Knee Solutions}
\label{sec:knee-des}
The MOEA generates a set of non-dominated solutions that approximate the Pareto front. However, not every non-dominated solutions can lead to balanced trade-off for SAS runtime optimization. Indeed, the most common purpose of MOEA is to search and visualize a set of non-dominated solutions that are as close to the true Pareto front as possible. Then, a human decision maker can pick whichever solution that s/he prefers. However, there is no such a human available in the SAS optimization problem. Therefore, a method is required to pick a sole solution from the resulted set of non-dominated solutions to execute adaptation.

A simple Pareto optimal front is shown in Figure~\ref{fig:knee-example} where the two objectives should be minimized. Clearly, solutions near the edges strongly favor one objective over the other but there is a visible bulge around the middle, which is the knee region. Those solutions in the knee region (or simply knee solutions) are characterized by the fact that a small improvement in either objective will cause a large deterioration in the other. In case where the human intervention is limited while the two objectives are equally important; or it is difficult to correctly weight them (which is common for SAS), the knee solutions are more balanced than the others and they are almost the most preferable ones. This is because the knee solutions achieve a good sense of compromise, while moving the solution in any direction from the knee region would create a bias towards an objective, leading to imbalanced adaptation results. Finding the knee solutions is challenging because real-world runtime SAS problems may not pose a perfect convex objective surface as shown in Figure~\ref{fig:knee-example}.

\subsection{Feature Model with Numeric Features}
\label{sec:fm}
The feature model~\cite{czarnecki2004staged}, expressed as the tree structure, is a widely used notation for software engineers to represent the functional variability of a software~\cite{Benavides2010615}. In the feature oriented domain analysis, the feature model is particularly important for expressing the possible variations under which a software system can operate in order to improve functional and non-functionary quality~\cite{fm1990}. In this perspective, features define the prominent or distinctive aspects between different variations of a software system~\cite{fm1990}, which range from high-level architectural elements (an entire component) to low-level configurations (a specific parameter). 

In the context of SAS, the inherited concept of a feature model allows it to define the extent to which the SAS is able to adapt at runtime, i.e., a range of variations that the SAS can achieve. Given such nature, there has been some successful attempts that apply the feature model to design SAS~\cite{fusion}\cite{Pascual2015392}.  Therefore, to correctly exploit the feature model for SAS, the software engineer must identify (i) the variations of different features that are supported by the SAS; and (ii) the dependency constraints that determine the validity of a given variation (adaptation solution). However, while the feature model is useful to express the variability of SAS, i.e., the search space of the adaptation decision making problem, it does not correlate the effects of those variations to the concerned quality attributes. Therefore, in this work, we exploit additional system model to evaluate how a variation can affect the quality of SAS, as we will discuss in Section~\ref{sec:subject}.

\begin{figure*}[t!]
  \includegraphics[width=\textwidth]{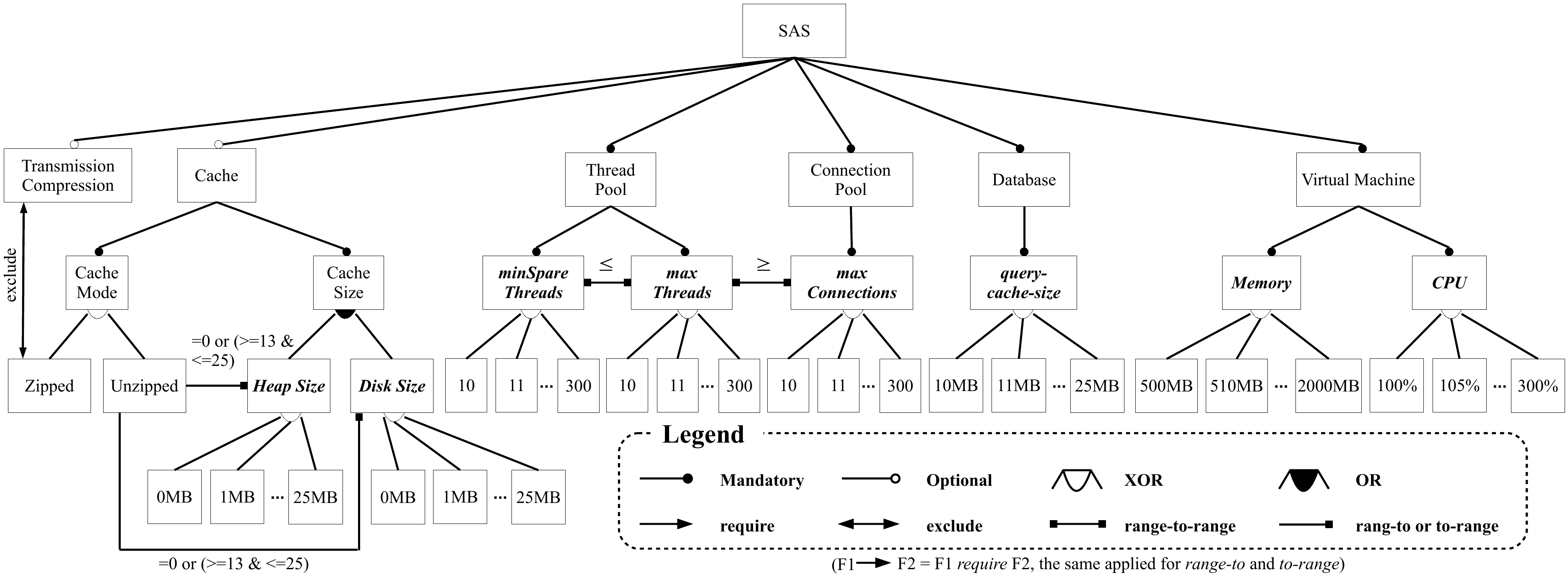}
  \caption{The feature model for the example SAS shown in Figure~\ref{fig:example}. (The numeric features are shown in bold and italic letters. A zipped cache mode means the cached data is compressed, thereby costing smaller amount of memory; otherwise, it is unzipped. CPU \% denotes to what extent the SAS can consume CPU capacity where each 100\% means a CPU core, e.g., 200\% means two cores; 150\% means one full core and 50\% of another)}
  \label{fig:fm}
  \vspace{-0.3cm}
  \end{figure*}

Figure~\ref{fig:fm} shows an example of a feature model for one of the SASs we study in this paper\footnote{In this paper, we use graphical figure of the feature model for more intuitive presentation. In practice, the feature model might be expressed in XML or conjunctive normal form, which can be parsed and analyzed directly by FEMOSAA.}. As we can see there are four types of in-branch relation between a feature and its parent: 
\begin{itemize}[leftmargin=0.5cm]

\item $\mathpzc{Optional}$ refers to the feature might be deselected, e.g., \emph{Cache}.
\item $\mathpzc{Mandatory}$ denotes core features, which cannot be deselceted, e.g., \emph{Thread Pool}.
\item $\mathcal{XOR}$ represents the feature in a group such that exactly one group member can be selected, e.g., \emph{Cache Mode}.
\item $\mathcal{OR}$ means a group that at least one group member needs to be selected,  e.g., \emph{Cache Size}.

\end{itemize}

When a feature is selected, it means that such a feature is \lq turned on\rq; similarly, deselection of a feature refers to it is \lq turned off\rq. Selecting a feature implies that its parent should be selected too. In this work,  we call a feature \textbf{\emph{deselectable}} if it has $\mathpzc{Optional}$, $\mathcal{OR}$ or $\mathcal{XOR}$ relation to its parent; or \textbf{\emph{conditionally deselectable}} if it has $\mathpzc{Mandatory}$ relation to its parent but there exist deselectable ancestors. On the other hand, common cross-branch relations include: 

\begin{itemize}[leftmargin=0.5cm]
\item $F_i \; \mathpzc{require} \; F_j$ means the former can only be selected if the latter is selected. 
\item $F_i \; \mathpzc{exclude} \; F_j$ denotes two features are symmetrically mutually exclusive. 
\item $F_i$ $\mathpzc{at}$-$\mathpzc{least}$-$\mathpzc{one}$-$\mathpzc{exist}$ $F_j$ is an implied relation between the members of an $\mathcal{OR}$ group. It represents the same notion as that of $\mathcal{OR}$.
\item $F_i$ $\mathpzc{at}$-$\mathpzc{least}$-$\mathpzc{one}$-$\mathpzc{require}$ $F_j$ is an implied relation between a member of an $\mathcal{OR}$ group and another external feature, which has $\mathpzc{require}$ to the root of the said $\mathcal{OR}$ group. It means $F_i$ can only be selected if at least one member of the $\mathcal{OR}$ group, which $F_j$ belongs to, is selected. 

\end{itemize}
All those relations constitute to the dependency chain(s) in the model. As in Figure~\ref{fig:fm}, the number of features in the above example is 1151 with a search space of more than a billion.

To better incorporate the feature model with SAS and simplify the design, we distinguish \textbf{\emph{categorical features}} and \textbf{\emph{numeric features}}. We define numeric features as: \emph{a feature is numeric if it has more than one child in its $\mathcal{XOR}$ group and all its children can be quantified by real numbers.} For example, in Figure~\ref{fig:fm}, the \emph{Memory} is clearly a numeric feature. Otherwise, the feature is categorical, e.g., \emph{Cache Mode}. Similarly, a dependency is numeric as long as it is linked to numeric features and it involves quantitative comparisons. As in Figure~\ref{fig:fm}, we propose the following cross-branch numeric dependencies for the engineers to specify in their design:

\begin{itemize}[leftmargin=0.5cm]

\item $\mathpzc{range}$-$\mathpzc{to}$-$\mathpzc{range}$. This is associated with two numeric features and it can be expressed as, e.g., $F_i$ $\mathpzc{range}$-$\mathpzc{to}$-$\mathpzc{range}$ $F_j$ ($F_i < F_j$), meaning that $F_i$'s selected child in its $\mathcal{XOR}$ needs to be smaller than that of $F_j$. It can be easily translated into categorical dependency: $F_i < F_j$ simply means $F_i$'s $\mathcal{XOR}$ child $C$ would have $\mathpzc{exclude}$ dependency on each $F_j$'s $\mathcal{XOR}$ child, which is larger than or equal to the value of $C$. Other quantitative comparisons (e.g., $>$) can be also applied. 

\item $\mathpzc{to}$-$\mathpzc{range}$. This constrains a categorical feature $F_i$ (dependent) with respect to a numeric feature $F_j$ (main), e.g., $F_i$ $\mathpzc{to}$-$\mathpzc{range}$ $F_j$ ($F_j<10$), meaning that $F_i$ can only be selected if $F_j$'s selected child in its $\mathcal{XOR}$ falls in the given range, as expressed by the mathematical formula. This can be translated to categorical dependency such that $F_i$ would have $\mathpzc{exclude}$ dependency on each of $F_j$'s $\mathcal{XOR}$ children that are not in the range.

\item $\mathpzc{range}$-$\mathpzc{to}$. This is the inverse of $\mathpzc{to}$-$\mathpzc{range}$ dependency where a numeric feature (dependent) is constrained by a categorical feature (main).

\end{itemize}

Clearly, numeric dependencies can only be cross-branched while categorical ones exist on both in-branch and cross-branch. When a dependency is associated with one categorical feature and one numeric feature (i.e., $\mathpzc{to}$-$\mathpzc{range}$ and $\mathpzc{range}$-$\mathpzc{to}$), we call it hybrid dependency which is a special case of numeric dependency. Note that numeric features might have all types of dependencies but categorical features cannot be linked to the $\mathpzc{range}$-$\mathpzc{to}$-$\mathpzc{range}$ numeric dependency. 


\subsubsection{The benefits of explicitly considering numeric features}

As mentioned, given that the feature model is discrete and statically defined at design time, it is possible to convert those numeric features and their dependencies into categorical ones without affecting the original variability of SAS. However, explicitly considering numeric features in the feature model will introduce the following benefits in terms of both design time analysis and runtime optimization in FEMOSAA:

\begin{itemize}[leftmargin=0.5cm]

\item Explicitly considering the numeric features provides simpler and more intuitive design of the feature model as numeric features can be interpreted directly by the software engineers.

\item Converting the numeric features into categorical ones will unnecessarily complicate the feature model, which can implicitly induce the software engineers to design the feature model in a way that the children of numeric features would need to be encoded as genes. As mentioned, this will largely increase the number of solution variables in the optimization, leading to the curse of dimensionality. Therefore, explicitly considering numeric features can provide us with the foundation to design novel and simpler encoding of chromosome representation in MOEA, as we will show in Section~\ref{sec:chro}.

\item Explicitly considering the numeric features results in less number of dependencies in contrast to the case where the numeric features are converted into categorical ones. As we will show in Section~\ref{sec:dep}, this simplifies our dependency extraction process for injecting the dependencies into mutation and crossover operators of MOEA. In addition, less number of dependencies implies simpler dependency structure, i.e., a dependent feature has less number of main features, which in turn reduces the running overhead of our dependency aware operators at runtime. 

\end{itemize}

 \begin{figure}[t!]
\centering
  \includegraphics[width=0.9\textwidth]{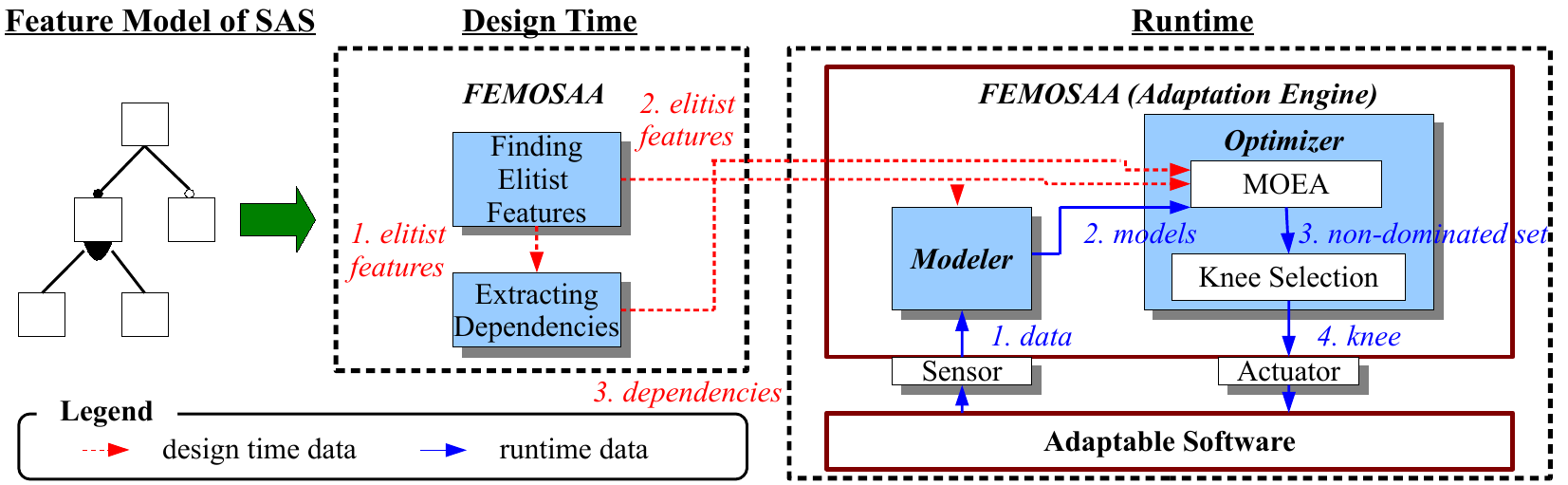}
  \caption{The architecture of FEMOSAA.}
  \label{fig:arch}
  \vspace{-0.5cm}
  \end{figure}


%

\section{FEMOSAA Overview}
 \label{sec:overview}
As shown in Figure~\ref{fig:arch}, a SAS generally consists of two parts: an adaptable software that is managed at runtime, and an engine that controls the adaptation. The adaptable software could be a software stack that contains different inter-connected software or middleware. 

Our FEMOSAA framework is deployed as the adaptation engine and it operates on both design time and runtime of the SAS. At design time, FEMOSAA analyzes and transposes the feature model of SAS, which is provided by the software engineers, to the context of MOEA. The transposition at first identifies the elitist features (see Section~\ref{sec:chro}), which are passed to the process for extracting the dependency to accommodate with the selected features (step 1), as we will explain in Section~\ref{sec:dep}. With the help of FEMOSAA, those elitist features and dependencies are stored and will be used directly by the MOEA at runtime (step 2 and 3). Given that only the elitist features would be encoded into the chromosome representation of MOEA, the identified elitist features can be used as the objective functions' inputs, and can serve as the indication of which sensors/actuators to use or to implement (step 2). 

FEMOSAA has two main components at runtime: a \emph{Modeler} which contains the objectives (fitness) functions that build the correlation between features and quality attributes. Those objectives functions can be created using analytical models~\cite{analytical}, simulation~\cite{6392599} or machine learning~\cite{Chen:2013}~\cite{Chen:2015:tse}~\cite{Chen:2014:ucc} in which they might be updated on-the-fly using the data from sensors; and an \emph{Optimizer} that realizes the MOEA (extended by our knee selection), which is guided by the transposed information from the feature model, to find a single optimized solution for adaptation via actuators (see Section~\ref{sec:moead-stm}).

Given the uncertain and dynamic environment, these two components constitute the feedback loop that continually adapts the SAS towards better quality, e.g., improve response time. The adaptation cycle starts from monitoring status of the SAS and the environment (step 1), which is then used to update the objective functions and model (step 2). Next, the feature guided MOEA optimizes and searches for a set of non-dominated solutions based on the updated objective functions (step 3), after which the knee selection selects the most balanced one for adaptation (step 4). The optimization can be triggered either by the violations on the quality requirements or, as what we did in this work, by a fixed frequency, e.g., at each point in time. Note that we consider the execution order of a solution as a separate issue from the optimization. Thus, given a valid and optimized solution, we assume that the valid order of execution, with respect to the dependency, is enforced in the actuators through analyzing the dependencies in the feature model. 


\section{Transposing Feature Model of Self-Adaptive Software to MOEA at Design Time}
 \label{sec:fm-to-ea-moea}
In this section, we present an automatic and systematic approach as part of FEMOSAA that transposes a feature model into MOEA's context. At design time, the approach finds the elitist features from the model, by which we refer to those that cannot be removed in the optimization without damaging the original variability of SAS while minimizing the length of encoding, to form chromosome representation; and then extracts the feature dependency with respect to the elitist features. Such information will be used at runtime to guide the evolutionary optimization.


To guarantee correctness of the transposition, it is imperative to ensure that the feature model has been fully tested and verified by existing tools~\cite{Benavides2010615}.  Henceforth, this ensures that faults, e.g., dead features, false options and contradictory relations, have been already dealt with before the transposition. The verification of a feature model is beyond the scope of this work, however. Unlike our work, the dependency related to numeric features is not treated explicitly in existing testing tools. However, as discussed in Section~\ref{sec:fm}, the numeric (and hybrid) dependency can be easily transferred into the categorical dependency, which can be then tested directly. We also assume that all possible children (including \emph{0}) of numeric features are discretized and predefined. It is worth noting that discretizing the numeric features is the firstly step to remove the unnecessary complexity of our SAS optimization problem, this is because many real-world features are often discrete and/or can be customized based on software engineers' knowledge, e.g., it can be known that changing memory allocation by less than 1MB does not affect the behaviors and quality of SAS, henceforth, instead of considering the memory feature as a continuous feature, the possible child features of the memory feature can be discretized at every 1MB.

While FEMOSAA is generic and can be applied on any cases as long as the feature model and MOEAs are involved, in the following, we specify the transposition approach in FEMOSAA for the general cases but refer to a concrete example for more intuitive illustration where appropriate. Specifically, in Section~\ref{sec:chro}, we introduce the approach to identify the elitist chromosome representation of a SAS's feature model. Subsequently, in Section~\ref{sec:dep}, we illustrate how the related dependency chains and the value trees can be extracted (Section~\ref{sec:dep-ex}) and merged (Section~\ref{sec:merge-dep}), according to the genes identified in Section~\ref{sec:chro}.

 \begin{figure}[t!]
\centering
  \includegraphics[width=0.7\textwidth]{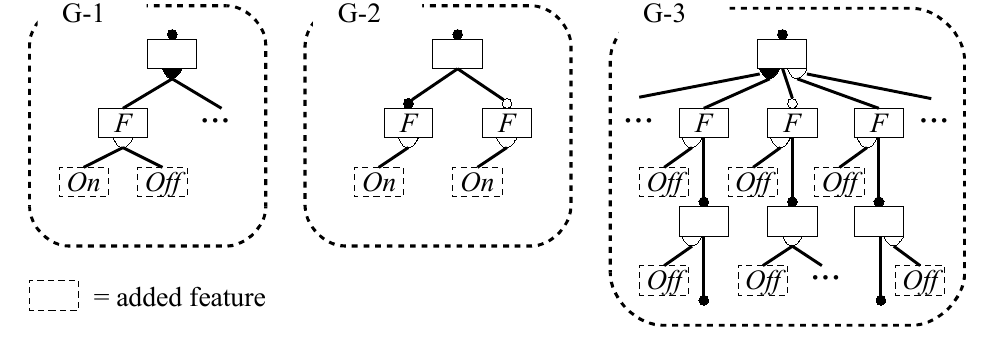}
  \caption{The growing process in SAS's feature model.}
  \label{fig:g}
  \end{figure}


\subsection{Finding Elitist Features for Chromosome Representation}
\label{sec:chro}
\subsubsection{Growing the Feature Model Tree}
Deselectable features in a feature model often do not explicitly indicate the `on' and `off' features as children, but they are important information for us to parse and understand the full variability of the model. Hence, to correctly transpose the feature model, we firstly grow the feature model tree for disclosing the hidden information inferred from the deselectable features. As illustrated in Figure~\ref{fig:g}, this is achieved by adding children representing \emph{On} and/or \emph{Off} to any given feature $F$ in the feature model using the following steps in order:


\begin{itemize}[leftmargin=0.5cm]

\item \textbf{G-1.} If $F$ is a leaf feature that has $\mathcal{OR}$ relation to its parent, we then add two children representing \emph{On} and \emph{Off} in a $\mathcal{XOR}$ group to $F$. This explicitly states that in such case, the leaf $F$ can have two mutually exclusive options, which is important to our encoding.

\item  \textbf{G-2.} If $F$ is a leaf feature that has neither $\mathcal{OR}$ nor $\mathcal{XOR}$ relation to its parent, we then add one child representing \emph{On} in a $\mathcal{XOR}$ group to $F$. This ensures that every feature has the option of 'on' (and translate them into branches to be parsed by \textbf{G-3}), except those with $\mathcal{OR}$ nor $\mathcal{XOR}$ relation to its parent, as the former has been considered in \textbf{G-1} while the latter's `on' option can be expressed by the parent. 

\item  \textbf{G-3.} If $F$ is a branch feature that has $\mathpzc{Optional}$, $\mathcal{OR}$ or $\mathcal{XOR}$ relation to its parent, we then add one child representing \emph{Off} in a $\mathcal{XOR}$ group to $F$ and to the descendants of $F$ that are branch features (if they do not currently have child representing \emph{Off}). This ensures that both the deselectable and conditionally deselectable features expose the option of `off'.

\end{itemize}


After growing the tree, the added features and the steps that create them are shown in Figure~\ref{fig:fm1}.

\subsubsection{Identifying Genes from the Feature Model Tree}
\label{sec:genes}
\begin{figure*}[t!]
  \includegraphics[width=\textwidth]{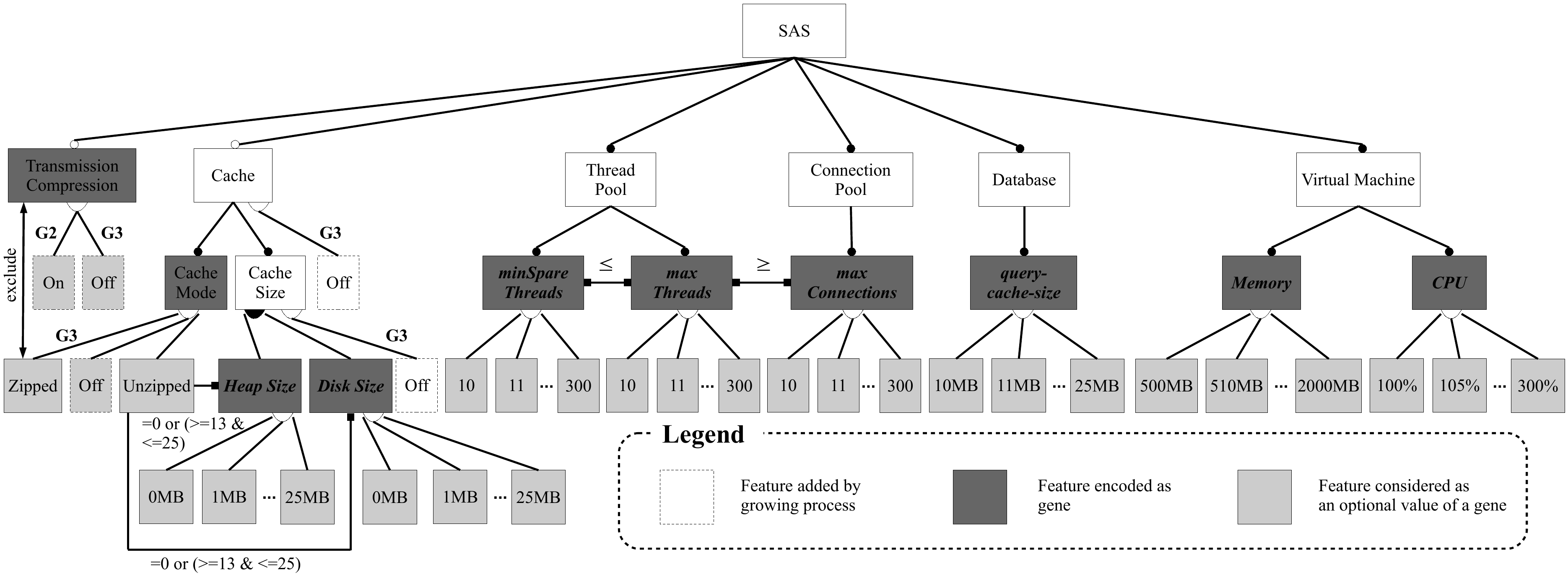}
  \caption{The example SAS's feature model after the process of finding elitist features.}
  \vspace{0.1cm}
  \label{fig:fm1}
  
  \end{figure*}

\begin{figure}[t!]
\centering
  \includegraphics[width=0.65\textwidth]{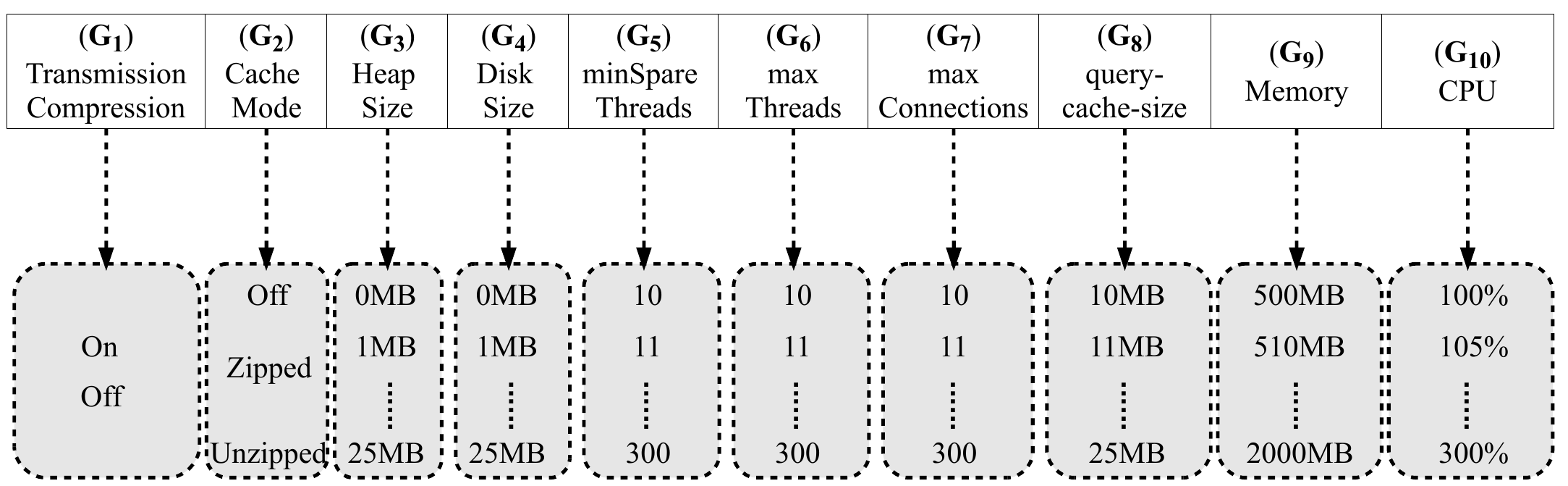}
  \caption{The resulted chromosome representation of the SAS studied in form of genes ($G_1$, $G_2$, ..., $G_{10}$).}
  \label{fig:chromo}
  \end{figure}

We have now obtained a model with no hidden information, the next phase is to find the elitist features for genotype encoding in MOEA, creating an \textbf{\emph{elitist chromosome representation}}. Intuitively, following the grown tree, our approach \emph{encodes a feature $F$ as gene in the chromosome, if and only if, it is the parent of a $\mathcal{XOR}$ group, which contains more than one group member.} Hence, $F$'s children within the $\mathcal{XOR}$ group constitute its set of alternative optional values to be chosen in MOEA, subject to the constraints in dependencies. Drawing on this, the representation can be simplified in three aspects without affecting the original variability:

\begin{enumerate}[leftmargin=0.5cm]

\item Eliminating features whose variability can be expressed by their parent, i.e., those with $\mathcal{XOR}$ relations to the parent, e.g., the variability of \emph{CPU}'s children can be represented by itself.

\item Eliminating features whose variability can be expressed by their descendants, e.g., the variability of \emph{Cache} feature can be represented by the combination of \emph{Cache Mode} and \emph{Cache Size} features; \emph{Cache Mode} can be represented by \emph{Heap Size} and \emph{Disk Size}.

\item Eliminating the features that have no implication on the variability, e.g., the \emph{Thread Pool} is always mandatory. This, however, does not mean that we simply remove all mandatory features (as in~\cite{sip}); instead, our approach retains those mandatory features that with a $\mathcal{XOR}$ group of children as they would often help us to considerably reduce the number of genes even more, as explained in (1) above. 
\end{enumerate}
From now on, we call those features, which are chosen to be encoded in the chromosome, as \textbf{\emph{genes}}. It is easy to see that numeric features are always chosen as genes. As shown in Figures~\ref{fig:fm1} and~\ref{fig:chromo}, there are 10 features being considered as genes in the example feature model of SAS\footnote{The other features, which are not genes, can be fixed to \emph{On} in the SAS}. To make the model informative, we prune the features that are the only member in their corresponding $\mathcal{XOR}$ group. For all genes, if they select \emph{Off} or \emph{0} as their value then it means they are deselected; any other values mean that they are selected. Note that when a gene selects \emph{0}, it implies that the numeric value is 0 and that the feature is \lq turned off\rq, which will have no further effects on the SAS.

In this way, the elitist chromosome representation is polyadic, elegant and free of unnecessary information (e.g., some unneeded relations to parent of a feature), which is otherwise unavoidable in the classic binary representation. This, as we will show in Section~\ref{eval-chro}, can bring non-trivial benefits for the optimization quality and runtime overhead.

\subsection{Extracting Feature Dependency for Guiding Evolution}
\label{sec:dep}
\subsubsection{Analyzing and Refactoring the Dependency}
\label{sec:dep-ex}
While the identified elitist chromosome representation can naturally prevent violation of the $\mathcal{XOR}$ relation, it does not contain any information about the other dependency constraints ($\mathpzc{require}$, $\mathcal{OR}$, \emph{etc}). This issue is non-trivial as leaving it without treatment could result in a high possibility of exploring invalid solutions, which negatively affect the quality of adaptation. To this end, our next step in the transposition is to extract and analyze the dependency chain(s) to accommodate with the genes, so that they can be injected into the mutation and crossover operators of MOEA to prevent the search from exploring the invalid solution. Here, a single dependency between two genes represents the constraint on the dependent gene with respect to the conditions of main gene. The extracted dependencies and their imposed constraints are shown in the Table~\ref{table:dependency-constraints}, which will be discussed in Section~\ref{sec:merge-dep}. Specifically, we distinguish two categories of dependency: \emph{in-branch} and \emph{cross-branch}.




 \begin{figure}[t!]
\centering
  \includegraphics[width=0.6\textwidth]{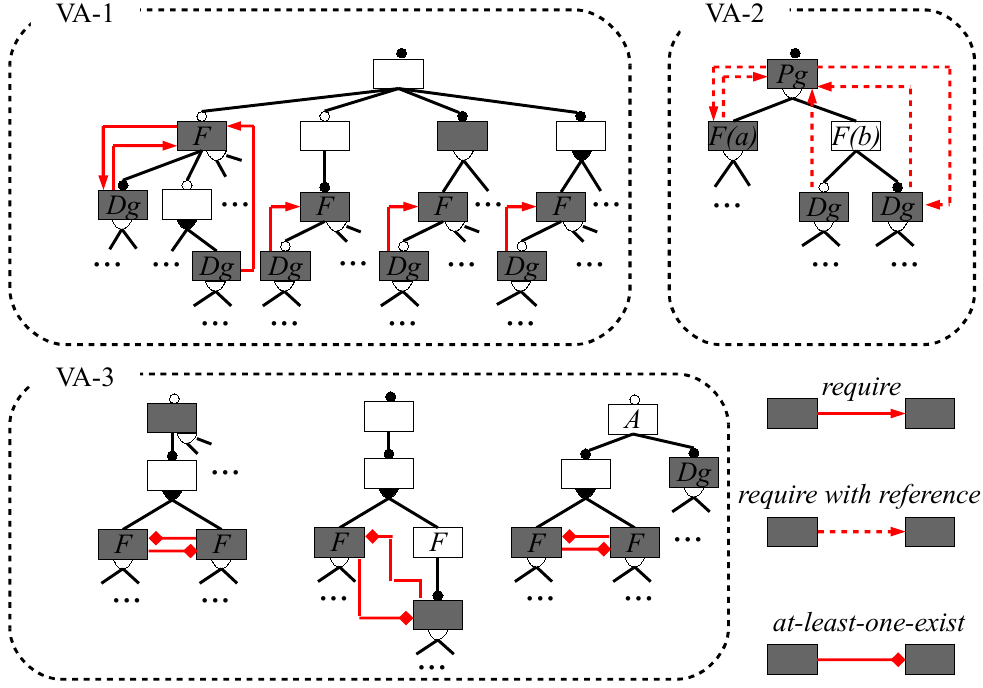}
  \caption{The vertical analysis for extracting in-branch dependencies in SAS's feature model with respect to the elitist genes.}
  \label{fig:va}
  \end{figure}

Extracting in-branch dependency chain(s) aims to handle the constraints introduced by $\mathpzc{Optional}$, $\mathpzc{Mandatory}$, $\mathcal{OR}$ and $\mathcal{XOR}$ relations with respect to the genes. To achieve this, the features' in-branch dependencies are extracted in both vertical and horizontal directions while considering all the four relations.

Vertical analysis for extracting in-branch dependency helps to ensure that the in-branch relation between feature and parent are captured. As shown in Figure~\ref{fig:va}, for any feature $F$ in the original feature model, we conduct the following vertical analysis:

\begin{itemize}[leftmargin=0.5cm]
\item \textbf{VA-1.} If $F$ is a gene and it is deselectable ($\mathpzc{Optional}$, $\mathpzc{OR}$ or $\mathcal{XOR}$ to its parent) or conditionally deselectable ($\mathpzc{Mandatory}$ to its parent but has deselectable ancestors), then for each path from $F$, the closest descendant gene $D_g$ of $F$ would have $\mathpzc{require}$ dependency on $F$, as $D_g$ cannot be selected without the presence of $F$. Additionally, if $D_g$ has $\mathpzc{Mandatory}$ relation to its parent and the path between $F$ and $D_g$ does not contain deselectable features, then $F$ would also have $\mathpzc{require}$ on $D_g$, as both features need to be selected at the same time.


\item \textbf{VA-2.} In addition to \textbf{VA-1}, if $F$ has $\mathcal{XOR}$ relation to its parent and it is a gene, then $F$ would have a $\mathpzc{require}$ on its parent, denoted as $P_g=\alpha$ ($F$'s parent $P_g$ would always be a gene as ensured by our gene identification process), where $\alpha$ is the reference of $F$ in $P_g$; similarly, $P_g=\alpha$ would also have $\mathpzc{require}$ on $F$, as both features need to be selected at the same time. On the other hand, if $F$ has $\mathcal{XOR}$ relation to its parent but it is not a gene, then for each path from $F$, the closest descendant gene $D_g$ of $F$ would have its own $\mathpzc{require}$ on $P_g=\alpha$. Under the same case, if $D_g$ has $\mathpzc{Mandatory}$ relation to its parent and the path between $F$ and $D_g$ does not contain deselectable features, then $P_g=\alpha$ would also need to have $\mathpzc{require}$ on $D_g$.


\item \textbf{VA-3.} In addition to \textbf{VA-1}, if $F$ has $\mathcal{OR}$ relation to its parent, we find the closest deselectable ancestor of $F$, denoted as $A$, if such a $A$ does exist. Now, if at least one ancestor of $F$ is gene; or $F$'s parent is neither deselectable nor conditionally deselectable; or there exist at least one closest descendant gene, $D_g$, of a path from $A$, such that $D_g$ has $\mathpzc{Mandatory}$ relation to its parent and there is no deselectable features in the path between $D_g$ and $A$, then this means that the $\mathcal{OR}$ group, for which $F$'s parent is the root, needs to select at least one group member. Thus, unless there already exist an $\mathpzc{at}$-$\mathpzc{least}$-$\mathpzc{one}$-$\mathpzc{exist}$ dependency, $F$ (if $F$ is a gene) or its closest descendant genes, each of which follows different paths (if $F$ is not a gene), would have $\mathpzc{at}$-$\mathpzc{least}$-$\mathpzc{one}$-$\mathpzc{exist}$ on (i) the other closest descendant genes of $F$ if it is not gene; (ii) those sibling genes of $F$ in the same $\mathcal{OR}$ group; and (iii) the closest descendant genes, each of which follows different paths from the $F$'s siblings that are not genes but in the same $\mathcal{OR}$ group as $F$.

\end{itemize}

\begin{figure}[t!]
\centering
  \includegraphics[width=0.6\textwidth]{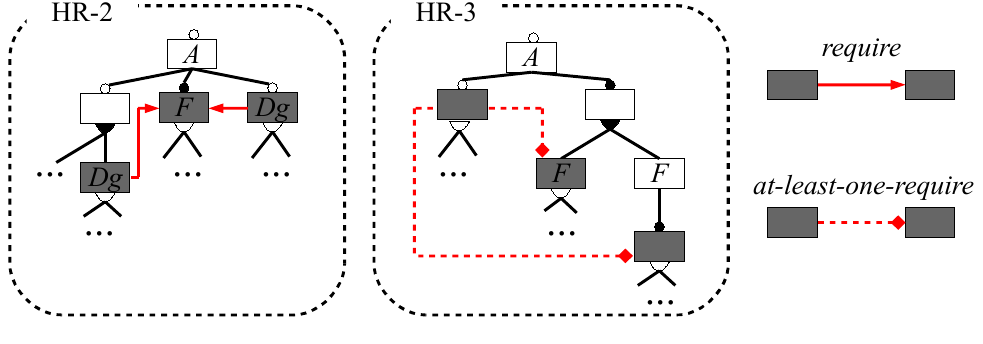}
  \caption{The horizontal refactoring for extracting in-branch dependencies in SAS's feature model with respect to the elitist genes.}
  \label{fig:hr}
  \end{figure}

The horizontal refactoring, on the other hand, is to ensure the elimination of some features does not mislead the dependencies implied by the original variability. Suppose that $F$ is a feature in the original feature model and that $A$ is the closest deselectable ancestor of $F$, if such a $A$ does exist. Now, assuming that $A$ is not a gene, and that there is no gene on the path from $A$ to $F$, we then conduct the following horizontal refactoring, illustrations can be found in Figure~\ref{fig:hr}:
\begin{itemize}[leftmargin=0.5cm]

\item \textbf{HR-1.} If $F$ has $\mathpzc{Optional}$ relation to its parent then we do nothing, even if it is a gene. This is because the selection of $F$ does not affect $A$'s closest descendant genes, each of which follows the other paths from $A$.
\item \textbf{HR-2.} If $F$ is a gene that has $\mathpzc{Mandatory}$ relation to its parent and it is conditionally deselectable, then for each path from $A$, the closest descendant gene $D_g$ of $A$ (excluding $F$ itself) would have $\mathpzc{require}$ on $F$. This can ensure when $D_g$ is selected, $F$ would be also selected.

\item \textbf{HR-3.} If $F$ has $\mathcal{OR}$ relation to its parent and it does not have $\mathpzc{at}$-$\mathpzc{least}$-$\mathpzc{one}$-$\mathpzc{exist}$ dependency for the group, then for each path from $A$ (except the paths that pass through $F$'s $\mathcal{OR}$ group), the closest descendant gene $D_g$ of $A$ would have $\mathpzc{at}$-$\mathpzc{least}$-$\mathpzc{one}$-$\mathpzc{require}$ on $F$, if $F$ is a gene; or on those closest descendant genes of $F$, each of which follows different paths from $F$, if $F$ is not a gene. Hence, when $D_g$ is selected, at least one member of the $\mathcal{OR}$ group of $F$ (or their closest descendant genes) would be also selected.

\item \textbf{HR-4.} If $F$ has $\mathcal{XOR}$ relation to its parent then we do nothing, even if it is a gene. This is because our gen identification process ensures that the parent of $F$ would always be a gene, which also express the selection of $F$. 



\end{itemize}

   \begin{figure}[t!]
\centering
  \includegraphics[width=0.7\textwidth]{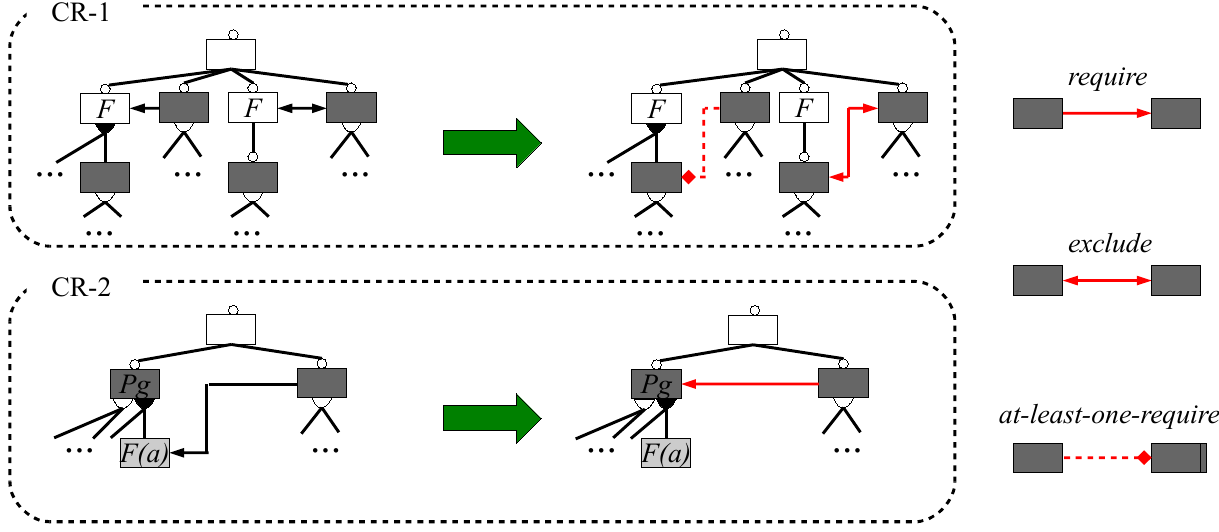}
  \caption{The refactoring for extracting cross-branch dependencies in SAS's feature model with respect to the elitist genes.}
  \label{fig:cr}
  \end{figure}

After considering the in-branch dependency, we now focus on refactoring the cross-branch dependency. If both sides of a cross-branch dependency are genes, then it can be extracted directly. However, if either side (or both) of the feature is not a gene, then a treatment is needed. Suppose that a feature $F$ is associated with one or more cross-branch dependency and that $F$ is not a gene, we then do the following refactoring as shown in Figure~\ref{fig:cr}:


\begin{itemize}[leftmargin=0.5cm]

\item  \textbf{CR-1.} If $F$ is a branch, then its cross-branch dependencies are migrated to those closest descendant genes of $F$, each of which follows different paths from $F$. Further, if $F$ is the root of a $\mathcal{OR}$ group and it is the main feature in any $\mathpzc{require}$ dependency, then those $\mathpzc{require}$ would be changed to $\mathpzc{at}$-$\mathpzc{least}$-$\mathpzc{one}$-$\mathpzc{require}$, which are migrated to the member genes of $F$'s $\mathcal{OR}$ group, and to the closest descendant genes, each of which follows different paths from those members that are not genes.

\item  \textbf{CR-2.} If $F$ is a leaf, then its cross-branch dependencies are migrated to the parent of $F$, denoted as $P_g$, where the dependency would remain the same but the main gene becomes $P_g=\alpha$, where $\alpha$ is the reference of $F$ in $P_g$. Here, $F$ would always have $\mathcal{XOR}$ relation to $P_g$, because if $F$ was to has $\mathpzc{Mandatory}$ relation to $P_g$, then there would be a contradiction as the main feature of a cross-branch dependency is mandatory. In addition, when $F$ is a leaf, our growing process has ensured that $F$ has neither $\mathcal{OR}$ nor $\mathpzc{Optional}$ relation to $P_g$ which would always be a gene.


\end{itemize}

\begin{figure*}[t!]
  \includegraphics[width=\textwidth]{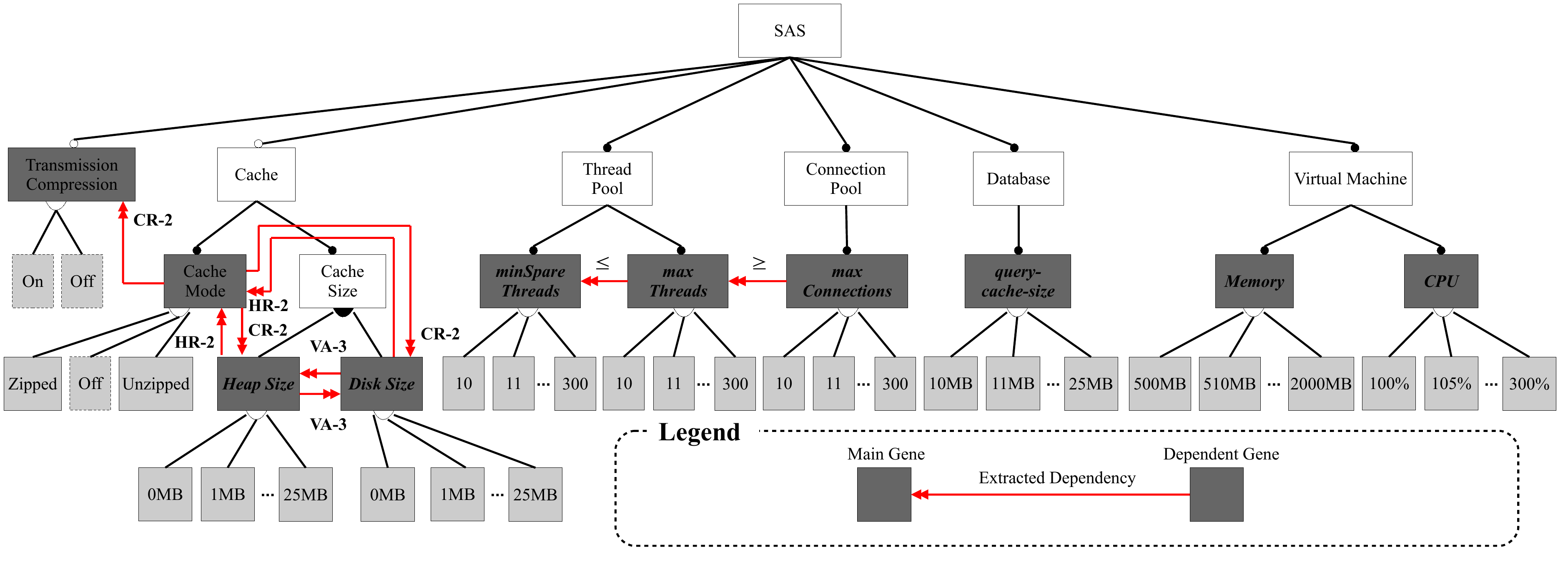}
  \caption{The example SAS's feature model and the extracted dependency after the process of VA, HR and CR.}
  \label{fig:fm2}
  \end{figure*}

\begin{figure}[t!]
\centering
  \includegraphics[width=0.5\textwidth]{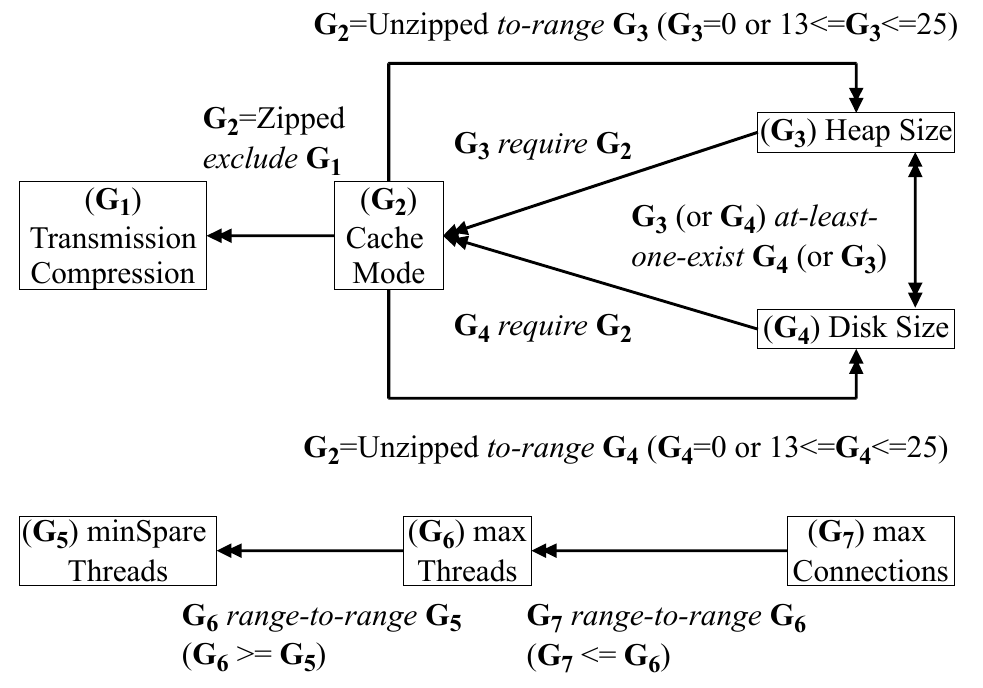}
  \caption{The extracted dependency chains of genes for the example feature model.}
  \label{fig:depchain}
  \end{figure}

\begin{figure*}[t!]
\centering
  \includegraphics[width=\textwidth]{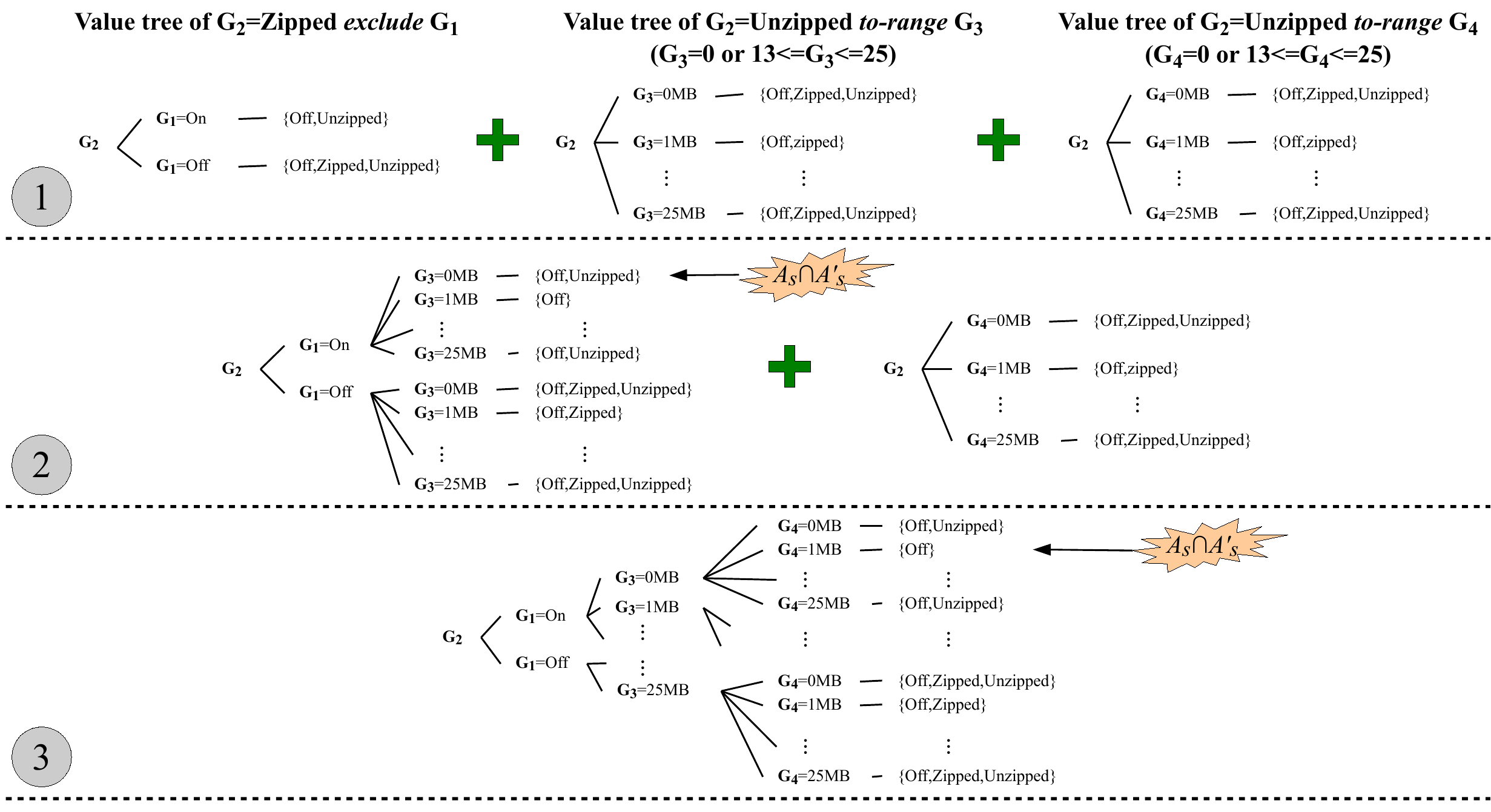}
  \caption{The example of dependency merging and the combined value tree for the gene $G_2$ (\emph{Cache Mode}), which is the dependent gene of three different main genes, i.e., $G_1$ (\emph{Transmission Compression}), $G_3$ (\emph{Heap Size}) and $G_4$ (\emph{Disk Size}). $A_s$ and $A'_s$ denote two sets of values (leaf-set) to be combined, as specified in Table~\ref{table:dependency-constraints}.}
  \label{fig:merge}
   \vspace{-0.3cm}
  \end{figure*}

Finally, putting everything together, the extraction which occurs on the model and the extracted dependency chain(s), with respect to the elitist chromosome, are shown in Figure~\ref{fig:fm2} and~\ref{fig:depchain} respectively. The constraint of a dependent gene imposed by a dependency, according to Table~\ref{table:dependency-constraints}, can be expressed using a \textbf{\emph{value tree}}, where each leaf is a set of optional values constrained by the corresponding condition in a branch, i.e., the selected values of the main gene. For example, in Figure~\ref{fig:merge}, the value tree for the dependency between \emph{Cache Mode ($G_2$)} gene and \emph{Transmission Compression ($G_1$)} gene constrains that the former can only be \emph{Off} or \emph{Unzipped} if the latter is \emph{On}; or any optional values, otherwise. Note that if a gene is not a dependent of any dependency, then it would have a value tree without any branches.


\begin{table}[t!]
\scriptsize
  \caption{The Extracted Dependency Constraints Between two Genes and The Related Set Operations for Merging Dependency.}
  
\label{table:dependency-constraints} 
\begin{tabularx}{\textwidth}{p{2.9cm}p{6.4cm}p{3.8cm}}

\hline 
\textbf{\emph{Dependency (denoted as $D$)}}&
\textbf{\emph{Constraints on $G_i$}}&
\textbf{\emph{Merge with Other Dependency $D'$}}
\\
\hline
$G_i$ $\mathpzc{require}$ $G_j$ &
if $G_j=Off\: or\: 0$, then $A_s=\{Off\}\: or\: \{0\}$. Otherwise, $A_s=A$. &
$A_s \cap A'_s$. \\
\hline 
$G_i=\alpha$ $\mathpzc{require}$ $G_j$ &
if $G_j=Off\: or\: 0$, then $A_s=A - \{\alpha\}$. Otherwise, $A_s=A$. &
$A_s \cap A'_s$. \\
\hline 
$G_i$ $\mathpzc{require}$ $G_j=\alpha$&
if $G_j \ne \alpha$, then $A_s=\{Off\}\: or\: \{0\}$. Otherwise, $A_s=A$. &
$A_s \cap A'_s$.
\\
\hline 
$G_i=\alpha_1$ $\mathpzc{require}$ $G_j=\alpha_2$&
if $G_j \ne \alpha_2$, then $A_s=A - \{\alpha_1\}$. Otherwise, $A_s=A$. &
$A_s \cap A'_s$.
\\
\hline 
$G_i$ $\mathpzc{exclude}$ $G_j$ &
if $G_j  \ne Off\: or\: 0$, then $A_s=\{Off\}\: or\: \{0\}$. Otherwise, $A_s=A$. &
$A_s \cap A'_s$. \\
\hline 
$G_i=\alpha$ $\mathpzc{exclude}$ $G_j$ &
if $G_j \ne Off\: or\: 0$, then $A_s=A - \{\alpha\}$. Otherwise, $A_s=A$. &
$A_s \cap A'_s$. \\
\hline 
$G_i=\alpha_1$ $\mathpzc{exclude}$ $G_j=\alpha_2$ &
if $G_j=\alpha_2$, then $A_s=A - \{\alpha_1\}$. Otherwise, $A_s=A$. &
$A_s \cap A'_s$.  \\
\hline 
$G_i$ $\mathpzc{at}$-$\mathpzc{least}$-$\mathpzc{one}$-$\mathpzc{require}$ $G_j$ &
if $G_j=Off\: or\: 0$, then $A_s=\{Off\}\: or\: \{0\}$. Otherwise, $A_s=A$.  &
if $D'$ is (or merged from only) $\mathpzc{at}$-$\mathpzc{least}$-$\mathpzc{one}$-$\mathpzc{require}$ that related to the same root of $\mathcal{OR}$ group as $D$'s, then $A_s \cup A'_s$. Otherwise, $A_s \cap A'_s$. \\
\hline 
$G_i=\alpha$ $\mathpzc{at}$-$\mathpzc{least}$-$\mathpzc{one}$-$\mathpzc{require}$ $G_j$ &
if $G_j=Off\: or\: 0$, then $A_s=A - \{\alpha\}$. Otherwise, $A_s=A$. &
if $D'$ is (or merged from only) $\mathpzc{at}$-$\mathpzc{least}$-$\mathpzc{one}$-$\mathpzc{require}$ that related to the same root of $\mathcal{OR}$ group as that of $D$, then $A_s \cup A'_s$. Otherwise, $A_s \cap A'_s$. \\
\hline 
$G_i$ $\mathpzc{at}$-$\mathpzc{least}$-$\mathpzc{one}$-$\mathpzc{exist}$ $G_j$ &
if $G_j=Off \: or \: 0$, then $A_s=A - \{Off\}\: or\: A - \{0\}$. Otherwise, $A_s=A$. &
if $D'$ is (or merged from only) $\mathpzc{at}$-$\mathpzc{least}$-$\mathpzc{one}$-$\mathpzc{exist}$ that related to the same root of $\mathcal{OR}$ group as that of $D$, then $A_s \cup A'_s$. Otherwise, $A_s \cap A'_s$.
\\
\hline 
$G_i$ $\mathpzc{range}$-$\mathpzc{to}$-$\mathpzc{range}$ $G_j$ (e.g., $G_i < G_j$) &
if $G_j=\alpha$, then $A_s=\{\alpha_1, ..., \alpha_n\}$ where $\forall \alpha_n \in A$, and $\forall \alpha_n$ meets the given condition with respect to $\alpha$, e.g., $\forall \alpha_n<\alpha$, etc.&
$A_s \cap A'_s$.  \\
\hline 
$G_i$ ($R$) $\mathpzc{range}$-$\mathpzc{to}$ $G_j$ &
if $G_j=Off \: or \: 0$, then $A_s=A - R$. Otherwise, $A_s=A$.&
$A_s \cap A'_s$.  \\
\hline 
$G_i$ ($R$) $\mathpzc{range}$-$\mathpzc{to}$ $G_j=\alpha$ &
if $G_j\ne \alpha$, then $A_s=A - R$. Otherwise, $A_s=A$. &
$A_s \cap A'_s$.  \\
\hline 
$G_i$ $\mathpzc{to}$-$\mathpzc{range}$ $G_j$ ($R$) &
if $G_j=\alpha$; $\alpha \notin R$, then $A_s=\{Off\} \: or \: \{0\}$. Otherwise, $A_s=A$. &
$A_s \cap A'_s$.  \\
\hline 
$G_i=\alpha_1$ $\mathpzc{to}$-$\mathpzc{range}$ $G_j$ ($R$) &
if $G_j=\alpha_2$; $\alpha_2 \notin R$, then $A_s=A - \{\alpha_1\}$. Otherwise, $A_s=A$.  &
$A_s \cap A'_s$. \\
\hline 
\hline 
\multicolumn{3}{p{17.5cm}}{Additionally, when the set operation leads to an empty set, we fix $A_s = \{Off\}\ or \{0\}$. }

 
\\
\hline 
\end{tabularx}
\begin{tablenotes}
\item\textit{$G_i$ and $G_j$ are dependent and main gene, respectively; $A$ is the entire set of optional values for $G_i$; $A_s$ denotes the set of values for $G_i$, given a selected value of $G_j$; $\alpha$, $\alpha_1$, $\alpha_2$ and $\alpha_n$ denote some selected values for $G_i$ or $G_j$; $R$ is a given constrained set of range, e.g., $G_j<10$, etc; $D'$ is another single or merged dependency for which $G_i$ is the dependent gene; $A'_s$ denotes the set of values for $G_i$, given the selected value(s) of the main gene(s) in $D'$.}
    \end{tablenotes}
  \vspace{-0.2cm}
\end{table}

\subsubsection{Merging the Dependency}
\label{sec:merge-dep}
After the extraction, we can see that a dependent gene might have multiple dependencies on the same or different main genes. To construct a combined value tree for a dependent gene, the dependencies, by which it is constrained, need to be merged one by one using set operators (union or intersection) to combine the leaves from their value trees. Table~\ref{table:dependency-constraints} shows what set operators are needed for each dependency type when merging with the others, which is derived from the conjuncture normal form of the related genes. 

Specifically, for every dependent gene, the merging process has the following steps:

\begin{itemize}[leftmargin=0.5cm]
\item[---]  \textbf{Step 1.} If it has two or more dependencies with identical main genes, then the leaves, which are constrained by the same condition in the branches, would be combined directly using the set operations shown in Table~\ref{table:dependency-constraints}.

\item[---]  \textbf{Step 2.} If it has dependencies on different main genes, all branch nodes of one single or already combined value tree are replicated and grafted (as a whole) to each right-most branch node of another single or already combined value tree, forming new levels for the newly combined value tree, representing the combinatorial conditions. Then, for the two value trees that were grafted, their leaves, whose original ancestors are now on the same path from root to a right-most node in the newly grafted tree, are combined using the set operations shown in Table~\ref{table:dependency-constraints} to create the new leaf-set. 

\end{itemize}

The process stops when all related dependencies are merged and their value trees are combined, resulting a finally combined value tree. As an example, Figure~\ref{fig:merge} illustrates the merging process and the finally combined value tree for \emph{Cache Mode ($G_2$)} gene, which originally contains dependencies on three different main genes. The same merging procedure is repeated for every dependent gene.

To ensure the correctness of merging, in both steps, the dependencies that require union are always merged ahead of the others, leading to a set of partially merged dependencies. Since each of them are merged from the identical dependency type, as shown in Table~\ref{table:dependency-constraints}, they can then be merged with each others (and the remaining single dependencies) using the same set operators as if they are single dependencies.

It is worth noting that, despite the given feature model is contradictions free, it might still be possible for the combined leaves to have an empty set under some conditions. For example, recall \emph{Disk Size} gene from Figure~\ref{fig:depchain}, when its $\mathpzc{require}$ dependency on \emph{Cache Mode} merges with its $\mathpzc{at}$-$\mathpzc{least}$-$\mathpzc{one}$-$\mathpzc{exist}$ on \emph{Heap Size}, then combing the leaves for the condition where \emph{Cache Mode} selects \emph{Off} and \emph{Heap Size} selects \emph{0 MB} would lead to an empty set. This is due to the fact that some dependencies have different priorities: in the above example, $\mathpzc{at}$-$\mathpzc{least}$-$\mathpzc{one}$-$\mathpzc{exist}$ would constrain \emph{Disk Size} only if it is allowed to be selected as indicated by the $\mathpzc{require}$ on \emph{Cache Mode}. In those cases, we fix $\{Off\}$ (or $\{0\}$) as the new leaf-set for the dependent gene's combined value tree, representing the fact that it needs to be \lq turned off\rq and cannot affect the SAS under the corresponding conditions. 

In Section~\ref{sec:moead-stm}, we will describe how the elitist chromosome, the extracted dependency chain(s) and the (combined) value trees of genes are seamlessly injected into the MOEA for optimizing SAS at runtime, regardless to the order of changes on genes.

\begin{algorithm}[!b]
\scriptsize
\caption{The Extended Feature Guided MOEA with Knee Selection}
\label{alg:moea-ext}
\begin{algorithmic}[1]
\Require the extracted dependency chain(s) $C$ and the (combined) value trees of the genes $VT$, in addition to the inputs in Algorithm~\ref{alg:moea}
\Ensure a single knee solution
\State \textbf{start evolution}
\State $P=\emptyset$
\State $eval=0$
\For{$i=1$ to $P_{size}$}
\State $S=$ \textproc{getRandomSolution()}
\State \textcolor{red}{\textproc{doDependencyAwareMutation($S$,$1$,$C$,$VT$)}} $\quad \rhd$mutation rate of 1 means mutating every gene.
\State \textproc{evaluateFitness($S$)}
\State $eval=eval+1$
\State $P=P+S$
\EndFor
\While{$eval<eval_{max}$}
\State $P_0=\emptyset$
\While{$|P_0|\leq P_{size}$}
\State $parents=$ \textproc{doMatingSelection($P$)}
\State $offspring=$ \textcolor{red}{\textproc{doDependencyAwareCrossover($parents$,$r_c$,$C$,$VT$)}}
\For{\textbf{each} solution $S$ in $offspring$}
\State \textcolor{red}{\textproc{doDependencyAwareMutation($S$,$r_m$,$C$,$VT$)}}
\EndFor
\State \textproc{evaluateFitness($offspring$)}
\State $eval:=eval+|offspring|$
\State $P_0=P_0\cup offspring$
\EndWhile
\State $P=P\cup P_0$
\State \textproc{doSurvivalSelection($P$, $P_{size}$)}
\EndWhile
\State $P=$ \textproc{getNonDominatedSolutions($P$)}
\State\Return \textcolor{red}{\textproc{getKneeSolution($P$)}}
\State \textbf{end evolution}
\end{algorithmic}
\end{algorithm}



\section{Feature Guided and Knee Driven MOEA at Runtime}
 \label{sec:moead-stm}

\begin{figure}[t!]
\centering
  \includegraphics[width=0.7\textwidth]{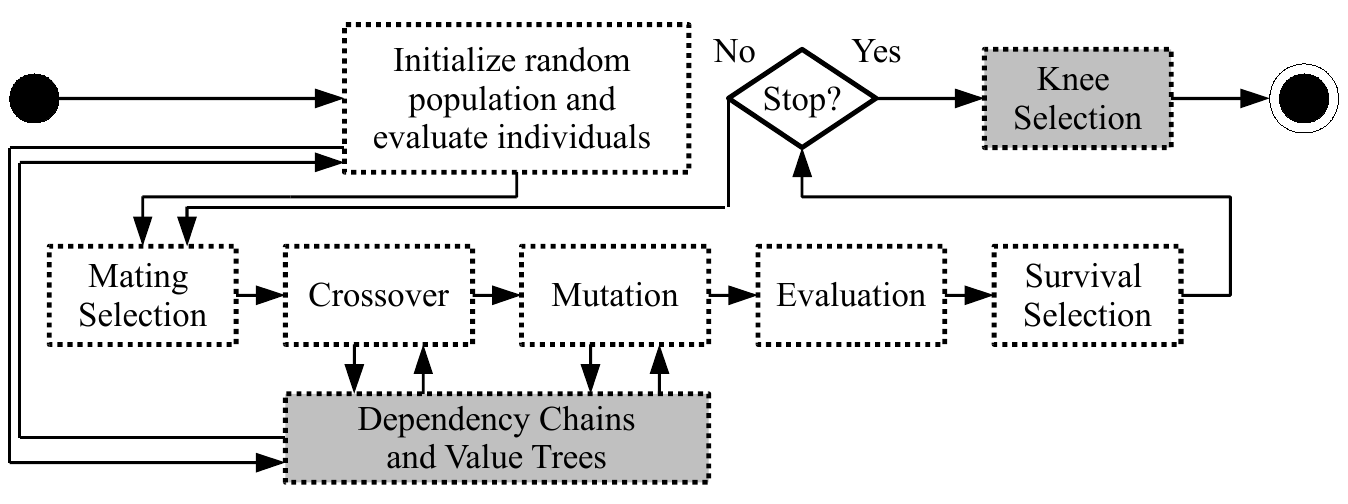}
  \caption{The workflow of the extended feature guided MOEA with knee selection.}
  \label{fig:new-flow}
   \vspace{-0.5cm}
  \end{figure}

We have now completed all the design time transposition and analysis in FEMOSAA. Next, to optimize the SAS at runtime using FEMOSAA, we extend MOEA with the domain information gathered from the feature model, creating a feature guided MOEA with knee selection, as shown in Figure~\ref{fig:new-flow} and Algorithm~\ref{alg:moea-ext}. In particular, the elitist features from Section~\ref{sec:fm-to-ea-moea} form the basic chromosome representation of the solution in MOEA. To avoid exploring invalid solutions, we explicitly inject the extracted dependencies (e.g., Figure~\ref{fig:depchain}) and the (combined) value trees of genes (e.g., Figure~\ref{fig:merge}) into the basic mutation and crossover phases of MOEA to create dependency aware operators (line 6, 15 and 17). Finally, we apply the knee selection proposed in Section~\ref{sec:kneeselection} to identify a single adaptation solution from the set of non-dominated solutions returned by MOEA (line 27).

\subsection{Objective Functions}

The elitist chromosome representation from Section~\ref{sec:fm-to-ea-moea} also helps to define the inputs of the objective (fitness) functions used in the optimization. It is worth noting that FEMOSAA works with a range of quantifiable quality objectives and is agnostic to the actual objective functions, as the framework itself does not rely on any assumptions about the internal structure of those objectives. FEMOSAA is also possible to consider more than two objectives but we use two in this paper to provide more intuitive illustration, since the fundamental principle of multi-objective optimization is the same regardless to the objective number.

The actual objective functions exploited by FEMOSAA can be built using various modeling approaches from the literature, e.g., machine learning based~\cite{Chen:2013}\cite{Chen:2014:ucc}\cite{Chen:2015:tse}, analytical~\cite{analytical} and simulation based~\cite{6392599}, as long as they are compatible with the genes identified by FEMOSAA. In Section~\ref{sec:subject}, we will elaborate the concrete objective functions, which are built by using different approaches, for each subject SAS.



\subsection{Dependency Aware Mutation Operator}

At runtime, to prevent generating invalid offspring when mutating the solutions, the extracted dependency chain(s) and the (combined) value trees of the genes are seamlessly injected into the mutation operation. In this work, we use the boundary mutation operator as the basis, in which each gene might be mutated subject to a mutation rate. Upon mutation of a gene, one of its optional values is randomly assigned. The reason why the boundary mutation operator was chosen is because: (i) it is one of the most commonly used mutation operator; and (ii) it works under discrete optimization problems while allowing to randomly select a value from a predefined value tree, which particularly fits with our SAS optimization problem. However, since the violation of dependency is prevented whenever a gene is changed, it is easy to modify FEMOSAA to work with any other operators that mutate the genes in a similar way as the boundary mutation operator. 

 \begin{figure}[t!]
\centering
  \includegraphics[width=0.6\textwidth]{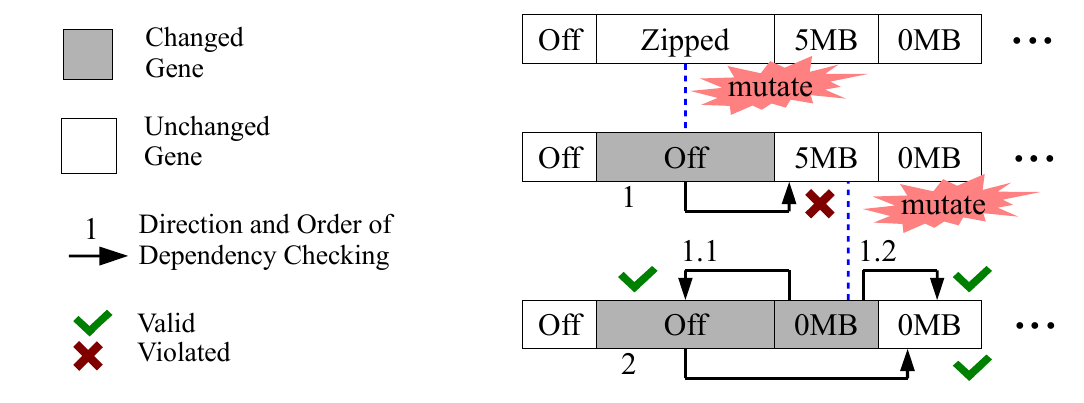}
  \caption{Workflow of dependency aware mutation operator. (The example genes from left to right represent \emph{Transmission Compression}, \emph{Cache Mode}, \emph{Heap Size} and \emph{Disk Size})}
  \label{fig:mutation}
  \end{figure}

\begin{algorithm}[!tp]
\scriptsize
\caption{Dependency aware mutation operator}
\label{alg:mutation}
\begin{algorithmic}[1]
\Require given a valid solution $S$, the extracted dependency chain(s) $C$ and the (combined) value trees of the genes $VT$
\Ensure a mutated valid solution (individual) $S$
\State \textbf{start mutation}
\State $G_{set}=$ the genes that requires mutation as identified by the basic crossover operator based on mutation rate
\For{\textbf{each} $G$ in $G_{set}$}
\State \textproc{mutateWithDependency($G$,$S$)}
\EndFor
\State\Return $S$
\State \textbf{end mutation}
\end{algorithmic}
\hspace{-9.5cm}\textbf{Function:} \textproc{mutateWithDependency($G$,$S$)}
\begin{algorithmic}[1]
\State $VT_g=$ the (combined) value tree of $G$ from $VT$
\State randomly \textit{choose} a new value for $G$ of $S$ from $VT_g$
\State $C_g=$ the chain that contains $G$ from $C$
\For{\textbf{each} dependent gene of $G$ ($G_d$) in $C_g$}
\State $VT_g=$ the (combined) value tree of $G_d$ from $VT$
\State \textit{validate} $G_d$ based on current conditions in $VT_g$
\If{$G_d$'s value in $S$ is invalid}
\State \textproc{mutateWithDependency($G_d$,$S$)}
\EndIf
\EndFor
\end{algorithmic}

\end{algorithm}


As illustrated in Algorithm~\ref{alg:mutation} and Figure~\ref{fig:mutation}, our extended dependency aware mutation operator has the following recursive steps:

\begin{itemize}[leftmargin=0.5cm]

\item[---]  \textbf{Step 1.} When a gene $G$ (e.g., \emph{Cache Mode}) needs to be mutated as identified by the basic mutation operator or due to its violation, we randomly select a value from its (combined) value tree with respect to the selected values of $G$'s main genes (e.g.,  \emph{Transmission Compression}, \emph{Heap Size} and \emph{Disk Size}).

\item[---]  \textbf{Step 2.} Then, we propagate, according to the dependency chain, to $G$'s dependent genes (e.g., \emph{Heap Size} and \emph{Disk Size}) and we then validate if those genes in the solution violate any dependency using their (combined) value trees. If violation found, we would mutate the corresponding gene and start from \textbf{Step 1} for it (e.g., \emph{Heap Size=5MB} in Figure~\ref{fig:mutation}), as shown from line 4 to 10 of the \textproc{mutateWithDependency} function.

\item[---]  \textbf{Step 3.} The process stops when all the genes, which are identified by the basic mutation operator, have been mutated and there is no further violation found.

\end{itemize}

Using this operator, we guarantee that the mutation process in MOEA would be better guided, such that only valid adaptation solutions can be explored regardless to the order of mutation. Since the given feature model has been fully validated and exhibits no design errors, there will be at least one state such that all genes can satisfy all dependencies, which in turn, preventing infinite loop in the presence of circular dependencies. As we will show in Section~\ref{eval-dep}, the dependency aware operators can lead to better quality of adaptations.

\subsection{Dependency Aware Crossover Operator}

Similar to the mutation process, it is necessary to eliminate invalid offspring when swapping elements of the solutions. To this end, the extracted dependency chain(s) and the (combined) value trees can be injected into the given crossover operator\footnote{In this work, we have used the most common type of crossover operators that takes two parents and produces two offsprings. However, our dependency can be injected with the other types of crossover operators (which are rare) in a similar way.} in the MOEA. In this work, we rely on the widely used uniform crossover where two genes, each of which from a different parent and both are at the same position in the chromosome, might be swapped subject to a crossover rate. Such a uniform crossover operator was chosen because it mitigates the problem of genes locus, i.e., the ability to explore the search space is less sensitive to the closeness of highly dependent genes (features) in the encoding, which helps to relax extra design requirements of the SAS. However, since the violation of dependency is prevented whenever a pair of genes is swapped, it is easy to modify FEMOSAA to work with any other operators in which each pair of the swapped genes would be always at the same position in the encoding. 


\begin{figure}[t!]
\centering
  \includegraphics[width=0.8\textwidth]{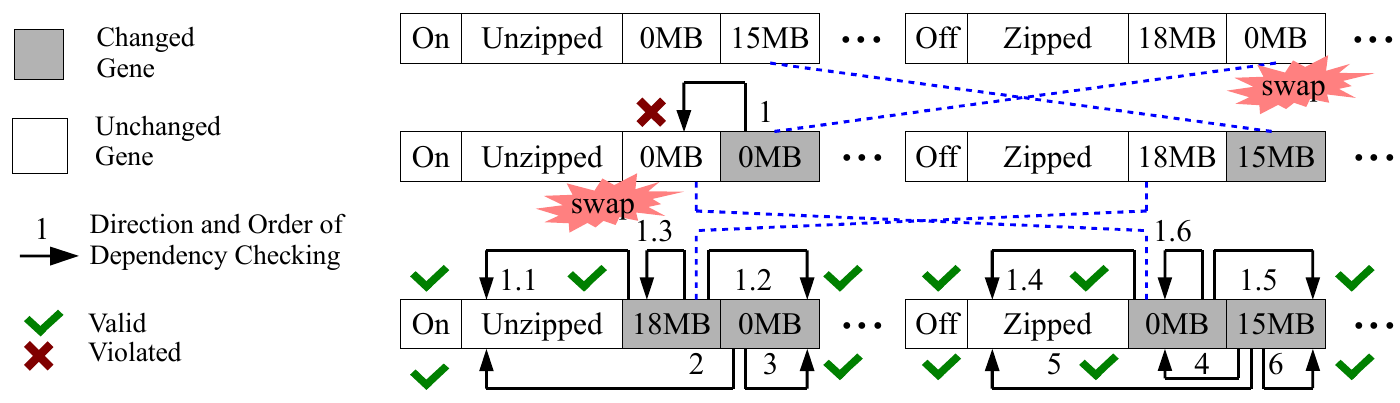}
  \caption{Workflow of dependency aware crossover operator. (The example genes from left to right represent \emph{Transmission Compression}, \emph{Cache Mode}, \emph{Heap Size} and \emph{Disk Size})}
  \label{fig:crossover}
  \end{figure}

\begin{algorithm}[!tp]
\scriptsize
\caption{Dependency aware crossover operator}
\label{alg:crossover}
\begin{algorithmic}[1]
\Require given two valid parent solutions $S_1$ and $S_2$, the extracted dependency chain(s) $C$ and the (combined) value trees of the genes $VT$
\Ensure two new valid solutions (individuals) $S_{n1}$ and $S_{n2}$
\State \textbf{start crossover}
\State $S_{n1}=S_1$
\State $S_{n2}=S_2$
\State $G_{set}=$ the genes that requires crossover as identified by the basic crossover operator based on crossover rate
\For{\textbf{each} $G$ in $G_{set}$}
\State \textproc{crossoverWithDependency($G$,$S_1$,$S_2$,$S_{n1}$,$S_{n2}$)}
\EndFor
\State\Return $S_{n1}$ and $S_{n2}$
\State \textbf{end crossover}
\end{algorithmic}
\hspace{-8cm}\textbf{Function:} \textproc{crossoverWithDependency($G$,$S_1$,$S_2$,$S_{n1}$,$S_{n2}$)}
\begin{algorithmic}[1]
\If{$G$ in $S_{n1}$ and $S_{n2}$ has already been swapped}
\State \textbf{return} 
\EndIf
\State \textit{swap} the values of $G$ in $S_{n1}$ and $S_{n2}$
\State $C_g=$ the chain that contains $G$ from $C$
\For{\textbf{each} solution $S$ in \{$S_{n1}$,$S_{n2}$\}}
\For{\textbf{each} dependent gene of $G$ ($G_d$) in $C_g$}
\State $VT_g=$ the (combined) value tree of $G_d$ from $VT$
\State \textit{validate} $G_d$ based on current conditions in $VT_g$
\If{$G_d$'s value in $S$ is invalid}
\State \textproc{crossoverWithDependency($G_d$,$S_1$,$S_2$,$S_{n1}$,$S_{n2}$)}
\EndIf
\EndFor
\State $VT_g=$ the (combined) value tree of $G$ from $VT$
\State \textit{validate} $G$ based on current conditions in $VT_g$
\If{$G$'s value in $S$ is invalid}
\For{\textbf{each} main gene of $G$ ($G_m$) in $C_g$}
\State \textproc{crossoverWithDependency($G_m$,$S_1$,$S_2$,$S_{n1}$,$S_{n2}$)}
\EndFor
\EndIf
\EndFor
\end{algorithmic}

\end{algorithm} 

After the injection, as shown in Algorithm~\ref{alg:crossover} and Figure~\ref{fig:crossover}, our dependency aware crossover operator uses the following recursive steps:
\begin{itemize}[leftmargin=0.5cm]

\item[---]  \textbf{Step 1.} When a gene $G$ (e.g., \emph{Disk Size}) needs to be swapped as identified by the basic crossover operator or due to violation, we swap it in the offspring if it has not been swapped already.

\item[---]  \textbf{Step 2.} We propagate, according to the dependency chain, to $G$'s dependent genes (\emph{Heap Size} and \emph{Cache Mode}). If a dependent gene in an offspring violates the dependency, we then attempt to swap the gene by repeating from \textbf{Step 1} for it (e.g., \emph{Heap Size=0MB} in Figure~\ref{fig:crossover}), as shown from line 7 to 13 of the \textproc{crossoverWithDependency} function.

\item[---]  \textbf{Step 3.} Next, we check if $G$ in an offspring violates dependencies; if violation exists, we then attempt to swap all the main genes of $G$ (e.g., \emph{Cache Mode} and \emph{Heap Size}, but this is not needed in the example shown) by repeating from \textbf{Step 1} for each of them, as shown from line 14 to 20 of the \textproc{crossoverWithDependency} function.
\item[---]  \textbf{Step 4.} The process terminates when there is no dependency violation or all genes in the parents have been swapped.

\end{itemize}

In this way, we guarantee that only valid adaptation solutions would be produced, given any order of the crossover. As we will show in Section~\ref{eval-dep}, the dependency aware operators can lead to better quality of adaptations.

 \begin{figure}[t!]
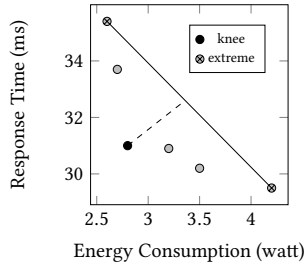

\centering
  \includestandalone[width=0.3\textwidth]{tikz/knee}

 \caption{Finding knee solution(s) in the non-dominated set.}
   \label{fig:knee}
  \end{figure}

\subsection{Knee Selection}
\label{sec:kneeselection}

As mentioned in Section~\ref{sec:knee-des}, knee points are those solutions that achieve well-balanced trade-off on all the objectives. This is particularly appealing as FEMOSAA aims for the cases where the relative importance between conflicting objectives of SAS is unknown and it is too difficult to quantify them. As we can see in Figure~\ref{fig:knee}, which shows the solutions that minimize both response time and energy consumption, those more balanced knee solutions are likely to be around the visible bulge, representing a good sense of compromise. Intuitively, the knee solutions tend to be the furthest away from all the solutions that have the worst result on each single objective, i.e., the extreme solutions. As a result, finding the knee solutions within a non-dominated set is equivalent to search for the solution(s) that has the largest general distance to the extreme solutions in the set. 

To achieve this, we developed a knee selection method for selecting single solution from the set returned by MOEA. As shown in Figure~\ref{fig:knee}, given the final non-dominated set obtained by the feature guided MOEA, we at first construct a line $\ell$ that connects the extreme solutions holding the worst value at each single objective. Then, we calculate the perpendicular distance from each non-dominated solution $\mathbf{x}$ to the $\ell$: 
\begin{equation}
d(\mathbf{x}, \ell)=
\left\{
  \begin{array}{lr}	
    \frac{|\epsilon|}{\sqrt{a^2+b^2}}, & {if}\quad \epsilon < 0\\
    -\frac{|\epsilon|}{\sqrt{a^2+b^2}}, & {otherwise}\\
  \end{array}
\right.
\end{equation}
where $\epsilon=a\times f(\mathbf{x})+b\times g(\mathbf{x})+c$ and the parameters $a$, $b$ and $c$ can be identified through the extreme solutions. In particular, $\epsilon<0$ means that $\mathbf{x}$ is on the left side of $\ell$; otherwise, $\mathbf{x}$ is on the right side of $\ell$. Clearly, solutions on the left side of $\ell$ are preferable than those on the right when considering a minimization problem. The solution(s), which has the largest perpendicular distance to $\ell$, is the knee solution(s) that we are seeking. When there are multiple knee solutions, we randomly select one for adaptation. As we will show in Section~\ref{eval-knee}, the knee selection can lead to more balanced trade-offs on the quality of adaptations. Note that, by replacing $\ell$ with a surface or a hyperplane, we can easily extend our knee selection for more than two objectives.


\subsection{The Concrete MOEAs}

Without loss of generality, FEMOSAA can work easily with a wide range of MOEAs. In this work, we run FEMOSAA with three distinct MOEAs (i.e., MOEA/D-STM, NSGA-II and IBEA), each of which is a widely-used representative of its own category as explained in Section~\ref{sec:MOEA}. In the following, we briefly explain their principles and ideas, while the details are beyond the scope of this paper.


\begin{itemize}[leftmargin=0.5cm]

    \item NSGA-II~\cite{nsgaii}\textemdash As one of the most popular MOEAs, NSGA-II at first uses the non-dominated sorting to divide the population into several non-domination levels. Solutions in the first several levels have a higher priority to survive to the next iteration. If the size of the current non-dominated solution set exceeds the pre-defined threshold, NSGA-II uses the crowding distance, a density estimation technique, to trim the population.

    \item IBEA~\cite{ZitzlerK04}\textemdash The basic idea of IBEA is to firstly define the optimization goal in terms of a binary performance measure/indicator, which is then used to guide the survival selection process. In this way, IBEA transfers a multi-objective optimization problem into a new single-objective optimization problem, with respect to the chosen indicator, to facilitate the fitness assignment procedure.

    \item MOEA/D-STM~\cite{LiZKLW14}\textemdash MOEA/D is a MOEA framework that combines the mathematical rigour of the classic multi-objective optimization method and the implicit parallelism of evolutionary algorithms in a single paradigm. Different from the classic multi-objective optimization method~\cite{nonlinear}, which can only obtain a single Pareto-optimal solution at a time by aggregating all objectives into a single-objective aggregation function, MOEA/D decomposes the original multi-objective optimization problem into a population of single-objective optimization subproblems. In particular, each subproblem corresponds to a pre-defined weight vector, generated in a systematic manner~\cite{LiDZK15}. Afterwards, MOEA/D uses a population-based technique to solve these subproblems in a collaborative manner. As a recent variant of MOEA/D, MOEA/D-STM achieves a balance between convergence and diversity by modifying the survival selection mechanism of the original MOEA/D. In a nutshell, MOEA/D-STM treats subproblems and solutions as two sets of agents. Each agent has its preferences over the agents on the other side. In particular, subproblems concern convergence while solutions concern diversity. The survival selection process is modelled as a matching process between subproblems and solutions. The stable matching between them finally turns out to be the selection result. Note that the stable matching achieves an equilibrium between the preferences of subproblems and solutions, thus the selection strikes the balance between convergence and diversity simultaneously. 
    

\end{itemize}

%

\section{Experiments and Evaluation}
 \label{sec:exp}
By using the actual running SAS that consists of a stack of real-world software, we have conducted comprehensive experiments to evaluate the effectiveness of FEMOSAA by means of comparing it with its variants and the state-of-the-art frameworks under different metrics. The source code of FEMOSAA, the comparative variants and state-of-the-art frameworks, the subject SAS and all the experiment data can be publicly accessed via GitHub\footnote{https://github.com/taochen/ssase} \footnote{https://github.com/JerryI00/Software-Adaptive-System}. Specifically, our evaluation aims to answer the following research questions:

\begin{itemize}[leftmargin=0.5cm]

\item  \textbf{RQ1.} What are the added values of the elitist chromosome representation in contrast to the conventional binary representation? 

\item  \textbf{RQ2.} What benefits do the dependency aware operators provide compared against the classical and widely-used operators 

\item  \textbf{RQ3.} What benefits does the knee selection mechanism provide in contrast to selecting arbitrary non-dominated solutions?
 

\item  \textbf{RQ4.} What is the effectiveness of FEMOSAA in contrast to other state-of-the-art search-based frameworks?

\item  \textbf{RQ5.} What is the overhead of FEMOSAA, in terms of execution time, compared to other state-of-the-art frameworks? Which part(s) cause the most overhead? Is the overhead suitable for SAS runtime?

\end{itemize}

\subsection{Verifiability and Methodology}

We have conducted experiments on a dedicated server, which runs Ubuntu Linux 14.04 on an Intel i5 2.8GHz Quad Core processor, 4GB RAM. To separate the adaptation engine and the adaptable software, we used Xen v3.0.3~\cite{xen} as the hypervisor to create a virtualized environment. We have implemented FEMOSAA as the adaptation engine using Java, JDK 1.6, and it is deployed on the \emph{Dom0} of Xen. For setting up all the MOEAs, we use a population size of 100 for 10 generations as the termination criteria; the mutation and crossover rate are 0.1 and 0.9 respectively. In particular, for IBEA, we use the $\epsilon$-indicator and an archive size of 500; while for MOEA/D-STM, we apply Tchebycheff aggregation for creating subproblems where the number of evenly distributed weight vectors is 100 (i.e., 100 subproblems) and the size of each subproblem's neighborhood is 20. Those settings are either common values or have been tailored for runtime optimization in our cases with respect to quality and overhead. All MOEAs used in the experiments are extended from the jMetal Framework~\cite{durillo2011jmetal}. To mitigate interference caused by the adaptation engine, we used one vCPU and 800MB RAM on \emph{Dom0}.



 \subsection{The Subject SAS}
\label{sec:subject}
We deploy two running SAS for our real-time experiments. The two diverse subject SAS aims to examine the generality and applicability of FEMOSAA under different domains. On the micro level, they help to demonstrate how FEMOSAA can be applied to different feature models, dependency structures, environmental factors, dimensions of quality objectives and degrees of objective conflicts:

\begin{itemize}[leftmargin=0.5cm]

\item[---] \textbf{\emph{RUBiS-SAS}}. This adaptable software is a software stack that contains RUBiS~\cite{rubis}, which is a well-known software benchmark that simulates the \emph{eBay} model, and a set of real-world software including Tomcat v6.0.28~\cite{tomcat}, MySQL v3.23.58~\cite{mysql} and Linux kernel v2.6.16.29 running on a configurable guest virtual machine. Ehcache v2.6~\cite{ehcache} is plugged to RUBiS as the cache management module. All the features that can be adapted at runtime are represented by the feature model in Figure~\ref{fig:fm} with a total of 1151 features and a search space of around $1.3\times 10^{16}$ using the elitist chromosome representation (including both valid and invalid solutions). We have also use two distinct workload patterns, a read-write patten and a read-only one, to create diverse runtime behaviors of the SAS. To simulate a realistic time-varying environmental conditions within the capacity of our testbed, we vary the number of clients according to the compressed FIFA98 workload trace~\cite{fifa98}, as shown in Figure~\ref{fig:fifa98}. This setup can generate up to 600 parallel requests, which offers sufficient dynamic and uncertainty. The workload is generated by another machine using the client emulator provided in RUBiS. In general, a heavier workload implies more pressure on the SAS, which reduces the number of effected solutions and thus make the problem harder. The goal is to continually optimize the following two conflicting quality attributes that need to be minimized:

\begin{enumerate}

\item \textbf{\emph{Response Time:}} In the \emph{RUBiS-SAS}, response time is measured as the elapsed time between a request's arrival and its response. To express the quality of \emph{RUBiS-SAS} by the end of a time point, we calculated the average response time for all monitored requests served in the past time interval.

\item \textbf{\emph{Energy Consumption:}} The energy consumed by software systems, which is accounted for 2\% of the global carbon emissions in 2007~\cite{mingay2007green}, is increasingly becoming an important quality concern that has clear conflicts with response time. To measure energy consumption, we leveraged PowerAPI~\cite{powerapi}, a tool that measures the actual energy (watt) incurred by the software's CPU and memory utilization through probing into the sensors of the hardware infrastructure. For each time interval, we computed the average of 30 measurement results of PowerAPI as the energy consumed by the \emph{RUBiS-SAS} for that interval.

\end{enumerate}

 \begin{figure}[t!]
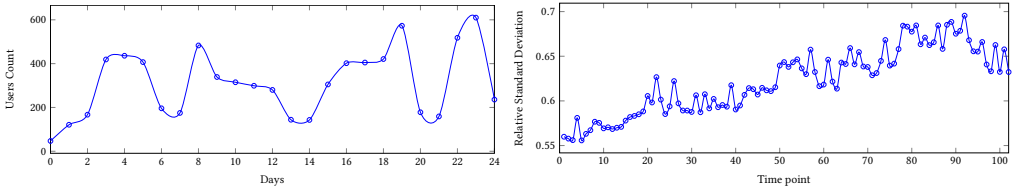

\centering
 \begin{subfigure}[t]{0.48\textwidth}
  \includestandalone[width=\textwidth]{tikz/workload}
    \subcaption{The compressed FIFA98 workload for all patterns in \emph{RUBiS-SAS} (from June 7 to July 1, 1998)}
      \label{fig:fifa98}
    \end{subfigure}
    ~
     \begin{subfigure}[t]{0.48\textwidth}
  \includestandalone[width=\textwidth]{tikz/soa-env}
    \subcaption{The mean relative standard deviation of all concrete services' throughput and cost in \emph{SOA-SAS} }
      \label{fig:rsd}
    \end{subfigure}

  \caption{The changing environment of subject SAS.}

  \end{figure}

These two quality attributes serve as the case of moderate degree of conflict. In other words, it is easier to find solutions that have better response time and energy consumption when the workload is lower. However, as the workload increases, the degree of conflict between them tends to amplify. Formally, the objective functions for \emph{RUBiS-SAS}'s quality attributes are defined as:
\begin{equation}
\label{rt-eq-rubis}
RT(t+1) = f(G_1(t+1), G_2(t+1), ..., G_n(t+1), \delta(t))
\end{equation}
\begin{equation}
\label{e-eq-rubis}
EC(t+1) = g(G_1(t+1), G_2(t+1), ..., G_n(t+1), \delta(t))
\end{equation}
\noindent whereby $G_1(t+1), G_2(t+1), ..., G_n(t+1)$ denote the genes and their selected values in a possible adaptation solution; $\delta(t)$ denotes a set of most recent environmental factors, e.g., workload in this paper. Given an adaptation solution, $RT(t+1)$ and $EC(t+1)$ are the expected fitness values for response time and energy consumption respectively. 

In \emph{RUBiS-SAS}, we adopt the machine learning based model from~\cite{Chen:2013}\cite{Chen:2014:ucc}\cite{Chen:2015:tse} to build the objective functions for \emph{RUBiS-SAS}, since it relies on few assumptions of the application domain. By using the actual data of quality performance, software status and environment monitored from the SAS, this model can be continually updated at runtime to provide sufficient accuracy~\cite{Chen:2015:tse}. To stabilize the objective functions in this work, we pre-trained these models by monitoring the data from SAS under random workload for 120 intervals, after which the models are gradually and dynamically updated at runtime. This step is similar to the testing phase before the actual production deployment happened in the industry. As we will show for the next subject SAS, other modeling approaches, e.g., analytical~\cite{analytical} or simulation based~\cite{6392599}, can be easily applied as long as they are compatible with the genes identified by FEMOSAA.

To emulate the behaviors of running software system, we run the SAS, empowered with FEMOSAA, under the entire FIFA98 workload trace where the sampling interval is 120s for a total of 102 timesteps, leading to around 5 hours per experiment run including the emulated end-users' thinking time. For all experiments, we trigger an optimization run by the end of each interval.



\item[---] \textbf{\emph{SOA-SAS}}. This is a Service Oriented Architecture (SOA) based adaptable software derived from~\cite{wada2012e3}. At the highest level, it is composed by 5 abstract services connected in parallel or in sequence, as shown in Figure~\ref{fig:soa-arch}. Each abstract service can select up to 2 or 5 concrete services, which are redundant parallel and associated with different quality values. Each concrete service could have up to 10 concurrent instances with the same quality of throughput and cost. Those abstract services, concrete services and their replicas are all considered as features of \emph{SOA-SAS} (e.g., a fragment in Figure~\ref{fig:soa-arch}), creating a total of 221 features in the feature model with a search space of around $5.6\times 10^{18}$ using the elitist chromosome representation (including both valid and invalid solutions). We have also placed various categorical and numeric dependencies on the feature model. To simulate dynamic and uncertainty under time-varying environmental conditions, we amended the throughput and cost of certain concrete services at each timestep, i.e., changing their diversity level according to Gaussion distribution, as shown in Figure~\ref{fig:rsd}. Generally, more diverse concrete services implies more sparse adaptation solutions, which reduces the number of effected solutions and thus make the problem harder.  Here, we consider the following conflicting objectives to be maximized and minimized for the entire composition:

\begin{enumerate}

\item \textbf{\emph{Throughput:}} In the \emph{SOA-SAS}, throughput of each concrete service is its maximum capacity under normal operation, expressed as number of requests per second. The throughput of the entire composition is calculated via an aggregate analytical model as shown in Eq.~\ref{t-eq-soa}.

\item \textbf{\emph{Cost:}} Each concrete service, when utilized, comes with a monetary cost. The cost of the entire composition is again calculated via an aggregate analytical model as shown in Eq.~\ref{c-eq-soa}.

\end{enumerate}

 \begin{figure}[t!]
\centering
 \begin{subfigure}[t]{0.5\textwidth}
  \includegraphics[width=\textwidth]{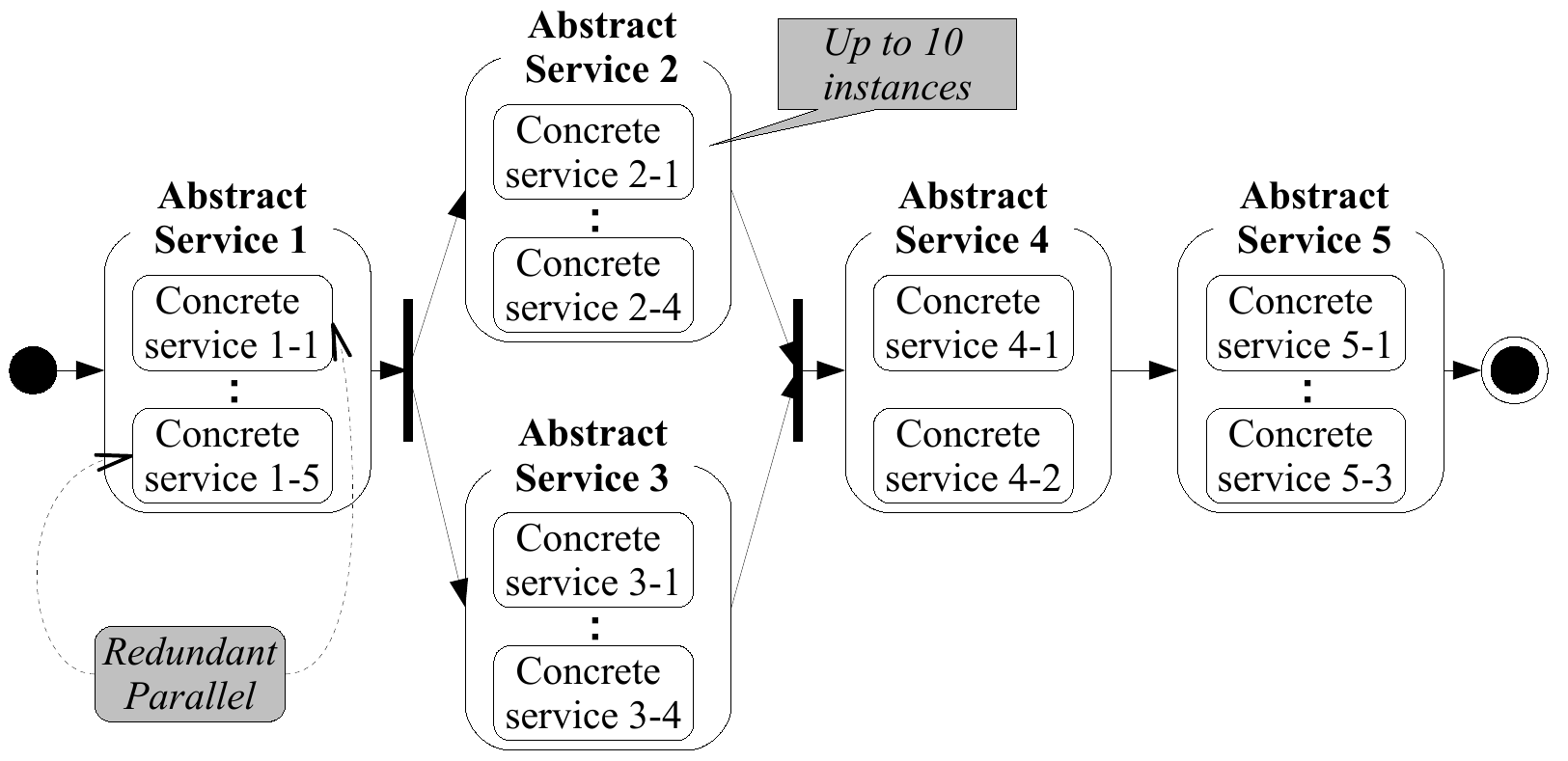}
    \subcaption{Composition workflow}
    \end{subfigure}
    ~
     \begin{subfigure}[t]{0.5\textwidth}
  \includegraphics[width=\textwidth]{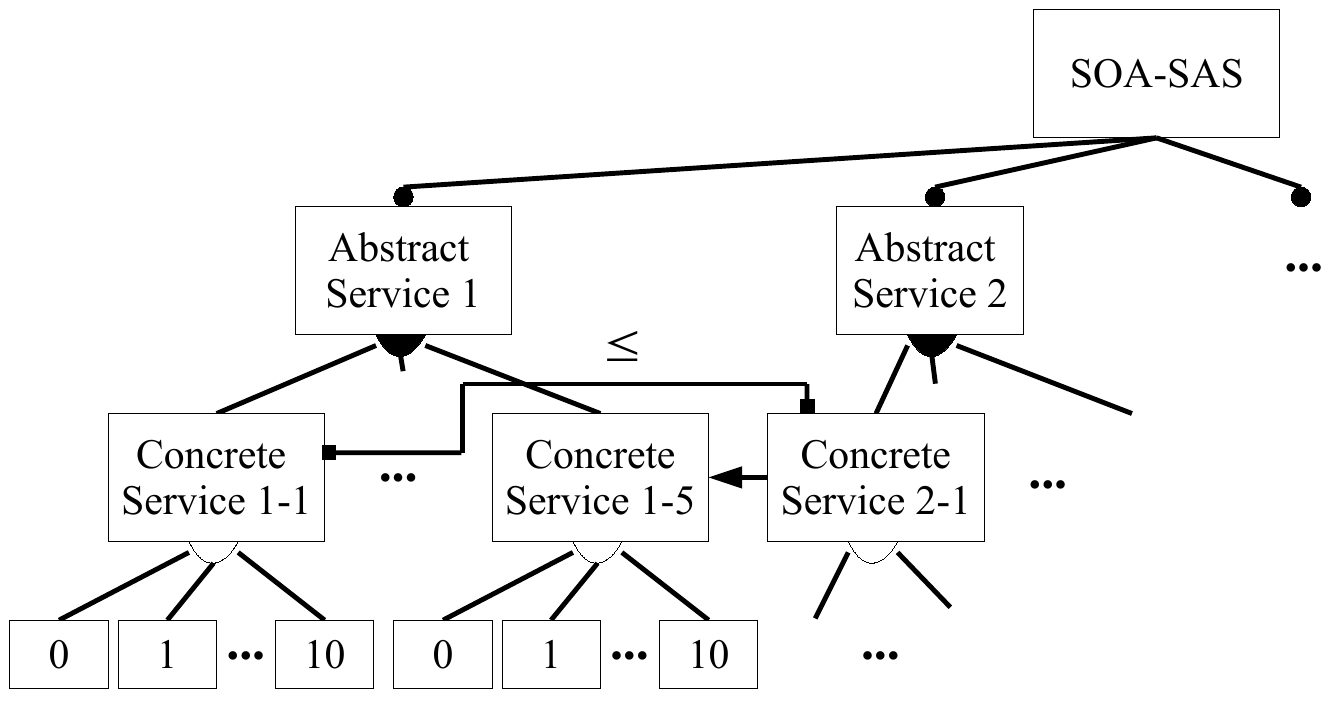}
    \subcaption{A fragment of the feature model}
    \end{subfigure}

  \caption{The architecture of \emph{SOA-SAS}.}
  \label{fig:soa-arch}
  \end{figure}


As suggested in~\cite{wada2012e3}, the basic throughput and cost for each concrete service in \emph{SOA-SAS} were specified following two different Gaussian distributions such that a concrete service with higher throughput would have higher cost too. In \emph{SOA-SAS}, we leverage analytical model to define the objective functions:
\begin{equation}
\label{t-eq-soa}
T(t+1) = \min \{a \in A : \sum_{i=1}^{a}{G_i(t+1) \times T_i(t)}\}
\end{equation}
\begin{equation}
\label{c-eq-soa}
C(t+1) = \sum_{i=1}^{n}{G_i(t+1) \times C_i(t)}
\end{equation}
where $G_i(t+1)$ is the $i$th gene (a concrete service) and its selected value (the actual number of instances for that concrete service) in a possible adaptation solution for the total of $n$ genes. $T_i(t)$ and $C_i(t)$ are the related throughout and cost value, respectively. $a$ refers to the genes that is associated with a given abstract service, denoted by $A$. Since the analytical model is customizable, we have forced the objective functions to produce a worst possible fitness value for any invalid solutions, creating a stronger pressure for them to be eliminated in the classic, non-dependency aware approaches.

Similar to \emph{RUBiS-SAS}, there is a total of 102 timesteps and we trigger an optimization run by the end of each interval.

\end{itemize}

As discussed, FEMOSAA finds elitist features and extracts their dependencies at design time, after which the outcomes are passed to the MOEA and knee selection for runtime optimization. 


 \subsection{The Metrics}

We consider the following metrics to evaluate various aspects of FEMOSAA:
\begin{itemize}[leftmargin=0.5cm]

\item[---] \textbf{Individual Quality Attribute:} In addition to the detailed plots of the results, we also report on the Geometric Mean (GM) for all the observed objective values over 102 intervals, as GM tends to be more resilient to the outliers than the arithmetic mean while still can be influenced by those outliers. This is important because outliers on the achieved quality are common in real-world scenarios and thus they cannot be eliminated, but we need to prevent them from strongly dominating the overall result.

\item[---] \textbf{Aggregate Quality of SAS:} To compare the overall quality for both objectives and provide an overall assessment, we apply a modified Hypervolume (HV)~\cite{797969}\cite{icse-nier} for two objectives and Euclidean Distance (ED)~\cite{Wang:2016} on GMs to assess the balance in trade-off and the extent to which both objectives are optimized, respectively. The GMs are normalized before computing those metrics so that they range from 0 to 1. Hence, the modified HV is computed as:
\begin{equation}
HV=\prod_{i=1}^{n}{(1 -  \frac {GM_{i}-GM_{i,b}} {GM_{i,w}-GM_{i,b}})}
\end{equation} 
and ED can be calculated by: 
\begin{equation}
ED=\frac{1}{n} \times \sqrt{\sum_{i=1}^{n}{(\frac {GM_{i}-GM_{i,b}} {GM_{i,w}-GM_{i,b}} - 0)^2}}
\end{equation} 
where $n$ is the number of objectives and $GM_{i}$ is the GM for the $i$th objective; $GM_{i,b}$ and $GM_{i,w}$ are the best and worst GMs that we observed for the $i$th objective, respectively. Notably, we have converted the maximizing objective (e.g., throughput) into a minimizing one by inverting the results.

\item[---] \textbf{Percentage of Valid Solutions Found:} For all the intervals, we also compare the average percentage of valid adaptation solutions found in the final population.

\item[---] \textbf{Running Overhead:} We report on the mean running overhead, in terms of the execution time, over all the intervals in the experiment runs.


\end{itemize}

To confirm statistical significance of the comparisons on the quality attributes, we performed \emph{Wilcoxon Signed-Rank} test (two-tailed) for all comparisons between FEMOSAA (or FEMOSAA-N) and the others, as our data does not follow Gaussian distributions. We use 95\% as the confidence interval ($\alpha=0.05$), which means that, if the test produces a $p$ that is smaller than 0.05, then we can reject the null hypothesis $H_0$, which states that the given two approaches cannot be statistically distinguished when optimizing a quality attribute of the SAS. The effect size for each test is also reported and we follow the categories in~\cite{Kampenes20071073} to measure the meaningfulness of effect size.



\subsection{Effectiveness of FEMOSAA} 
\label{sec:exp-effectiveness}

To ensure generality of the evaluation, we run each of the three MOEAs for optimizing the SAS under each case which results in a total of 9 scenarios. Further, to evaluate the effectiveness of FEMOSAA in each scenario, we compare FEMOSAA with numbers of its variants:

\textbf{FEMOSAA-K}\textemdash This is similar to our FEMOSAA except that it relies on the classic, non-dependency aware operators. Therefore, if the final population contains invalid solutions, it automatically filters them and work on the valid ones only. When no valid solutions found in the final non-dominated set, it corrects the invalid solutions via dependency aware mutation operator. FEMOSAA-K aims to examine whether the specifically tailored dependency aware operators can create benefit over the classic operators that are widely-used in SBSE.

\textbf{FEMOSAA-D}\textemdash Another variant of FEMOSAA without knee selection. Hence, in the final non-dominated set, one solution is randomly selected for adaptation. FEMOSAA-D aims to examine the importance of considering knee in runtime SAS optimization. Notably, the use of randomized baseline is strongly recommended by existing SBSE community~\cite{arcuri2011practical}.

\textbf{FEMOSAA-N}\textemdash This variant neither considers dependency nor knee selection in the evolutionary optimization. Hence, it uses the same ad hoc strategies from FEMOSAA-D and FEMOSAA-K. FEMOSAA-N is designed to evaluate the combinatorial benefit of dependency aware operators and the knee selection.

\textbf{FEMOSAA-0/1}\textemdash A baseline variant, built from FEMOSAA-N, that exploits the conventional binary chromosome representation using all the features from the feature model. Thus, it works under a search space of $2^{1151}$ and $2^{221}$ for the two subject SAS, as opposed to the search space of $1.3\times 10^{16}$ and $5.6\times 10^{18}$ when using the elitist chromosome representations. FEMOSAA-0/1 neither considers dependency nor knee in the evolutionary optimization. Note that if more than one members in a $\mathcal{XOR}$ group are selected, we then randomly choose one among them to create a computable solution for the fitness functions. FEMOSAA-0/1 aims to evaluate whether the elitist chromosome representation outperforms the commonly used binary one.


We report the results for all the cases from Figures~\ref{fig:moead-rw-quality} to~\ref{fig:ibea-soa-quality}, we also plot the achieved quality for all timesteps with respect to the environmental conditions in Figures~\ref{fig:femosaa-details} and~\ref{fig:soa-femosaa-details}, as well as the results of an example optimization run in Figure~\ref{fig:femosaa-pred-details}. Note that, in Figures~\ref{fig:femosaa-details} and~\ref{fig:soa-femosaa-details}, the area near the bottom-left line of the cube is the ideal area; the closer the points converge to that line, the better the overall result is.

\subsubsection{Evaluating the elitist chromosome representation} 
\label{eval-chro}
\begin{figure}[!t]
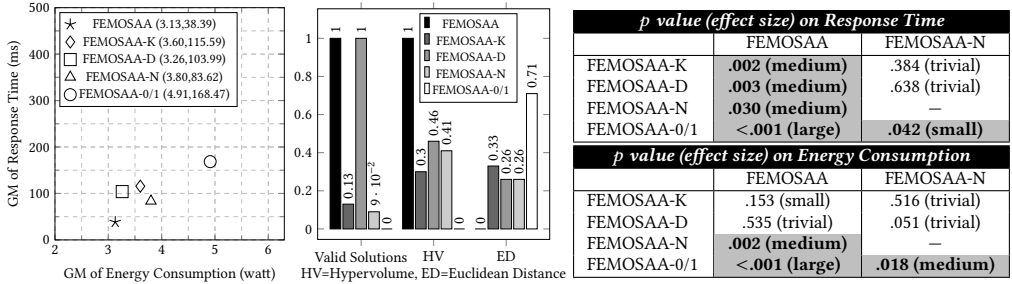
 
   \begin{minipage}{\textwidth}
  \begin{minipage}[t]{0.45\textwidth}
    \centering
\includestandalone[width=0.6\textwidth]{tikz/moead-rw-quality}
~
\includestandalone[width=0.6\textwidth]{tikz/moead-rw-sta}
  \end{minipage}
  \hspace{0.3cm}
  \begin{minipage}[b]{0.55\textwidth}
    \centering
    \scriptsize     
    \begin{tabularx}{0.76\textwidth}{|p{1.5cm}|Y|Y|}\hline
     \multicolumn{3}{|c|}{\cellcolor{black}\textbf{\emph{\textcolor{white}{$p$ value (effect size) on Response Time}}}} \\\hline
      & FEMOSAA & FEMOSAA-N\\ \hline
        FEMOSAA-K&\cellcolor{gray!50}\bfseries.002 (medium)& .384 (trivial) \\
        FEMOSAA-D&\cellcolor{gray!50}\bfseries.003 (medium)& .638 (trivial)  \\
        FEMOSAA-N&\cellcolor{gray!50}\bfseries.030 (medium)& --- \\
        FEMOSAA-0/1&\cellcolor{gray!50}\bfseries $<$.001 (large)&\cellcolor{gray!50}\bfseries.042 (small) \\ \hline
      \end{tabularx} 
            \vspace{0.12cm}     
     \begin{tabularx}{0.76\textwidth}{|p{1.5cm}|Y|Y|}\hline
     \multicolumn{3}{|c|}{\cellcolor{black}\textbf{\emph{\textcolor{white}{$p$ value (effect size) on Energy Consumption}}}} \\\hline
      & FEMOSAA & FEMOSAA-N\\ \hline
          FEMOSAA-K & .153 (small)  & .516 (trivial) \\
        FEMOSAA-D & .535 (trivial)  & .051 (trivial) \\
         FEMOSAA-N&\cellcolor{gray!50}\bfseries.002 (medium)& ---  \\
        FEMOSAA-0/1 & \cellcolor{gray!50}\bfseries $<$.001 (large) & \cellcolor{gray!50}\bfseries.018 (medium) \\ \hline
      \end{tabularx}  
    \end{minipage} 
     \vspace{-0.8cm}  
      \captionof{figure}{Results with MOEA/D-STM under read-write pattern on RUBiS-SAS. (GM denotes Geometric Mean. The significant statistics of comparisons, i.e., $p<$ 0.05, are highlighted and shown in bold.)}
       \label{fig:moead-rw-quality}
  \end{minipage}
  \vspace{-0.5cm}
\end{figure}

\begin{figure}[!t]
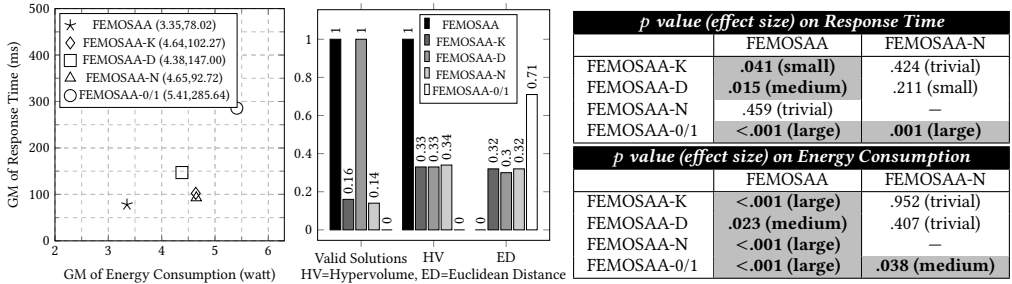
 
   \begin{minipage}{\textwidth}
  \begin{minipage}[t]{0.45\textwidth}
    \centering
\includestandalone[width=0.6\textwidth]{tikz/ibea-rw-quality}
~
\includestandalone[width=0.6\textwidth]{tikz//ibea-rw-sta}
  \end{minipage}
  \hspace{0.3cm}
  \begin{minipage}[b]{0.55\textwidth}
    \centering
    \scriptsize     
    \begin{tabularx}{0.76\textwidth}{|p{1.5cm}|Y|Y|}\hline
     \multicolumn{3}{|c|}{\cellcolor{black}\textbf{\emph{\textcolor{white}{$p$ value (effect size) on Response Time}}}} \\\hline
      & FEMOSAA & FEMOSAA-N\\ \hline
        FEMOSAA-K&\cellcolor{gray!50}\bfseries.041 (small)& .424 (trivial) \\
        FEMOSAA-D&\cellcolor{gray!50}\bfseries.015 (medium)& .211 (small)  \\
        FEMOSAA-N&.459 (trivial)& --- \\
        FEMOSAA-0/1&\cellcolor{gray!50}\bfseries $<$.001 (large)&\cellcolor{gray!50}\bfseries.001 (large) \\ \hline
      \end{tabularx} 
            \vspace{0.12cm}     
     \begin{tabularx}{0.76\textwidth}{|p{1.5cm}|Y|Y|}\hline
     \multicolumn{3}{|c|}{\cellcolor{black}\textbf{\emph{\textcolor{white}{$p$ value (effect size) on Energy Consumption}}}} \\\hline
      & FEMOSAA & FEMOSAA-N\\ \hline
          FEMOSAA-K & \cellcolor{gray!50}\bfseries$<$.001 (large)  & .952 (trivial) \\
        FEMOSAA-D &\cellcolor{gray!50}\bfseries.023 (medium) & .407 (trivial) \\
         FEMOSAA-N& \cellcolor{gray!50}\bfseries$<$.001 (large)& ---  \\
        FEMOSAA-0/1 & \cellcolor{gray!50}\bfseries $<$.001 (large) & \cellcolor{gray!50}\bfseries.038 (medium) \\ \hline
      \end{tabularx}  
    \end{minipage} 
     \vspace{-0.8cm}  
    \captionof{figure}{Results with IBEA under read-write pattern on RUBiS-SAS. (GM denotes Geometric Mean. The significant statistics of comparisons, i.e., $p<$ 0.05, are highlighted and shown in bold.)}
     \label{fig:ibea-rw-quality}
  \end{minipage}
\end{figure}

\begin{figure}[!t]
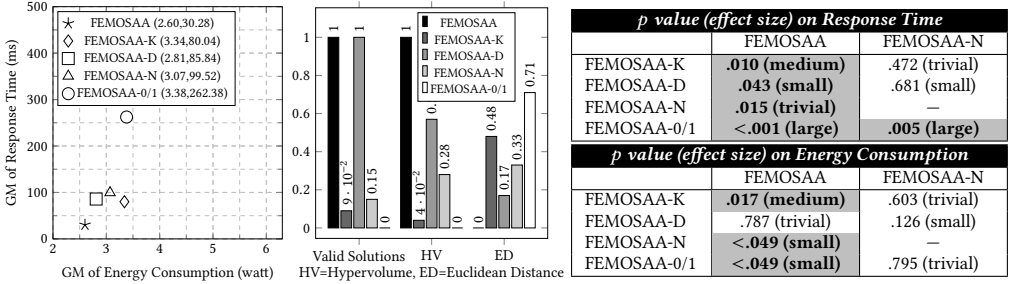
 
   \begin{minipage}{\textwidth}
  \begin{minipage}[t]{0.45\textwidth}
    \centering
\includestandalone[width=0.6\textwidth]{tikz/moead-r-quality}
~
\includestandalone[width=0.6\textwidth]{tikz/moead-r-sta}
  \end{minipage}
  \hspace{0.3cm}
  \begin{minipage}[b]{0.55\textwidth}
    \centering
    \scriptsize     
    \begin{tabularx}{0.76\textwidth}{|p{1.5cm}|Y|Y|}\hline
     \multicolumn{3}{|c|}{\cellcolor{black}\textbf{\emph{\textcolor{white}{$p$ value (effect size) on Response Time}}}} \\\hline
      & FEMOSAA & FEMOSAA-N\\ \hline
        FEMOSAA-K&\cellcolor{gray!50}\bfseries.010 (medium)& .472 (trivial) \\
        FEMOSAA-D&\cellcolor{gray!50}\bfseries.043 (small)& .681 (small)  \\
        FEMOSAA-N&\cellcolor{gray!50}\bfseries.015 (trivial)& --- \\
        FEMOSAA-0/1&\cellcolor{gray!50}\bfseries $<$.001 (large)&\cellcolor{gray!50}\bfseries.005 (large) \\ \hline
      \end{tabularx} 
            \vspace{0.12cm}     
     \begin{tabularx}{0.76\textwidth}{|p{1.5cm}|Y|Y|}\hline
     \multicolumn{3}{|c|}{\cellcolor{black}\textbf{\emph{\textcolor{white}{$p$ value (effect size) on Energy Consumption}}}} \\\hline
      & FEMOSAA & FEMOSAA-N\\ \hline
          FEMOSAA-K &\cellcolor{gray!50}\bfseries.017 (medium)& .603 (trivial) \\
        FEMOSAA-D &.787 (trivial) & .126 (small) \\
         FEMOSAA-N& \cellcolor{gray!50}\bfseries$<$.049 (small)& ---  \\
        FEMOSAA-0/1 & \cellcolor{gray!50}\bfseries $<$.049 (small) & .795 (trivial) \\ \hline
      \end{tabularx}  
    \end{minipage} 
     \vspace{-0.8cm}  
    \captionof{figure}{Results with MOEA/D-STM under read-only pattern on RUBiS-SAS. (GM denotes Geometric Mean. The significant statistics of comparisons, i.e., $p<$ 0.05, are highlighted and shown in bold.)}
     \label{fig:moead-r-quality}
  \end{minipage}
    \vspace{-0.5cm}
\end{figure}

\begin{figure}[!t]
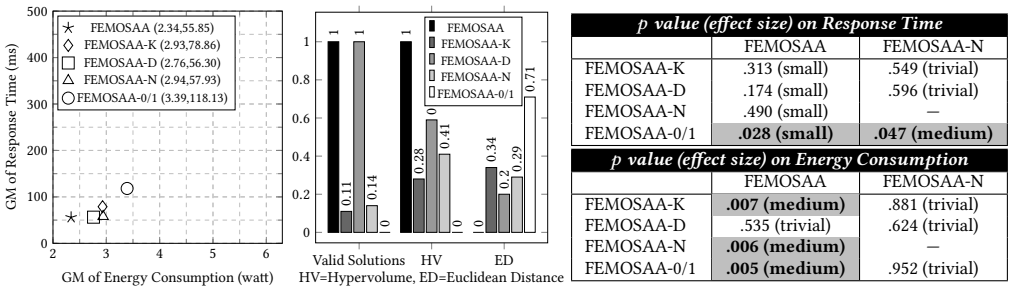
 
   \begin{minipage}{\textwidth}
  \begin{minipage}[t]{0.45\textwidth}
    \centering
\includestandalone[width=0.6\textwidth]{tikz/ibea-r-quality}
~
\includestandalone[width=0.6\textwidth]{tikz/ibea-r-sta}
  \end{minipage}
  \hspace{0.3cm}
  \begin{minipage}[b]{0.55\textwidth}
    \centering
    \scriptsize     
    \begin{tabularx}{0.76\textwidth}{|p{1.5cm}|Y|Y|}\hline
     \multicolumn{3}{|c|}{\cellcolor{black}\textbf{\emph{\textcolor{white}{$p$ value (effect size) on Response Time}}}} \\\hline
      & FEMOSAA & FEMOSAA-N\\ \hline
        FEMOSAA-K&.313 (small)& .549 (trivial) \\
        FEMOSAA-D&.174 (small)& .596 (trivial)  \\
        FEMOSAA-N&.490 (small)& --- \\
        FEMOSAA-0/1&\cellcolor{gray!50}\bfseries.028 (small)&\cellcolor{gray!50}\bfseries.047 (medium) \\ \hline
      \end{tabularx} 
            \vspace{0.12cm}     
     \begin{tabularx}{0.76\textwidth}{|p{1.5cm}|Y|Y|}\hline
     \multicolumn{3}{|c|}{\cellcolor{black}\textbf{\emph{\textcolor{white}{$p$ value (effect size) on Energy Consumption}}}} \\\hline
      & FEMOSAA & FEMOSAA-N\\ \hline
          FEMOSAA-K &\cellcolor{gray!50}\bfseries.007 (medium)& .881 (trivial) \\
        FEMOSAA-D &.535 (trivial) & .624 (trivial) \\
         FEMOSAA-N&\cellcolor{gray!50}\bfseries.006 (medium)& ---  \\
        FEMOSAA-0/1 & \cellcolor{gray!50}\bfseries.005 (medium) &.952 (trivial) \\ \hline
      \end{tabularx}  
    \end{minipage} 
     \vspace{-0.8cm}  
    \captionof{figure}{Results with IBEA under read-only pattern on RUBiS-SAS. (GM denotes Geometric Mean. The significant statistics of comparisons, i.e., $p<$ 0.05, are highlighted and shown in bold.)}
     \label{fig:ibea-r-quality}
  \end{minipage}
\end{figure}

\begin{figure}[!t]
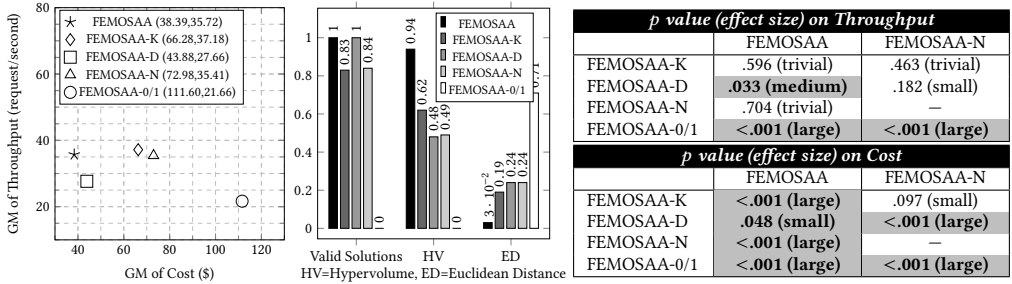
 
   \begin{minipage}{\textwidth}
  \begin{minipage}[t]{0.45\textwidth}
    \centering
\includestandalone[width=0.6\textwidth]{tikz/moead-sas-quality}
~
\includestandalone[width=0.6\textwidth]{tikz/moead-sas-sta}
  \end{minipage}
  \hspace{0.3cm}
  \begin{minipage}[b]{0.55\textwidth}
    \centering
    \scriptsize     
    \begin{tabularx}{0.76\textwidth}{|p{1.5cm}|Y|Y|}\hline
     \multicolumn{3}{|c|}{\cellcolor{black}\textbf{\emph{\textcolor{white}{$p$ value (effect size) on Throughput}}}} \\\hline
      & FEMOSAA & FEMOSAA-N\\ \hline
        FEMOSAA-K&.596 (trivial)& .463 (trivial) \\
        FEMOSAA-D&\cellcolor{gray!50}\bfseries.033 (medium)& .182 (small)  \\
        FEMOSAA-N&.704 (trivial)& --- \\
        FEMOSAA-0/1&\cellcolor{gray!50}\bfseries$<$.001 (large)&\cellcolor{gray!50}\bfseries$<$.001 (large) \\ \hline
      \end{tabularx} 
            \vspace{0.12cm}     
     \begin{tabularx}{0.76\textwidth}{|p{1.5cm}|Y|Y|}\hline
     \multicolumn{3}{|c|}{\cellcolor{black}\textbf{\emph{\textcolor{white}{$p$ value (effect size) on Cost}}}} \\\hline
      & FEMOSAA & FEMOSAA-N\\ \hline
          FEMOSAA-K &\cellcolor{gray!50}\bfseries$<$.001 (large)& .097 (small) \\
        FEMOSAA-D &\cellcolor{gray!50}\bfseries.048 (small)&\cellcolor{gray!50}\bfseries$<$.001 (large)\\
         FEMOSAA-N&\cellcolor{gray!50}\bfseries$<$.001 (large)& ---  \\
        FEMOSAA-0/1 & \cellcolor{gray!50}\bfseries$<$.001 (large) &\cellcolor{gray!50}\bfseries$<$.001 (large) \\ \hline
      \end{tabularx}  
    \end{minipage} 
     \vspace{-0.8cm}  
    \captionof{figure}{Results with MOEA/D-STM on SOA-SAS. (GM denotes Geometric Mean. The significant statistics of comparisons, i.e., $p<$ 0.05, are highlighted and shown in bold.)}
     \label{fig:moead-soa-quality}
  \end{minipage}
    \vspace{-0.5cm}
\end{figure}

\begin{figure}[!t]
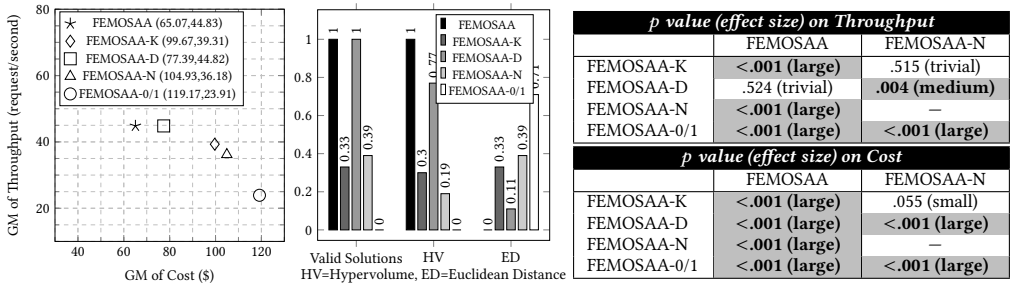
 
   \begin{minipage}{\textwidth}
  \begin{minipage}[t]{0.45\textwidth}
    \centering
\includestandalone[width=0.6\textwidth]{tikz/ibea-sas-quality}
~
\includestandalone[width=0.6\textwidth]{tikz/ibea-sas-sta}
  \end{minipage}
  \hspace{0.3cm}
  \begin{minipage}[b]{0.55\textwidth}
    \centering
    \scriptsize     
    \begin{tabularx}{0.76\textwidth}{|p{1.5cm}|Y|Y|}\hline
     \multicolumn{3}{|c|}{\cellcolor{black}\textbf{\emph{\textcolor{white}{$p$ value (effect size) on Throughput}}}} \\\hline
      & FEMOSAA & FEMOSAA-N\\ \hline
        FEMOSAA-K&\cellcolor{gray!50}\bfseries$<$.001 (large)& .515 (trivial) \\
        FEMOSAA-D&.524 (trivial)& \cellcolor{gray!50}\bfseries.004 (medium)  \\
        FEMOSAA-N& \cellcolor{gray!50}\bfseries$<$.001 (large)& --- \\
        FEMOSAA-0/1&\cellcolor{gray!50}\bfseries$<$.001 (large)&\cellcolor{gray!50}\bfseries$<$.001 (large) \\ \hline
      \end{tabularx} 
            \vspace{0.12cm}     
     \begin{tabularx}{0.76\textwidth}{|p{1.5cm}|Y|Y|}\hline
     \multicolumn{3}{|c|}{\cellcolor{black}\textbf{\emph{\textcolor{white}{$p$ value (effect size) on Cost}}}} \\\hline
      & FEMOSAA & FEMOSAA-N\\ \hline
          FEMOSAA-K &\cellcolor{gray!50}\bfseries$<$.001 (large)& .055 (small) \\
        FEMOSAA-D &\cellcolor{gray!50}\bfseries$<$.001 (large)&\cellcolor{gray!50}\bfseries$<$.001 (large)\\
         FEMOSAA-N&\cellcolor{gray!50}\bfseries$<$.001 (large)& ---  \\
        FEMOSAA-0/1 & \cellcolor{gray!50}\bfseries$<$.001 (large) &\cellcolor{gray!50}\bfseries$<$.001 (large) \\ \hline
      \end{tabularx}  
    \end{minipage} 
     \vspace{-0.8cm}  
    \captionof{figure}{Results with IBEA on SOA-SAS. (GM denotes Geometric Mean. The significant statistics of comparisons, i.e., $p<$ 0.05, are highlighted and shown in bold.)}
    \label{fig:ibea-soa-quality}
  \end{minipage}
\end{figure}

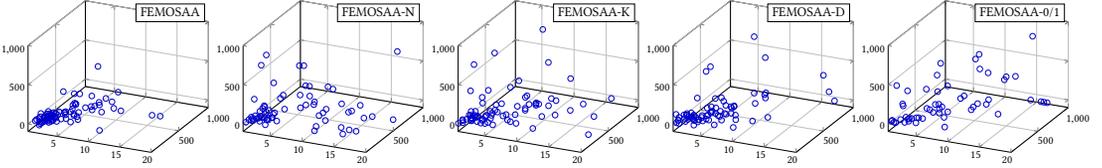
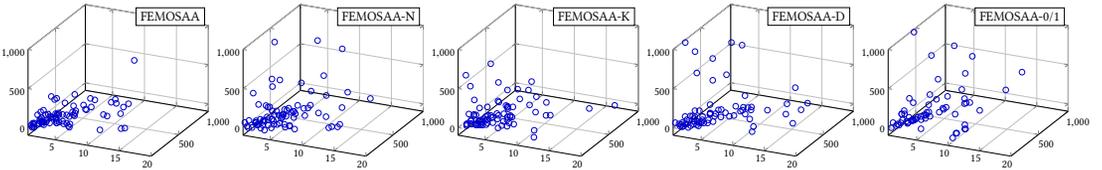
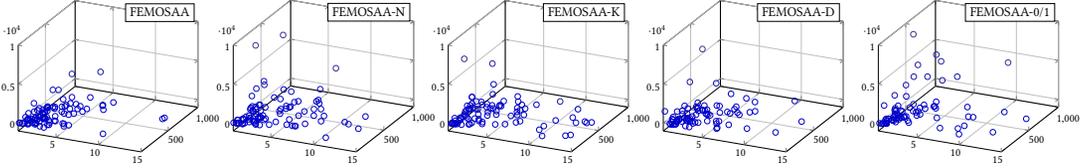
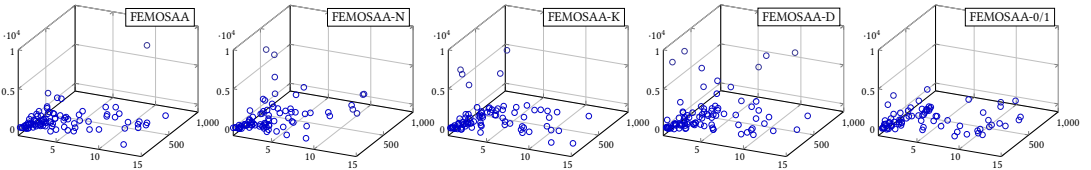
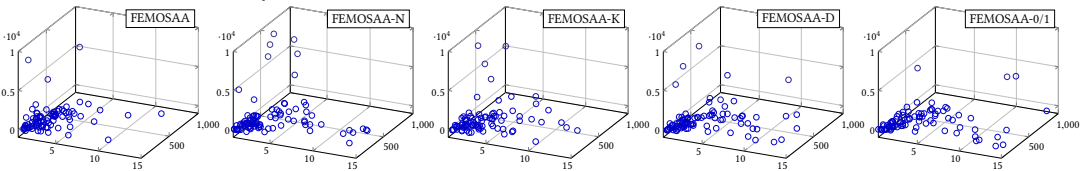
\begin{figure}[t!]

\hspace{-0.4cm}
\centering
 \begin{subfigure}[t]{0.2\textwidth}
    \begin{tikzpicture}[scale=0.35]
    \begin{axis}[point meta max=100000,grid=both,zmax=1000,xmax=20,ymax=1000,yticklabel style = {font=\Large},xticklabel style = {font=\Large},zticklabel style = {font=\Large},legend style={empty legend,font=\LARGE}]
      \addplot3[only marks,scatter,mark=o,mark size=3pt] file {moead.txt};
      \addlegendentry{FEMOSAA}
    \end{axis}
  \end{tikzpicture}
  \vspace{-0.5cm}
    \end{subfigure}
     \begin{subfigure}[t]{0.2\textwidth}
    \begin{tikzpicture}[scale=0.35]
    \begin{axis}[point meta max=100000,grid=both,zmax=1000,xmax=20,ymax=1000,yticklabel style = {font=\Large},xticklabel style = {font=\Large},zticklabel style = {font=\Large},legend style={empty legend,font=\LARGE}]
      \addplot3[only marks,scatter,mark=o,mark size=3pt] file {moead-plain.txt};
      \addlegendentry{FEMOSAA-N}
    \end{axis}
  \end{tikzpicture}
    \vspace{-0.5cm}
    \end{subfigure}
     \begin{subfigure}[t]{0.2\textwidth}
    \begin{tikzpicture}[scale=0.35]
    \begin{axis}[point meta max=100000,grid=both,zmax=1000,xmax=20,ymax=1000,yticklabel style = {font=\Large},xticklabel style = {font=\Large},zticklabel style = {font=\Large},legend style={empty legend,font=\LARGE}]
      \addplot3[only marks,scatter,mark=o,mark size=3pt] file {moead-k.txt};
      \addlegendentry{FEMOSAA-K}
    \end{axis}
  \end{tikzpicture}
    \vspace{-0.5cm}
    \end{subfigure}
     \begin{subfigure}[t]{0.2\textwidth}
    \begin{tikzpicture}[scale=0.35]
    \begin{axis}[point meta max=100000,grid=both,zmax=1000,xmax=20,ymax=1000,yticklabel style = {font=\Large},xticklabel style = {font=\Large},zticklabel style = {font=\Large},legend style={empty legend,font=\LARGE}]
      \addplot3[only marks,scatter,mark=o,mark size=3pt] file {moead-d.txt};
      \addlegendentry{FEMOSAA-D}
    \end{axis}
  \end{tikzpicture}
    \vspace{-0.5cm}
    \end{subfigure}
     \begin{subfigure}[t]{0.2\textwidth}
    \begin{tikzpicture}[scale=0.35]
    \begin{axis}[point meta max=100000,grid=both,zmax=1000,xmax=20,ymax=1000,yticklabel style = {font=\Large},xticklabel style = {font=\Large},zticklabel style = {font=\Large},legend style={empty legend,font=\LARGE}]
      \addplot3[only marks,scatter,mark=o,mark size=3pt] file {moead-01.txt};
      \addlegendentry{FEMOSAA-0/1}
    \end{axis}
  \end{tikzpicture}
    \vspace{-0.5cm}
    \end{subfigure}

    \vspace{-0.5cm}
    \begin{subfigure}[t]{\textwidth}
     \subcaption{MOEA/D-STM and read-write pattern [the axis from left to right are Response Time (ms), Energy Consumption (watt) and Workload (requests/min)]}
      \end{subfigure}
      
      \hspace{-0.4cm}
\centering
       \begin{subfigure}[t]{0.2\textwidth}
    \begin{tikzpicture}[scale=0.35]
    \begin{axis}[point meta max=100000,grid=both,zmax=1000,xmax=20,ymax=1000,yticklabel style = {font=\Large},xticklabel style = {font=\Large},zticklabel style = {font=\Large},legend style={empty legend,font=\LARGE}]
      \addplot3[only marks,scatter,mark=o,mark size=3pt] file {nsgaii.txt};
        \addlegendentry{FEMOSAA}
    \end{axis}
  \end{tikzpicture}
     \vspace{-0.5cm}
    \end{subfigure}
     \begin{subfigure}[t]{0.2\textwidth}
    \begin{tikzpicture}[scale=0.35]
    \begin{axis}[point meta max=100000,grid=both,zmax=1000,xmax=20,ymax=1000,yticklabel style = {font=\Large},xticklabel style = {font=\Large},zticklabel style = {font=\Large},legend style={empty legend,font=\LARGE}]
      \addplot3[only marks,scatter,mark=o,mark size=3pt] file {nsgaii-plain.txt};
        \addlegendentry{FEMOSAA-N}
    \end{axis}
  \end{tikzpicture}
     \vspace{-0.5cm}
    \end{subfigure}
     \begin{subfigure}[t]{0.2\textwidth}
    \begin{tikzpicture}[scale=0.35]
    \begin{axis}[point meta max=100000,grid=both,zmax=1000,xmax=20,ymax=1000,yticklabel style = {font=\Large},xticklabel style = {font=\Large},zticklabel style = {font=\Large},legend style={empty legend,font=\LARGE}]
      \addplot3[only marks,scatter,mark=o,mark size=3pt] file {nsgaii-k.txt};
        \addlegendentry{FEMOSAA-K}
    \end{axis}
  \end{tikzpicture}
     \vspace{-0.5cm}
    \end{subfigure}
     \begin{subfigure}[t]{0.2\textwidth}
    \begin{tikzpicture}[scale=0.35]
    \begin{axis}[point meta max=100000,grid=both,zmax=1000,xmax=20,ymax=1000,yticklabel style = {font=\Large},xticklabel style = {font=\Large},zticklabel style = {font=\Large},legend style={empty legend,font=\LARGE}]
      \addplot3[only marks,scatter,mark=o,mark size=3pt] file {nsgaii-d.txt};
        \addlegendentry{FEMOSAA-D}
    \end{axis}
  \end{tikzpicture}
     \vspace{-0.5cm}
    \end{subfigure}
     \begin{subfigure}[t]{0.2\textwidth}
    \begin{tikzpicture}[scale=0.35]
    \begin{axis}[point meta max=100000,grid=both,zmax=1000,xmax=20,ymax=1000,yticklabel style = {font=\Large},xticklabel style = {font=\Large},zticklabel style = {font=\Large},legend style={empty legend,font=\LARGE}]
      \addplot3[only marks,scatter,mark=o,mark size=3pt] file {nsgaii-01.txt};
        \addlegendentry{FEMOSAA-0/1}
    \end{axis}
  \end{tikzpicture}
     \vspace{-0.5cm}
    \end{subfigure}
    
     \vspace{-0.5cm}
        \begin{subfigure}[t]{\textwidth}
     \subcaption{NSGA-II and read-write pattern [the axis from left to right are Response Time (ms), Energy Consumption (watt) and Workload (requests/min)]}
      \end{subfigure}
      
      \hspace{-0.4cm}
\centering
             \begin{subfigure}[t]{0.2\textwidth}
    \begin{tikzpicture}[scale=0.35]
    \begin{axis}[point meta max=100000,grid=both,zmax=1000,xmax=20,ymax=1000,yticklabel style = {font=\Large},xticklabel style = {font=\Large},zticklabel style = {font=\Large},legend style={empty legend,font=\LARGE}]
      \addplot3[only marks,scatter,mark=o,mark size=3pt] file {ibea.txt};
        \addlegendentry{FEMOSAA}
    \end{axis}
  \end{tikzpicture}
     \vspace{-0.5cm}
    \end{subfigure}
     \begin{subfigure}[t]{0.2\textwidth}
    \begin{tikzpicture}[scale=0.35]
    \begin{axis}[point meta max=100000,grid=both,zmax=1000,xmax=20,ymax=1000,yticklabel style = {font=\Large},xticklabel style = {font=\Large},zticklabel style = {font=\Large},legend style={empty legend,font=\LARGE}]
      \addplot3[only marks,scatter,mark=o,mark size=3pt] file {ibea-plain.txt};
        \addlegendentry{FEMOSAA-N}
    \end{axis}
  \end{tikzpicture}
     \vspace{-0.5cm}
    \end{subfigure}
     \begin{subfigure}[t]{0.2\textwidth}
    \begin{tikzpicture}[scale=0.35]
    \begin{axis}[point meta max=100000,grid=both,zmax=1000,xmax=20,ymax=1000,yticklabel style = {font=\Large},xticklabel style = {font=\Large},zticklabel style = {font=\Large},legend style={empty legend,font=\LARGE}]
      \addplot3[only marks,scatter,mark=o,mark size=3pt] file {ibea-k.txt};
        \addlegendentry{FEMOSAA-K}
    \end{axis}
  \end{tikzpicture}
     \vspace{-0.5cm}
    \end{subfigure}
     \begin{subfigure}[t]{0.2\textwidth}
    \begin{tikzpicture}[scale=0.35]
    \begin{axis}[point meta max=100000,grid=both,zmax=1000,xmax=20,ymax=1000,yticklabel style = {font=\Large},xticklabel style = {font=\Large},zticklabel style = {font=\Large},legend style={empty legend,font=\LARGE}]
      \addplot3[only marks,scatter,mark=o,mark size=3pt] file {ibea-d.txt};
        \addlegendentry{FEMOSAA-D}
    \end{axis}
  \end{tikzpicture}
     \vspace{-0.5cm}
    \end{subfigure}
     \begin{subfigure}[t]{0.2\textwidth}
    \begin{tikzpicture}[scale=0.35]
    \begin{axis}[point meta max=100000,grid=both,zmax=1000,xmax=20,ymax=1000,yticklabel style = {font=\Large},xticklabel style = {font=\Large},zticklabel style = {font=\Large},legend style={empty legend,font=\LARGE}]
      \addplot3[only marks,scatter,mark=o,mark size=3pt] file {ibea-01.txt};
        \addlegendentry{FEMOSAA-0/1}
    \end{axis}
  \end{tikzpicture}
     \vspace{-0.5cm}
    \end{subfigure}
    
     \vspace{-0.5cm}
        \begin{subfigure}[t]{\textwidth}
     \subcaption{IBEA and read-write pattern [the axis from left to right are Response Time (ms), Energy Consumption (watt) and Workload (requests/min)]}
      \end{subfigure}
      
      \hspace{-0.4cm}
\centering
             \begin{subfigure}[t]{0.2\textwidth}
    \begin{tikzpicture}[scale=0.35]
    \begin{axis}[point meta max=100000,grid=both,zmax=10000,xmax=15,ymax=1000,yticklabel style = {font=\Large},xticklabel style = {font=\Large},zticklabel style = {font=\Large},legend style={empty legend,font=\LARGE}]
      \addplot3[only marks,scatter,mark=o,mark size=3pt] file {moead-r.txt};
        \addlegendentry{FEMOSAA}
    \end{axis}
  \end{tikzpicture}
     \vspace{-0.5cm}
    \end{subfigure}
     \begin{subfigure}[t]{0.2\textwidth}
    \begin{tikzpicture}[scale=0.35]
    \begin{axis}[point meta max=100000,grid=both,zmax=10000,xmax=15,ymax=1000,yticklabel style = {font=\Large},xticklabel style = {font=\Large},zticklabel style = {font=\Large},legend style={empty legend,font=\LARGE}]
      \addplot3[only marks,scatter,mark=o,mark size=3pt] file {moead-r-plain.txt};
        \addlegendentry{FEMOSAA-N}
    \end{axis}
  \end{tikzpicture}
     \vspace{-0.5cm}
    \end{subfigure}
     \begin{subfigure}[t]{0.2\textwidth}
    \begin{tikzpicture}[scale=0.35]
    \begin{axis}[point meta max=100000,grid=both,zmax=10000,xmax=15,ymax=1000,yticklabel style = {font=\Large},xticklabel style = {font=\Large},zticklabel style = {font=\Large},legend style={empty legend,font=\LARGE}]
      \addplot3[only marks,scatter,mark=o,mark size=3pt] file {moead-r-k.txt};
        \addlegendentry{FEMOSAA-K}
    \end{axis}
  \end{tikzpicture}
     \vspace{-0.5cm}
    \end{subfigure}
     \begin{subfigure}[t]{0.2\textwidth}
    \begin{tikzpicture}[scale=0.35]
    \begin{axis}[point meta max=100000,grid=both,zmax=10000,xmax=15,ymax=1000,yticklabel style = {font=\Large},xticklabel style = {font=\Large},zticklabel style = {font=\Large},legend style={empty legend,font=\LARGE}]
      \addplot3[only marks,scatter,mark=o,mark size=3pt] file {moead-r-d.txt};
        \addlegendentry{FEMOSAA-D}
    \end{axis}
  \end{tikzpicture}
     \vspace{-0.5cm}
    \end{subfigure}
     \begin{subfigure}[t]{0.2\textwidth}
    \begin{tikzpicture}[scale=0.35]
    \begin{axis}[point meta max=100000,grid=both,zmax=10000,xmax=15,ymax=1000,yticklabel style = {font=\Large},xticklabel style = {font=\Large},zticklabel style = {font=\Large},legend style={empty legend,font=\LARGE}]
      \addplot3[only marks,scatter,mark=o,mark size=3pt] file {moead-r-01.txt};
        \addlegendentry{FEMOSAA-0/1}
    \end{axis}
  \end{tikzpicture}
     \vspace{-0.5cm}
    \end{subfigure}
    
     \vspace{-0.5cm}
        \begin{subfigure}[t]{\textwidth}
     \subcaption{MOEA/D-STM and read-only pattern [the axis from left to right are Response Time (ms), Energy Consumption (watt) and Workload (requests/min)]}
      \end{subfigure}
      
      \hspace{-0.4cm}
\centering
             \begin{subfigure}[t]{0.2\textwidth}
    \begin{tikzpicture}[scale=0.35]
    \begin{axis}[point meta max=100000,grid=both,zmax=10000,xmax=15,ymax=1000,yticklabel style = {font=\Large},xticklabel style = {font=\Large},zticklabel style = {font=\Large},legend style={empty legend,font=\LARGE}]
      \addplot3[only marks,scatter,mark=o,mark size=3pt] file {nsgaii-r.txt};
        \addlegendentry{FEMOSAA}
    \end{axis}
  \end{tikzpicture}
     \vspace{-0.5cm}
    \end{subfigure}
     \begin{subfigure}[t]{0.2\textwidth}
    \begin{tikzpicture}[scale=0.35]
    \begin{axis}[point meta max=100000,grid=both,zmax=10000,xmax=15,ymax=1000,yticklabel style = {font=\Large},xticklabel style = {font=\Large},zticklabel style = {font=\Large},legend style={empty legend,font=\LARGE}]
      \addplot3[only marks,scatter,mark=o,mark size=3pt] file {nsgaii-r-plain.txt};
        \addlegendentry{FEMOSAA-N}
    \end{axis}
  \end{tikzpicture}
     \vspace{-0.5cm}
    \end{subfigure}
     \begin{subfigure}[t]{0.2\textwidth}
    \begin{tikzpicture}[scale=0.35]
    \begin{axis}[point meta max=100000,grid=both,zmax=10000,xmax=15,ymax=1000,yticklabel style = {font=\Large},xticklabel style = {font=\Large},zticklabel style = {font=\Large},legend style={empty legend,font=\LARGE}]
      \addplot3[only marks,scatter,mark=o,mark size=3pt] file {nsgaii-r-k.txt};
        \addlegendentry{FEMOSAA-K}
    \end{axis}
  \end{tikzpicture}
     \vspace{-0.5cm}
    \end{subfigure}
     \begin{subfigure}[t]{0.2\textwidth}
    \begin{tikzpicture}[scale=0.35]
    \begin{axis}[point meta max=100000,grid=both,zmax=10000,xmax=15,ymax=1000,yticklabel style = {font=\Large},xticklabel style = {font=\Large},zticklabel style = {font=\Large},legend style={empty legend,font=\LARGE}]
      \addplot3[only marks,scatter,mark=o,mark size=3pt] file {nsgaii-r-d.txt};
        \addlegendentry{FEMOSAA-D}
    \end{axis}
  \end{tikzpicture}
     \vspace{-0.5cm}
    \end{subfigure}
     \begin{subfigure}[t]{0.2\textwidth}
    \begin{tikzpicture}[scale=0.35]
    \begin{axis}[point meta max=100000,grid=both,zmax=10000,xmax=15,ymax=1000,yticklabel style = {font=\Large},xticklabel style = {font=\Large},zticklabel style = {font=\Large},legend style={empty legend,font=\LARGE}]
      \addplot3[only marks,scatter,mark=o,mark size=3pt] file {nsgaii-r-01.txt};
        \addlegendentry{FEMOSAA-0/1}
    \end{axis}
  \end{tikzpicture}
     \vspace{-0.5cm}
    \end{subfigure}
    
    \vspace{-0.5cm}
        \begin{subfigure}[t]{\textwidth}
     \subcaption{NSGA-II and read-only pattern [the axis from left to right are Response Time (ms), Energy Consumption (watt) and Workload (requests/min)]}
      \end{subfigure}

      \hspace{-0.4cm}
\centering
        \begin{subfigure}[t]{0.2\textwidth}
    \begin{tikzpicture}[scale=0.35]
    \begin{axis}[point meta max=100000,grid=both,zmax=10000,xmax=15,ymax=1000,yticklabel style = {font=\Large},xticklabel style = {font=\Large},zticklabel style = {font=\Large},legend style={empty legend,font=\LARGE}]
      \addplot3[only marks,scatter,mark=o,mark size=3pt] file {ibea-r.txt};
        \addlegendentry{FEMOSAA}
    \end{axis}
  \end{tikzpicture}
     \vspace{-0.5cm}
    \end{subfigure}
     \begin{subfigure}[t]{0.2\textwidth}
    \begin{tikzpicture}[scale=0.35]
    \begin{axis}[point meta max=100000,grid=both,zmax=10000,xmax=15,ymax=1000,yticklabel style = {font=\Large},xticklabel style = {font=\Large},zticklabel style = {font=\Large},legend style={empty legend,font=\LARGE}]
      \addplot3[only marks,scatter,mark=o,mark size=3pt] file {ibea-r-plain.txt};
        \addlegendentry{FEMOSAA-N}
    \end{axis}
  \end{tikzpicture}
     \vspace{-0.5cm}
    \end{subfigure}
     \begin{subfigure}[t]{0.2\textwidth}
    \begin{tikzpicture}[scale=0.35]
    \begin{axis}[point meta max=100000,grid=both,zmax=10000,xmax=15,ymax=1000,yticklabel style = {font=\Large},xticklabel style = {font=\Large},zticklabel style = {font=\Large},legend style={empty legend,font=\LARGE}]
      \addplot3[only marks,scatter,mark=o,mark size=3pt] file {ibea-r-k.txt};
        \addlegendentry{FEMOSAA-K}
    \end{axis}
  \end{tikzpicture}
     \vspace{-0.5cm}
    \end{subfigure}
     \begin{subfigure}[t]{0.2\textwidth}
    \begin{tikzpicture}[scale=0.35]
    \begin{axis}[point meta max=100000,grid=both,zmax=10000,xmax=15,ymax=1000,yticklabel style = {font=\Large},xticklabel style = {font=\Large},zticklabel style = {font=\Large},legend style={empty legend,font=\LARGE}]
      \addplot3[only marks,scatter,mark=o,mark size=3pt] file {ibea-r-d.txt};
        \addlegendentry{FEMOSAA-D}
    \end{axis}
  \end{tikzpicture}
     \vspace{-0.5cm}
    \end{subfigure}
     \begin{subfigure}[t]{0.2\textwidth}
    \begin{tikzpicture}[scale=0.35]
    \begin{axis}[point meta max=100000,grid=both,zmax=10000,xmax=15,ymax=1000,yticklabel style = {font=\Large},xticklabel style = {font=\Large},zticklabel style = {font=\Large},legend style={empty legend,font=\LARGE}]
      \addplot3[only marks,scatter,mark=o,mark size=3pt] file {ibea-r-01.txt};
        \addlegendentry{FEMOSAA-0/1}
    \end{axis}
  \end{tikzpicture}
     \vspace{-0.5cm}
    \end{subfigure}

     \vspace{-0.5cm}
        \begin{subfigure}[t]{\textwidth}
     \subcaption{IBEA and read-only pattern [the axis from left to right are Response Time (ms), Energy Consumption (watt) and Workload (requests/min)]}
      \end{subfigure}

 \caption{Measured quality results with respect to workload for all the timesteps for RUBiS-SAS.}
  \label{fig:femosaa-details}
  \end{figure}

\begin{figure}[t!]

 \hspace{-0.4cm}
\centering
 \begin{subfigure}[t]{0.2\textwidth}
    \begin{tikzpicture}[scale=0.35]
    \begin{axis}[point meta max=100000,grid=both,zmax=0.3,xmin=30,xmax=150,ymax=0.7,yticklabel style = {font=\Large},xticklabel style = {font=\Large},zticklabel style = {font=\Large},legend style={empty legend,font=\LARGE}]
      \addplot3[only marks,scatter,mark=o,mark size=3pt] file {soa/soa-moead.txt};
        \addlegendentry{FEMOSAA}
    \end{axis}
  \end{tikzpicture}
  \vspace{-0.5cm}
    \end{subfigure}
     \begin{subfigure}[t]{0.2\textwidth}
    \begin{tikzpicture}[scale=0.35]
    \begin{axis}[point meta max=100000,grid=both,zmax=0.3,xmin=30,xmax=150,ymax=0.7,yticklabel style = {font=\Large},xticklabel style = {font=\Large},zticklabel style = {font=\Large},legend style={empty legend,font=\LARGE}]
      \addplot3[only marks,scatter,mark=o,mark size=3pt] file {soa/soa-moead-nothing.txt};
        \addlegendentry{FEMOSAA-N}
    \end{axis}
  \end{tikzpicture}
    \vspace{-0.5cm}
    \end{subfigure}
     \begin{subfigure}[t]{0.2\textwidth}
    \begin{tikzpicture}[scale=0.35]
    \begin{axis}[point meta max=100000,grid=both,zmax=0.3,xmin=30,xmax=150,ymax=0.7,yticklabel style = {font=\Large},xticklabel style = {font=\Large},zticklabel style = {font=\Large},legend style={empty legend,font=\LARGE}]
      \addplot3[only marks,scatter,mark=o,mark size=3pt] file {soa/soa-moead-k.txt};
        \addlegendentry{FEMOSAA-K}
    \end{axis}
  \end{tikzpicture}
    \vspace{-0.5cm}
    \end{subfigure}
     \begin{subfigure}[t]{0.2\textwidth}
    \begin{tikzpicture}[scale=0.35]
    \begin{axis}[point meta max=100000,grid=both,zmax=0.3,xmin=30,xmax=150,ymax=0.7,yticklabel style = {font=\Large},xticklabel style = {font=\Large},zticklabel style = {font=\Large},legend style={empty legend,font=\LARGE}]
      \addplot3[only marks,scatter,mark=o,mark size=3pt] file {soa/soa-moead-d.txt};
        \addlegendentry{FEMOSAA-D}
    \end{axis}
  \end{tikzpicture}
    \vspace{-0.5cm}
    \end{subfigure}
     \begin{subfigure}[t]{0.2\textwidth}
    \begin{tikzpicture}[scale=0.35]
    \begin{axis}[point meta max=100000,grid=both,zmax=0.3,xmin=30,xmax=150,ymax=0.7,yticklabel style = {font=\Large},xticklabel style = {font=\Large},zticklabel style = {font=\Large},legend style={empty legend,font=\LARGE}]
      \addplot3[only marks,scatter,mark=o,mark size=3pt] file {soa/soa-moead-01.txt};
        \addlegendentry{FEMOSAA-0/1}
    \end{axis}
  \end{tikzpicture}
    \vspace{-0.5cm}
    \end{subfigure}

    \vspace{-0.5cm}
    \begin{subfigure}[t]{\textwidth}
     \subcaption{MOEA/D-STM [the axis from left to right are \textbf{inverted} Throughput (request/second), Cost (\$) and mean relative standard deviation of the candidate concrete services' quality]}
      \end{subfigure}
      
      \hspace{-0.4cm}
\centering
       \begin{subfigure}[t]{0.2\textwidth}
    \begin{tikzpicture}[scale=0.35]
    \begin{axis}[point meta max=100000,grid=both,zmax=0.3,xmin=30,xmax=150,ymax=0.7,yticklabel style = {font=\Large},xticklabel style = {font=\Large},zticklabel style = {font=\Large},legend style={empty legend,font=\LARGE}]
      \addplot3[only marks,scatter,mark=o,mark size=3pt] file {soa/soa-nsgaii.txt};
        \addlegendentry{FEMOSAA}
    \end{axis}
  \end{tikzpicture}
     \vspace{-0.5cm}
    \end{subfigure}
     \begin{subfigure}[t]{0.2\textwidth}
    \begin{tikzpicture}[scale=0.35]
    \begin{axis}[point meta max=100000,grid=both,zmax=0.3,xmin=30,xmax=150,ymax=0.7,yticklabel style = {font=\Large},xticklabel style = {font=\Large},zticklabel style = {font=\Large},legend style={empty legend,font=\LARGE}]
      \addplot3[only marks,scatter,mark=o,mark size=3pt] file {soa/soa-nsgaii-plain.txt};
        \addlegendentry{FEMOSAA-N}
    \end{axis}
  \end{tikzpicture}
     \vspace{-0.5cm}
    \end{subfigure}
     \begin{subfigure}[t]{0.2\textwidth}
    \begin{tikzpicture}[scale=0.35]
    \begin{axis}[point meta max=100000,grid=both,zmax=0.3,xmin=30,xmax=150,ymax=0.7,yticklabel style = {font=\Large},xticklabel style = {font=\Large},zticklabel style = {font=\Large},legend style={empty legend,font=\LARGE}]
      \addplot3[only marks,scatter,mark=o,mark size=3pt] file {soa/soa-nsgaii-k.txt};
        \addlegendentry{FEMOSAA-K}
    \end{axis}
  \end{tikzpicture}
     \vspace{-0.5cm}
    \end{subfigure}
     \begin{subfigure}[t]{0.2\textwidth}
    \begin{tikzpicture}[scale=0.35]
    \begin{axis}[point meta max=100000,grid=both,zmax=0.3,xmin=30,xmax=150,ymax=0.7,yticklabel style = {font=\Large},xticklabel style = {font=\Large},zticklabel style = {font=\Large},legend style={empty legend,font=\LARGE}]
      \addplot3[only marks,scatter,mark=o,mark size=3pt] file {soa/soa-nsgaii-d.txt};
        \addlegendentry{FEMOSAA-D}
    \end{axis}
  \end{tikzpicture}
     \vspace{-0.5cm}
    \end{subfigure}
     \begin{subfigure}[t]{0.2\textwidth}
    \begin{tikzpicture}[scale=0.35]
    \begin{axis}[point meta max=100000,grid=both,zmax=0.3,xmin=30,xmax=150,ymax=0.7,yticklabel style = {font=\Large},xticklabel style = {font=\Large},zticklabel style = {font=\Large},legend style={empty legend,font=\LARGE}]
      \addplot3[only marks,scatter,mark=o,mark size=3pt] file {soa/soa-nsgaii-01.txt};
        \addlegendentry{FEMOSAA-0/1}
    \end{axis}
  \end{tikzpicture}
     \vspace{-0.5cm}
    \end{subfigure}
    
     \vspace{-0.5cm}
        \begin{subfigure}[t]{\textwidth}
     \subcaption{NSGA-II [the axis from left to right are \textbf{inverted} Throughput (request/second), Cost (\$) and mean relative standard deviation of the candidate concrete services' quality]}
      \end{subfigure}
      
      \hspace{-0.4cm}
\centering
             \begin{subfigure}[t]{0.2\textwidth}
    \begin{tikzpicture}[scale=0.35]
    \begin{axis}[point meta max=100000,grid=both,zmax=0.3,xmin=30,xmax=150,ymax=0.7,yticklabel style = {font=\Large},xticklabel style = {font=\Large},zticklabel style = {font=\Large},legend style={empty legend,font=\LARGE}]
      \addplot3[only marks,scatter,mark=o,mark size=3pt] file {soa/soa-ibea.txt};
        \addlegendentry{FEMOSAA}
    \end{axis}
  \end{tikzpicture}
     \vspace{-0.5cm}
    \end{subfigure}
     \begin{subfigure}[t]{0.2\textwidth}
    \begin{tikzpicture}[scale=0.35]
    \begin{axis}[point meta max=100000,grid=both,zmax=0.3,xmin=30,xmax=150,ymax=0.7,yticklabel style = {font=\Large},xticklabel style = {font=\Large},zticklabel style = {font=\Large},legend style={empty legend,font=\LARGE}]
      \addplot3[only marks,scatter,mark=o,mark size=3pt] file {soa/soa-ibea-plain.txt};
        \addlegendentry{FEMOSAA-N}
    \end{axis}
  \end{tikzpicture}
     \vspace{-0.5cm}
    \end{subfigure}
     \begin{subfigure}[t]{0.2\textwidth}
    \begin{tikzpicture}[scale=0.35]
    \begin{axis}[point meta max=100000,grid=both,zmax=0.3,xmin=30,xmax=150,ymax=0.7,yticklabel style = {font=\Large},xticklabel style = {font=\Large},zticklabel style = {font=\Large},legend style={empty legend,font=\LARGE}]
      \addplot3[only marks,scatter,mark=o,mark size=3pt] file {soa/soa-ibea-k.txt};
        \addlegendentry{FEMOSAA-K}
    \end{axis}
  \end{tikzpicture}
     \vspace{-0.5cm}
    \end{subfigure}
     \begin{subfigure}[t]{0.2\textwidth}
    \begin{tikzpicture}[scale=0.35]
    \begin{axis}[point meta max=100000,grid=both,zmax=0.3,xmin=30,xmax=150,ymax=0.7,yticklabel style = {font=\Large},xticklabel style = {font=\Large},zticklabel style = {font=\Large},legend style={empty legend,font=\LARGE}]
      \addplot3[only marks,scatter,mark=o,mark size=3pt] file {soa/soa-ibea-d.txt};
        \addlegendentry{FEMOSAA-D}
    \end{axis}
  \end{tikzpicture}
     \vspace{-0.5cm}
    \end{subfigure}
     \begin{subfigure}[t]{0.2\textwidth}
    \begin{tikzpicture}[scale=0.35]
    \begin{axis}[point meta max=100000,grid=both,zmax=0.3,xmin=30,xmax=150,ymax=0.7,yticklabel style = {font=\Large},xticklabel style = {font=\Large},zticklabel style = {font=\Large},legend style={empty legend,font=\LARGE}]
      \addplot3[only marks,scatter,mark=o,mark size=3pt] file {soa/soa-ibea-01.txt};
        \addlegendentry{FEMOSAA-0/1}
    \end{axis}
  \end{tikzpicture}
     \vspace{-0.5cm}
    \end{subfigure}
    
     \vspace{-0.5cm}
        \begin{subfigure}[t]{\textwidth}
     \subcaption{IBEA [the axis from left to right are \textbf{inverted} Throughput (request/second), Cost (\$) and mean relative standard deviation of the candidate concrete services' quality]}
      \end{subfigure}

 \caption{Measured quality results with respect to changes for all the timesteps for SOA-SAS.}
  \label{fig:soa-femosaa-details}
  \end{figure}
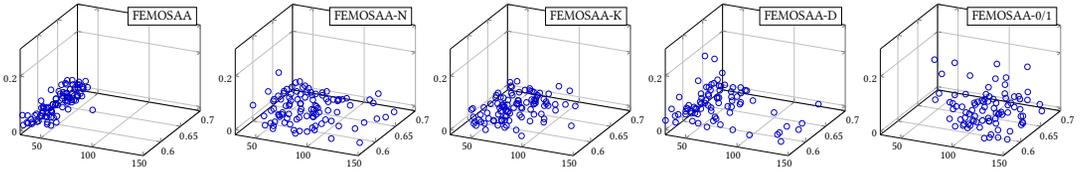
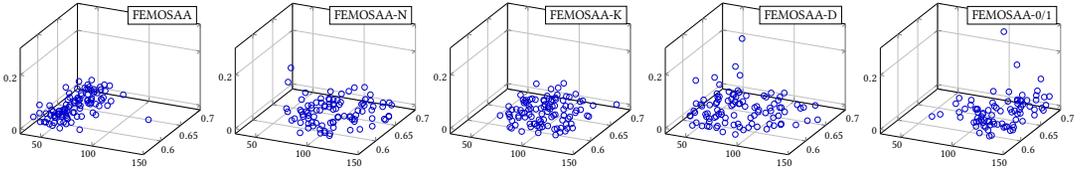
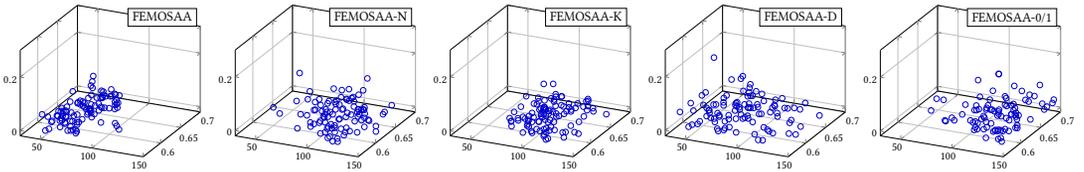

To evaluate the effectiveness of our elitist chromosome representation in contrast to the conventional binary representation, we firstly compare FEMOSAA with FEMOSAA-0/1 for all cases on \emph{RUBiS-SAS}, as shown in Figures~\ref{fig:moead-rw-quality} to~\ref{fig:ibea-r-quality}. Clearly, we see that FEMOSAA largely outperform FEMOSAA-0/1 on both response time and energy consumption with $p<0.05$ and non-trivial effect sizes. Further, FEMOSAA yields better HV and ED values for all cases. In Figure~\ref{fig:femosaa-details}, we also note that, in contrast to FEMOSAA-0/1 under all cases, the quality achieved by FEMOSAA tends to be much more convergent to the left-bottom line of the cube even under heavy workload, meaning that it leads to better quality results and more balanced trade-off. For \emph{SOA-SAS}, we can observe similar results on throughput and cost, as shown in Figures~\ref{fig:moead-soa-quality} to~\ref{fig:ibea-soa-quality} and Figure~\ref{fig:soa-femosaa-details}.

 \begin{figure}[!t] 
   \begin{minipage}{\textwidth}
  \begin{minipage}[t]{0.45\textwidth}
    \centering
\includestandalone[width=0.6\textwidth]{tikz/nsgaii-rw-quality}
~
\includestandalone[width=0.6\textwidth]{tikz/nsgaii-rw-sta}
  \end{minipage}
  \hspace{0.3cm}
  \begin{minipage}[b]{0.55\textwidth}
    \centering
    \scriptsize     
    \begin{tabularx}{0.76\textwidth}{|p{1.5cm}|Y|Y|}\hline
     \multicolumn{3}{|c|}{\cellcolor{black}\textbf{\emph{\textcolor{white}{$p$ value (effect size) on Response Time}}}} \\\hline
      & FEMOSAA & FEMOSAA-N\\ \hline
        FEMOSAA-K&\cellcolor{gray!50}\bfseries.046 (small)& .960 (trivial) \\
        FEMOSAA-D&\cellcolor{gray!50}\bfseries.025 (medium)& .968 (trivial)  \\
        FEMOSAA-N&\cellcolor{gray!50}\bfseries.043 (small)& --- \\
        FEMOSAA-0/1&\cellcolor{gray!50}\bfseries $<$.001 (large)&\cellcolor{gray!50}\bfseries.003 (large) \\ \hline
      \end{tabularx} 
            \vspace{0.12cm}     
     \begin{tabularx}{0.76\textwidth}{|p{1.5cm}|Y|Y|}\hline
     \multicolumn{3}{|c|}{\cellcolor{black}\textbf{\emph{\textcolor{white}{$p$ value (effect size) on Energy Consumption}}}} \\\hline
      & FEMOSAA & FEMOSAA-N\\ \hline
          FEMOSAA-K & \cellcolor{gray!50}\bfseries.034 (medium)  & .810 (trivial) \\
        FEMOSAA-D & .110 (small)  & .562 (trivial) \\
         FEMOSAA-N&\cellcolor{gray!50}\bfseries.032 (medium)& ---  \\
        FEMOSAA-0/1 & \cellcolor{gray!50}\bfseries $<$.001 (large) & \cellcolor{gray!50}\bfseries.019 (large) \\ \hline
      \end{tabularx}  
    \end{minipage} 
     \vspace{-0.8cm}  
    \captionof{figure}{Results with NSGA-II under read-write pattern on RUBiS-SAS. (GM denotes Geometric Mean. The significant statistics of comparisons, i.e., $p<$ 0.05, are highlighted and shown in bold.)}
     \label{fig:nsgaii-rw-quality}
  \end{minipage}
    \vspace{-0.5cm}
\end{figure}

However, the above comparison only demonstrates the combinatorial effectiveness of elitist chromosome representation, the dependency aware operators and the knee selection, while it is not clear whether the superiority over the conventional binary representation is truly resulted from the elitist chromosome representation. To answer this, we then compare FEMOSAA-N with FEMOSAA-0/1 for all cases on \emph{RUBiS-SAS}, as shown in Figures~\ref{fig:moead-rw-quality} to~\ref{fig:ibea-r-quality}. We see that, again, FEMOSAA-N yields much better results on both quality attributes than FEMOSAA-0/1 under all cases while such improvements are statistically significant on at least one attribute with non-trivial effect sizes. It also achieves better HV and ED results. Further, although it is less than 15\%, FEMOSAA-N does produce more valid solutions than FEMOSAA-0/1 which cannot identify any valid solution and thus affect the quality of optimization. Next, we take a more detailed comparison between FEMOSAA-N and FEMOSAA-0/1 through Figure~\ref{fig:femosaa-details}, which reveals that, for all cases, the results of FEMOSAA-N are more closed to left-bottom line of the cube when the workload is low, e.g., less than around 500 requests/min. For \emph{SOA-SAS}, we observe even better results: FEMOSAA-N largely outperforms FEMOSAA-0/1 on all the metrics, and the comparisons on quality attributes exhibit $p<0.05$ and large effect sizes for all cases, as shown in Figures~\ref{fig:moead-soa-quality} to~\ref{fig:ibea-soa-quality} and Figure~\ref{fig:soa-femosaa-details}. Those improvements of elitist chromosome representation over the conventional binary representation are mainly due to the fact that it fundamentally reduces the search space without affecting the original variability of SAS, so that the search of MOEA has a larger chance to identified the ideal solutions, leading to more valid solutions and better quality. In addition, as we will show in Section~\ref{sec:exp-overhead}, the elitist chromosome representation can significantly shorten the running time of MOEAs.

 \begin{figure}[!t] 
   \begin{minipage}{\textwidth}
  \begin{minipage}[t]{0.45\textwidth}
    \centering
\includestandalone[width=0.6\textwidth]{tikz/nsgaii-r-quality}
~
\includestandalone[width=0.6\textwidth]{tikz/nsgaii-r-sta}
  \end{minipage}
  \hspace{0.3cm}
  \begin{minipage}[b]{0.55\textwidth}
    \centering
    \scriptsize     
    \begin{tabularx}{0.76\textwidth}{|p{1.5cm}|Y|Y|}\hline
     \multicolumn{3}{|c|}{\cellcolor{black}\textbf{\emph{\textcolor{white}{$p$ value (effect size) on Response Time}}}} \\\hline
      & FEMOSAA & FEMOSAA-N\\ \hline
        FEMOSAA-K&\cellcolor{gray!50}\bfseries.010 (medium)& .242 (small) \\
        FEMOSAA-D&\cellcolor{gray!50}\bfseries.004 (medium)& .549 (trivial)  \\
        FEMOSAA-N&\cellcolor{gray!50}\bfseries$<$.001 (large)& --- \\
        FEMOSAA-0/1&\cellcolor{gray!50}\bfseries $<$.001 (large)&.080 (medium) \\ \hline
      \end{tabularx} 
            \vspace{0.12cm}     
     \begin{tabularx}{0.76\textwidth}{|p{1.5cm}|Y|Y|}\hline
     \multicolumn{3}{|c|}{\cellcolor{black}\textbf{\emph{\textcolor{white}{$p$ value (effect size) on Energy Consumption}}}} \\\hline
      & FEMOSAA & FEMOSAA-N\\ \hline
          FEMOSAA-K &.968 (trivial)& .904 (trivial) \\
        FEMOSAA-D &.992 (trivial) & .697 (trivial) \\
         FEMOSAA-N& .624 (trivial)& ---  \\
        FEMOSAA-0/1 & \cellcolor{gray!50}\bfseries.002 (medium) & \cellcolor{gray!50}\bfseries.002 (medium) \\ \hline
      \end{tabularx}  
    \end{minipage} 
     \vspace{-0.8cm}  
    \captionof{figure}{Results with NSGA-II under read-only pattern on RUBiS-SAS. (GM denotes Geometric Mean. The significant statistics of comparisons, i.e., $p<$ 0.05, are highlighted and shown in bold.)}
     \label{fig:nsgaii-r-quality}
  \end{minipage}
    \vspace{-0.5cm}
\end{figure}

Notably, as shown in Figure~\ref{fig:femosaa-details} for \emph{RUBiS-SAS}, when the workload increases (i.e., more than 500 requests/min), we can see that the achieved quality between FEMOSAA-N and FEMOSAA-0/1 are barely different. This is because the number of good and effected solutions tends to be limited when the workload is heavy, causing the benefits of reducing search space introduced by the elitist chromosome representation less obvious. Those results indicate that, overall, the elitist chromosome representation can better guide the search towards ideal solutions but such improvement tends to blur as the workload increases. Similarly, for \emph{SOA-SAS} in Figure~\ref{fig:soa-femosaa-details}, we see that although FEMOSAA-N creates better results than FEMOSAA-0/1 across different levels of diversity on the concrete services, the improvement tends to degrade under high diversity.

 \begin{figure}[!t] 
   \begin{minipage}{\textwidth}
  \begin{minipage}[t]{0.45\textwidth}
    \centering
\includestandalone[width=0.6\textwidth]{tikz/nsgaii-sas-quality}
~
\includestandalone[width=0.6\textwidth]{tikz/nsgaii-sas-sta}
  \end{minipage}
  \hspace{0.3cm}
  \begin{minipage}[b]{0.55\textwidth}
    \centering
    \scriptsize     
    \begin{tabularx}{0.76\textwidth}{|p{1.5cm}|Y|Y|}\hline
     \multicolumn{3}{|c|}{\cellcolor{black}\textbf{\emph{\textcolor{white}{$p$ value (effect size) on Throughput}}}} \\\hline
      & FEMOSAA & FEMOSAA-N\\ \hline
        FEMOSAA-K&.352 (small)& .831 (trivial) \\
        FEMOSAA-D&\cellcolor{gray!50}\bfseries.016 (medium)&\cellcolor{gray!50}\bfseries.002 (medium)  \\
        FEMOSAA-N&.300 (small)& --- \\
        FEMOSAA-0/1&\cellcolor{gray!50}\bfseries$<$.001 (large)&\cellcolor{gray!50}\bfseries$<$.001 (large) \\ \hline
      \end{tabularx} 
            \vspace{0.12cm}     
     \begin{tabularx}{0.76\textwidth}{|p{1.5cm}|Y|Y|}\hline
     \multicolumn{3}{|c|}{\cellcolor{black}\textbf{\emph{\textcolor{white}{$p$ value (effect size) on Cost}}}} \\\hline
      & FEMOSAA & FEMOSAA-N\\ \hline
          FEMOSAA-K &\cellcolor{gray!50}\bfseries$<$.001 (large)&\cellcolor{gray!50}\bfseries.015 (medium) \\
        FEMOSAA-D &\cellcolor{gray!50}\bfseries$<$.001 (large)&\cellcolor{gray!50}\bfseries$<$.001 (large)\\
         FEMOSAA-N&\cellcolor{gray!50}\bfseries$<$.001 (large)& ---  \\
        FEMOSAA-0/1 & \cellcolor{gray!50}\bfseries$<$.001 (large) &\cellcolor{gray!50}\bfseries$<$.001 (large) \\ \hline
      \end{tabularx}  
    \end{minipage} 
     \vspace{-0.8cm}  
    \captionof{figure}{Results with NSGA-II on SOA-SAS. (GM denotes Geometric Mean. The significant statistics of comparisons, i.e., $p<$ 0.05, are highlighted and shown in bold.)}
    \label{fig:nsgaii-soa-quality}
  \end{minipage}
    \vspace{-0.5cm}
\end{figure}

In summary, the results conclude that:

\begin{tcolorbox}
\textbf{Answering \emph{RQ1}}\textemdash In contrast to the binary encoding, the elitist chromosome representation helps to produce better optimized quality for SAS with statistically significant results and non-trivial effect sizes on at least one quality attribute. It can also shorten the running time of MOEAs (in Section~\ref{sec:exp-overhead}). However, the improvements might be hardly observed when the number of effected solutions is limited, e.g., under heavy workload.
\end{tcolorbox}


\subsubsection{Evaluating the dependency aware operators}
\label{eval-dep}
To evaluate the benefit of dependency aware operators, we firstly compare FEMOSAA with FEMOSAA-N for \emph{RUBiS-SAS} as shown in Figures~\ref{fig:moead-rw-quality} to~\ref{fig:ibea-r-quality}. It is easy to see that for all cases, FEMOSAA yields better results on both quality attributes with $p<0.05$ and non-trivial effect sizes on at least one attribute. FEMOSAA also outperforms FEMOSAA-N in terms of HV, ED and the number of valid solutions. In Figure~\ref{fig:femosaa-details}, we can see that the achieved quality of FEMOSAA better converges to the ideal area of the cube on different degrees of workload. Similar observations can be obtained for \emph{SOA-SAS} as illustrated in Figures~\ref{fig:moead-soa-quality} to~\ref{fig:ibea-soa-quality} and Figure~\ref{fig:soa-femosaa-details}.

However, the comparison above does not indicate whether the improvement is mainly introduced by the dependency aware operators or the knee selection method, thus, we then further compare FEMOSAA with FEMOSAA-K, which omitted the dependency aware operators. In Figures~\ref{fig:moead-rw-quality} to~\ref{fig:ibea-r-quality} for \emph{RUBiS-SAS}, we see that FEMOSAA still exhibits superior quality results under all cases which are statistically significant on at least one attribute with non-trivial effect sizes; it also comes with better HV and ED values while the number of valid solutions are 100\% versus less than 20\%. In all cases, the achieved results of FEMOSAA, as shown in Figure~\ref{fig:femosaa-details}, are more convergent around the ideal area of the cube, especially when the workload is heavy, i.e., more than around 500 requests/min. Similar results can be seen for \emph{SOA-SAS} as shown in Figures~\ref{fig:moead-soa-quality} to~\ref{fig:ibea-soa-quality} and Figure~\ref{fig:soa-femosaa-details}. But, FEMOSAA-K is better than FEMOSAA on throughput when using MOEA/D-STM, the comparisons are not statistically significant though. This means that the degree of objective conflicts in \emph{SOA-SAS} is much stronger than the case of \emph{RUBiS-SAS}.

\begin{figure}[!t]
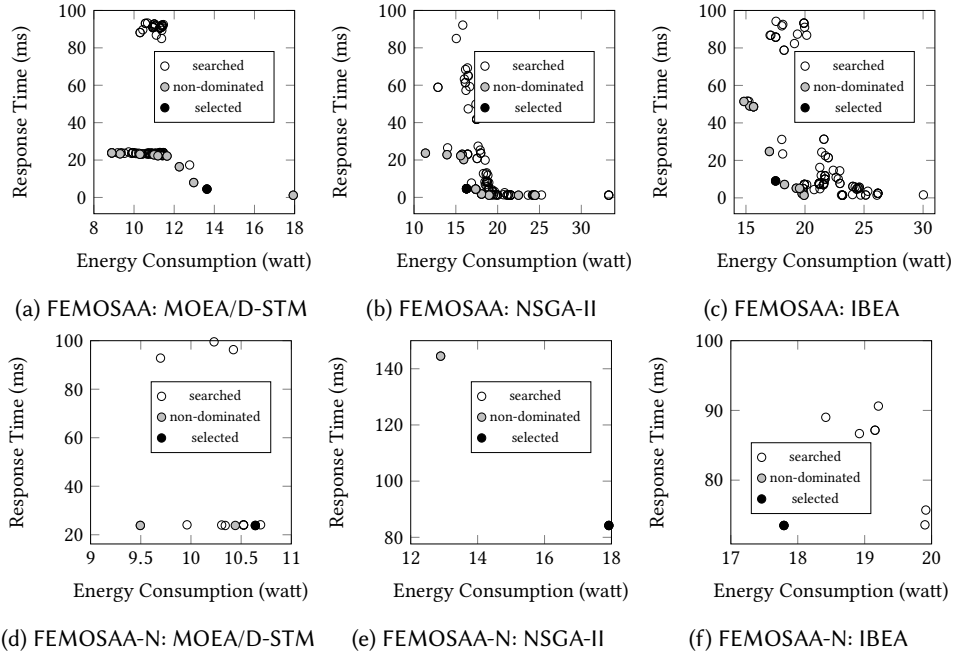

{
\centering
 \begin{subfigure}[t]{0.3\textwidth}
  \includestandalone[width=\columnwidth]{tikz/moead-femosaa}
    \subcaption{FEMOSAA: MOEA/D-STM}
    \end{subfigure}
    ~
 \begin{subfigure}[t]{0.3\textwidth}
  \includestandalone[width=\columnwidth]{tikz/nsgaii-femosaa}
   \subcaption{FEMOSAA: NSGA-II}
    \end{subfigure}
        ~
 \begin{subfigure}[t]{0.3\textwidth}
  \includestandalone[width=\columnwidth]{tikz/ibea-femosaa}
   \subcaption{FEMOSAA: IBEA}
    \end{subfigure}
        
     \begin{subfigure}[t]{0.3\textwidth}
  \includestandalone[width=\columnwidth]{tikz/moead-femosaa-nothing}
   \subcaption{FEMOSAA-N: MOEA/D-STM}
    \end{subfigure}
        ~
     \begin{subfigure}[t]{0.3\textwidth}
  \includestandalone[width=\columnwidth]{tikz/nsgaii-femosaa-nothing}
    \subcaption{FEMOSAA-N: NSGA-II}
    \end{subfigure}
    ~
     \begin{subfigure}[t]{0.3\textwidth}
  \includestandalone[width=\columnwidth]{tikz/ibea-femosaa-nothing}
    \subcaption{FEMOSAA-N: IBEA}
    \end{subfigure}
}
  \caption{Objective space of searched, non-dominated and selected valid solution(s) on a timestep of \emph{RUBiS-SAS}.}
 \label{fig:femosaa-pred-details}
 \vspace{-0.5cm}
  \end{figure}

To gain a better understand about why our dependency aware operators can improve the quality of optimization, in Figure~\ref{fig:femosaa-pred-details}, we plot the objective space of all searched, non-dominated and selected valid solution(s) in the final population of an example run. We can clearly see that, under all studied MOEAs, the dependency aware operators (i.e., FEMOSAA) can help to find solutions with better convergence and diversity, when compared with the case where dependencies are ignored (i.e., FEMOSAA-N). Such a benefit eventually leads to a better set of non-dominated solutions within which a final one can be selected for adaptation. Fundamentally, this is because the dependency aware operators can better guide the search to prevent MOEAs from exploring the unneeded invalid solutions, which virtually reduces the search space and provides a larger chance to find better solutions.

Interestingly, when we compare FEMOSAA-D with FEMOSAA-N in Figures~\ref{fig:moead-rw-quality} to~\ref{fig:ibea-r-quality} for \emph{RUBiS-SAS}, we did not observe significant differences between them in terms of all the metrics and their comparison has also failed in all statistical significance tests, which differs from our expectation that FEMOSAA-D should achieve statistically better results. The same can be also registered in Figure~\ref{fig:femosaa-details}. We believe the reason for this is due to although FEMOSAA-D can guide the search for better solutions, the fact that a solution from the final non-dominated solutions is randomly selected for adaptation has obscured the benefits of preventing the exploration of invalid solutions, as such a solution might be highly imbalanced for the conflicting objectives. On contrary, for \emph{SOA-SAS} in Figures~\ref{fig:moead-soa-quality} to~\ref{fig:ibea-soa-quality} and Figure~\ref{fig:soa-femosaa-details}, we see that FEMOSAA-D generally outperforms FEMOSAA-N with better HV and ED values. In particular, under NSGA-II and IBEA, FEMOSAA-D is better on both attributes with $p<0.05$ and non-trivial effect sizes. This implies that for a feature model with relatively more complex dependencies as in the case of \emph{SOA-SAS}, the dependency aware operators can guide the search process to evolve many highly effected solutions in the final non-dominated set, thus the randomly selected adaptation solution, although might be imbalanced, could still be much better than those of the case when dependencies are omitted.

In conclusion, the results suggest that:

\begin{tcolorbox}
\textbf{Answering \emph{RQ2}}\textemdash In contrast to the classic and widely-used operators, the benefit of dependency aware operators is that they help to find solution with better convergence and diversity, leading to better optimized quality for SAS with statistically significant results and non-trivial effect sizes on at least one quality attribute. Such improvement can be more obvious when the number of effected solutions is limited, e.g., when the workload is heavy. However, applying dependency aware operators without ensuring balance of the selected adaptation solution might obscure its effectiveness.
\end{tcolorbox}

\subsubsection{Evaluating the knee selection}
\label{eval-knee}
We have already shown that the combination of dependency aware operators and knee selection can lead to results that outperform the case where both of them are omitted, we now evaluate whether the knee selection method itself can introduce benefit for runtime optimization of SAS. From Figures~\ref{fig:moead-rw-quality} to~\ref{fig:ibea-r-quality} for \emph{RUBiS-SAS}, when comparing FEMOSAA with FEMOSAA-D in which knee selecting has been omitted, we see that FEMOSAA achieves superior HV, ED and better results on both quality attributes. The comparisons are statistically significant with non-trivial effect sizes on at least one attribute for 5 out of 6 cases. In Figure~\ref{fig:femosaa-details}, we can see that the results of FEMOSAA are more balanced, i.e., points are more converge to the bottom-left, and such improvement is more obvious when the workload is heavy under all cases. For \emph{SOA-SAS}, we observed similar outcomes overall as illustrated in Figures~\ref{fig:moead-soa-quality} to~\ref{fig:ibea-soa-quality} and Figure~\ref{fig:soa-femosaa-details}, except that FEMOSAA-D has better throughput than FEMOSAA under NSGA-II due to the strong conflicts of the objectives in \emph{SOA-SAS}. These results show that our knee selection method in FEMOSAA contributes to the overall quality of adaptation by automatically selecting the solution, which has balanced trade-off on the objectives (i.e., a good sense of compromise), to execute the adaptation. The resulted quality achieved by those knee solutions are naturally the most preferable ones, when the relative preferences between objectives are unknown or it is too difficult to quantify them (which is common for SAS).

For all the studied MOEAs in Figure~\ref{fig:femosaa-pred-details} of \emph{RUBiS-SAS}, we can see that, when comparing with randomly selecting an adaptation solution from the non-dominated set (i.e., FEMOSAA-N), incorporating knee selection can indeed select more balanced solution for adaptation (i.e., FEMOSAA) as it is closer to the bulge near the bottom-left area of the objective space.

Nevertheless, unlike our expectation, FEMOSAA-K and FEMOSAA-N do not yield statistically significant differences as shown in Figures~\ref{fig:moead-rw-quality} to~\ref{fig:ibea-r-quality} and Figure~\ref{fig:femosaa-details} for \emph{RUBiS-SAS}. This is because the effectiveness of knee selection is fundamentally dependent on the quality of searched valid solution; henceforth, as we have shown in the previous section, when the search process wastes efforts to explore invalid solutions, the quality of the solutions in the final population might be compromised, which would negatively affect the benefit that could have been introduced by our knee selection method. Due to the same reason, for \emph{SOA-SAS} in Figures~\ref{fig:moead-soa-quality} to~\ref{fig:ibea-soa-quality} and Figure~\ref{fig:soa-femosaa-details}, we see that although FEMOSAA-K slightly outperforms FEMOSAA-N on both attributes under all the cases, the majority of the differences (5 out of 6) are not significant statistically.

Overall, the results indicate that:

\begin{tcolorbox}
\textbf{Answering \emph{RQ3}}\textemdash In contrast to the randomly selected non-dominated adaptation solution, the knee selection helps to find a more balanced solution for adaptation, leading to better optimized quality for SAS with statistically significant results and non-trivial effect sizes on at least one attribute. Such improvement can be more obvious when the number of effected solutions is limited, e.g., when the workload is heavy. However, applying knee selection without ensuring the quality of searched valid solution can obscure its effectiveness.
\end{tcolorbox}

\subsection{Comparing FEMOSAA with State-of-the-art Frameworks} 

To further evaluate the effectiveness of FEMOSAA, we compare it with the following state-of-the-art search-based frameworks from the literature:


\textbf{DUSE}\footnote{DUSE is actually the same as FEMOSAA-N with NSGA-II which we have evaluated in the previous sections.}~\cite{duse}\textemdash A representative framework that optimizes SAS using NSGA-II. Since it does not consider dependency and knee in the optimization, we adapt the ad hoc strategies applied by FEMOSAA-D and FEMOSAA-K. As DUSE relies on manual transposition, we have used elitist chromosome representations to ensure fair comparison and to eliminate bias in reproducibility.

\textbf{PLATO}~\cite{plato}\textemdash An approach that applies weighted sum of objectives and Genetic Algorithm, a popular single-objective evolutionary algorithm, in the optimization. It also does not consider dependency in the optimization and thus we use the same ad hoc strategy as FEMOSAA-K. We specify equal weights on the objectives to find the single best solution. As PLATO relies on manual transposition, we have used elitist chromosome representations to ensure fair comparison and to eliminate bias in reproducibility.

\textbf{FUSION}~\cite{fusion}\textemdash A well-known feature model based framework that formulates the SAS optimization as an Integer Programming problem, which assumes aggregate objective function and is resolved by an exact algorithm (we use \emph{Branch-and-Bound} in the experiments). FUSION uses binary representation of the solutions and it considers categorical dependency only, thus when no valid solution found, we then fix the violations on numeric dependency. To avoid unacceptable execution time, we forcibly return the best solution so far when it hits a time threshold, i.e., 40s.

Since the source codes of these state-of-the-art frameworks are not openly accessible, we have reproduced the implementation of the optimization algorithms, which are core to these frameworks, following the exact guidance and setups as mentioned in the work. Relevant open-sourced frameworks (e.g.,  jMetal~\cite{durillo2011jmetal}) are exploited when possible. In particular, we have reproduced the representation of the solution and dependency handling with respect to these algorithms as realized by the state-of-the-art frameworks. The other parts, e.g., the modeling component, are assumed to be identical to FEMOSAA in all experiments. We apply the same set of metrics, statistical test and method for categorizing effect size as discussed earlier.

\begin{figure*}[!t]
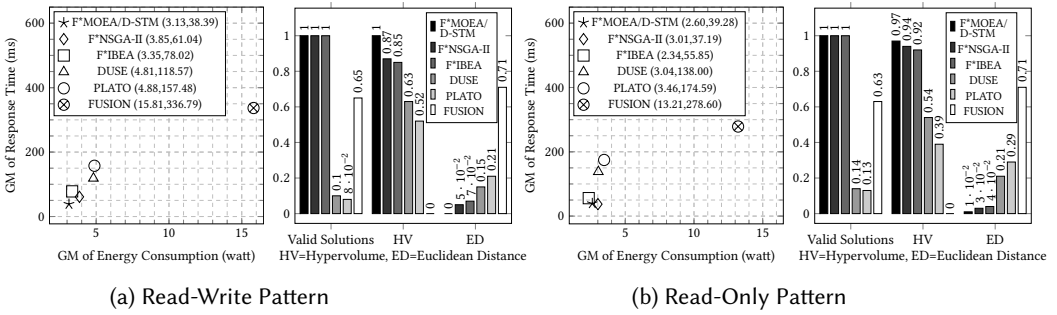

\centering
 \hspace{-1.45cm}
 \begin{subfigure}[t]{0.42\textwidth}
 \centering
\includestandalone[width=0.6\textwidth]{tikz/all-rw-quality}
~
\includestandalone[width=0.6\textwidth]{tikz/all-rw-sta}
    \subcaption{Read-Write Pattern}
    \end{subfigure}
   \hspace{0.9cm}
 \begin{subfigure}[t]{0.42\textwidth}
 \centering
\includestandalone[width=0.6\textwidth]{tikz/all-r-quality}
~
\includestandalone[width=0.6\textwidth]{tikz/all-r-sta}
   \subcaption{Read-Only Pattern}
    \end{subfigure}
  \caption{Comparing FEMOSAA (under different MOEAs) with state-of-the-art frameworks for RUBiS-SAS.. (GM denotes Geometric Mean. F*MOEA/D-STM, F*NSGA-II and F*IBEA denote FEMOSAA with MOEA/D-STM, NSGA-II and IBEA, respectively.)}
 \label{fig:rubis-results}
 \vspace{-0.5cm}
  \end{figure*}

\begin{figure*}[!t]
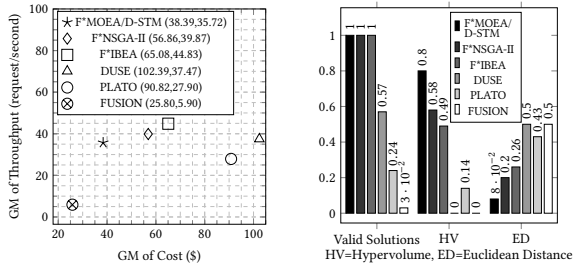

\centering
 \hspace{-1.45cm}
 \begin{subfigure}[t]{0.42\textwidth}
 \centering
\includestandalone[width=0.6\textwidth]{tikz/all-sas-quality}
~ \hspace{0.5cm}
\includestandalone[width=0.6\textwidth]{tikz/all-sas-sta}
    \end{subfigure}
  \caption{Comparing FEMOSAA (under different MOEAs) with state-of-the-art frameworks for SOA-SAS. (GM denotes Geometric Mean. F*MOEA/D-STM, F*NSGA-II and F*IBEA denote FEMOSAA with MOEA/D-STM, NSGA-II and IBEA, respectively.)}
 \label{fig:soa-results}
 \vspace{-0.5cm}
  \end{figure*}

As shown in Figure~\ref{fig:rubis-results}, for both workload patterns on \emph{RUBiS-SAS}, we can see that FEMOSAA with MOEA/D-STM, NSGA-II and IBEA obtain better quality results than the other state-of-the-art frameworks. They also yield better results in terms of HV and ED. The statistical tests and effect size for the comparisons have been shown in Table~\ref{table:state-of-the-arts-statistics}, in which we can see that the improvements obtained by FEMOSAA with all three MOEAs over the other state-of-the-arts frameworks have been statistically significant in general, resulting $p<0.05$ with non-trivial effect sizes on at least one quality attribute. The superiority of FEMOSAA when compared with DUSE again conforms that the combination of elitist representation, the dependency aware operators and knee selection can guide the MOEA to achieve better quality results, even when using different underlying MOEA. PLATO, on the other hand, is further constrained by its nature of weighted sum objective functions, preventing it from finding some of the better trade-off points. Although FUSION applies an exact optimization algorithm which should obtain the optimal solution, the fact of the given large search space has prevented it from reaching such an optimality in reasonable time, henceforth forcing it to terminate when the time threshold is hit. Moreover, since the state-of-the-art frameworks do not consider all dependency, their ratio of valid solutions ranges from 8.39\% to 64.71\% only. For the case of \emph{SOA-SAS}, as shown in Figure~\ref{fig:soa-results} and Table~\ref{table:soa-state-of-the-arts-statistics}, we see that FEMOSAA with all three MOEAs yield significantly better results ($p<0.05$) than PLATO on both quality attribute. Through spending the highest cost, DUSE achieves higher value than that of FEMOSAA on throughput under MOEA/D-STM, but the comparison is statistically insignificant. Further, FEMOSAA is superior to and more balanced than DUSE overall as evidenced by the greatly improved, statistically significant results on rest of the cases, as well as better HV and ED values. We can also observe that FUSION has very competitive result on cost, which is the best for this quality. However, this is due to the fact that the two quality attributes are strongly conflicting, which casues the exact search in FUSION to trap at a local area of the search space within the given time. This eventually resulted in an unwisely strong bias towards the improvement of cost, as evident by the even larger degradation on throughput in contrast to FEMOSAA, as well as the poor HV and ED values. FUSION leads to only 3\% valid solutions as it considers categorical dependencies only while most of the cross-branched dependencies in \emph{SOA-SAS} are numeric.

\begin{table}[t!]
\scriptsize
  \caption{Wilcoxon Signed-Rank Test Results Between FEMOSAA with Different MOEAs and State-of-the-art Frameworks for RUBiS-SAS. (The Significant Statistics, i.e., $p<0.05$, are Highlighted and Shown in Bold)}
\label{table:state-of-the-arts-statistics}

  \subcaption{Read-Write Pattern}
\begin{tabularx}{\textwidth}{p{1cm}YYY|YYY}
\hline 
&
\multicolumn{3}{c|}{\textbf{\emph{$p$ value (effect size) for Response Time (ms)}}} &
\multicolumn{3}{c}{\textbf{\emph{$p$ value (effect size) for Energy Consumption (watt)}}} 

\\ \cline{2-7}


&

FEMOSAA with MOEA/D-STM &
FEMOSAA with NSGA-II &
FEMOSAA with IBEA &
FEMOSAA with MOEA/D-STM &
FEMOSAA with NSGA-II &
FEMOSAA with IBEA 
 \\ \hline

DUSE &
\cellcolor{gray!50}\bfseries .004 (medium) &
\cellcolor{gray!50}\bfseries .043 (small) &
.327 (small) &
\cellcolor{gray!50}\bfseries $<$.001 (large) &
\cellcolor{gray!50}\bfseries .032 (medium) &
\cellcolor{gray!50}\bfseries $<$.001 (large)
 \\

PLATO &
\cellcolor{gray!50}\bfseries $<$.001 (large) &
\cellcolor{gray!50}\bfseries .001 (large) &
\cellcolor{gray!50}\bfseries .015 (medium) &
\cellcolor{gray!50}\bfseries $<$.001 (large) &
\cellcolor{gray!50}\bfseries $<$.001 (large)  &
\cellcolor{gray!50}\bfseries $<$.001 (large)
 \\

FUSION &
\cellcolor{gray!50}\bfseries $<$.001 (large)&
\cellcolor{gray!50}\bfseries $<$.001 (large) &
\cellcolor{gray!50}\bfseries $<$.001 (large)  &
\cellcolor{gray!50}\bfseries $<$.001 (large) &
\cellcolor{gray!50}\bfseries $<$.001 (large)  &
\cellcolor{gray!50}\bfseries $<$.001 (large) 
 \\

\hline 
\\
\end{tabularx}

    \subcaption{Read-Only Pattern}
\begin{tabularx}{\textwidth}{p{1cm}YYY|YYY}
\hline 
&
\multicolumn{3}{c|}{\textbf{\emph{$p$ value (effect size) for Response Time (ms)}}} &
\multicolumn{3}{c}{\textbf{\emph{$p$ value (effect size) for Energy Consumption (watt)}}} 

\\ \cline{2-7}


&

FEMOSAA with MOEA/D-STM &
FEMOSAA with NSGA-II &
FEMOSAA with IBEA &
FEMOSAA with MOEA/D-STM &
FEMOSAA with NSGA-II &
FEMOSAA with IBEA 
 \\ \hline

DUSE &
\cellcolor{gray!50}\bfseries .004 (medium) &
\cellcolor{gray!50}\bfseries $<$.001 (large) &
\cellcolor{gray!50}\bfseries .029 (medium) &
.246 (small) &
.624 (trivial) &
\cellcolor{gray!50}\bfseries .005 (medium)
 \\

PLATO &
\cellcolor{gray!50}\bfseries $<$.001 (large) &
\cellcolor{gray!50}\bfseries $<$.001 (large) &
\cellcolor{gray!50}\bfseries $<$.001 (large) &
\cellcolor{gray!50}\bfseries .010 (medium) &
.303 (small) &
\cellcolor{gray!50}\bfseries .001 (large)
 \\

FUSION &
\cellcolor{gray!50}\bfseries $<$.001 (large)&
\cellcolor{gray!50}\bfseries $<$.001 (large) &
\cellcolor{gray!50}\bfseries $<$.001 (large)  &
\cellcolor{gray!50}\bfseries $<$.001 (large) &
\cellcolor{gray!50}\bfseries $<$.001 (large)  &
\cellcolor{gray!50}\bfseries $<$.001 (large) 
 \\

\hline 
\end{tabularx}

\end{table}

In conclusion, those results suggest that:

\begin{tcolorbox}
\textbf{Answering \emph{RQ4}}\textemdash In contrast to the state-of-the-art frameworks (DUSE, PLATO and FUSION), FEMOSAA achieves better optimized quality for SAS with statistically significant results and non-trivial effect sizes on at least one quality attribute, even when using different underlying MOEAs.
\end{tcolorbox}

\begin{table}[t!]
\scriptsize
  \caption{Wilcoxon Signed-Rank Test Results Between FEMOSAA with Different MOEAs and State-of-the-art Frameworks for SOA-SAS. (The Significant Statistics, i.e., $p<0.05$, are Highlighted and Shown in Bold)}
\label{table:soa-state-of-the-arts-statistics}

\begin{tabularx}{\textwidth}{p{1cm}YYY|YYY}
\hline 
&
\multicolumn{3}{c|}{\textbf{\emph{$p$ value (effect size) for Throughput (request/second)}}} &
\multicolumn{3}{c}{\textbf{\emph{$p$ value (effect size) for Cost (\$)}}} 

\\ \cline{2-7}


&

FEMOSAA with MOEA/D-STM &
FEMOSAA with NSGA-II &
FEMOSAA with IBEA &
FEMOSAA with MOEA/D-STM &
FEMOSAA with NSGA-II &
FEMOSAA with IBEA 
 \\ \hline

DUSE &
.451 (trivial)&
.300 (small) &
\cellcolor{gray!50}\bfseries .007 (medium)  &
\cellcolor{gray!50}\bfseries $<$.001 (large) &
\cellcolor{gray!50}\bfseries $<$.001 (large)  &
\cellcolor{gray!50}\bfseries $<$.001 (large) 
 \\

PLATO &
\cellcolor{gray!50}\bfseries $<$.001 (large)&
\cellcolor{gray!50}\bfseries $<$.001 (large) &
\cellcolor{gray!50}\bfseries $<$.001 (large)  &
\cellcolor{gray!50}\bfseries $<$.001 (large) &
\cellcolor{gray!50}\bfseries $<$.001 (large)  &
\cellcolor{gray!50}\bfseries $<$.001 (large) 
 \\

FUSION &
\cellcolor{gray!50}\bfseries $<$.001 (large)&
\cellcolor{gray!50}\bfseries $<$.001 (large) &
\cellcolor{gray!50}\bfseries $<$.001 (large)  &
\cellcolor{gray!50}\bfseries $<$.001 (large) &
\cellcolor{gray!50}\bfseries $<$.001 (large)  &
\cellcolor{gray!50}\bfseries $<$.001 (large) 
 \\

\hline 
\\
\end{tabularx}

   \vspace{-0.3cm}
  
\end{table}

 \begin{table}[t!]
  \vspace{-0.2cm} 
\scriptsize
  \caption{Comparing Mean Running Time For Producing An Adaptation Solution.}
\label{table:overhead}

\begin{tabularx}{\textwidth}{p{5cm}YYY}
\hline 
&
\textbf{\emph{RUBiS-SAS with Read-Write Pattern}}&
\textbf{\emph{RUBiS-SAS with Read-Only Pattern}}&
\textbf{\emph{SOA-SAS}}

 \\ \hline

FEMOSAA with MOEA/D-STM &
1.96s&
1.88s &
0.30s
 \\
 
 FEMOSAA-N with MOEA/D-STM &
2.04s&
2.5s &
0.43s
 \\

 FEMOSAA-0/1 with MOEA/D-STM &
22.02s&
22.94s &
1.06s
 \\
 
 FEMOSAA with NSGA-II &
0.9s&
0.86s &
0.08s
 \\
 
 FEMOSAA-N with NSGA-II &
0.94s&
3.53s &
0.04s
 \\
 
   FEMOSAA-0/1 with NSGA-II &
10.65s&
10.48s &
0.24s
 \\
 
  FEMOSAA with IBEA &
1.09s&
1.12s &
0.20s
 \\
 
 FEMOSAA-N with IBEA &
1.13s&
1.12s &
0.14s
 \\
 
   FEMOSAA-0/1 with IBEA &
20.90s&
20.59s &
1.45s
 \\

DUSE &
0.94s&
3.53s &
0.04s
 \\
 
PLATO &
0.99s&
0.79s &
0.02s
 \\
 
FUSION&
37.98s&
55.09s &
54.54s
 \\

\hline 
\\
\end{tabularx}

 \vspace{-0.2cm} 
\end{table}

\subsection{Runtime Overhead} 
\label{sec:exp-overhead}
Since we are interested in optimizing SAS at runtime, the running overhead of the MOEAs guided by FEMOSAA is an important aspect to evaluate. In Table~\ref{table:overhead}, we compare the runtime overhead of FEMOSAA, FEMOSAA-N, FEMOSAA-0/1, DUSE, PLATO and FUSION under \emph{RUBiS-SAS}'s two different workload patterns and \emph{SOA-SAS}. For FEMOSAA and its variants, we examine them using all the studied MOEAs. We report on the mean overhead over all the timesteps. As we can see, under all MOEAs and both subject SAS, FEMOSAA (less than 2s) yields much smaller overhead than FEMOSAA-0/1 (up to 22.94s) as the former encodes the elitist features only, which fundamentally reduces the search space and also simplifies the process of generating new solutions (individuals) in MOEAs. Interestingly, we note that for \emph{RUBiS-SAS}, FEMOSAA (0.9s to 1.96s) has slightly smaller overhead than FEMOSAA-N (0.94s to 3.53s), which is quite surprising as we expected that the former should introduce slightly bigger overhead as it exploits additional processes in the reproduction operators and the selection of final solution for adaptation. These results could be attributed to two reasons: (i) the extra efforts spent in dependency aware operators and knee selection are negligible, and (ii) the dependency aware operators tend to produce solutions that can affect the running time of MOEAs, for example, in NSGA-II, if the number of solutions in most of the higher ranked fronts is smaller, then the calculation of their crowding distances would yield less running time. 

In contrast to the state-of-the-art frameworks, FEMOSAA achieves competitive results on runtime overhead, the actual time taken by FEMOSAA depends on the underlying MOEA though. Notably, FUSION has the biggest overhead as exact algorithm fails to scale with large search space and the optimization runs are often forcibly returned as they hit our predefined threshold of 40s.

To summary, the results reveal that:

\begin{tcolorbox}
\textbf{Answering \emph{RQ5}}\textemdash While the running time of FEMOSAA depends on the underlying MOEA, FEMOSAA has very competitive runtime overhead in contrast to the state-of-the-art frameworks. In addition, the extra efforts spent in dependency aware operators and knee selection are negligible; sometimes, they can even slightly speedup the running time of MOEAs.
\end{tcolorbox}



\subsection{Discussion}

\subsubsection{FEMOSAA Benefits and Applicability}
The most notable benefit of FEMOSAA is that it advances the synergy between software engineering for SAS and evolutionary computation. Without in-depth expertise on evolutionary algorithms in general, software engineers are granted the ability to influence the behaviors of a MOEA in a way that they are familiar with, i.e., the feature model design of a SAS. On the other hand, such a design serves as strong domain knowledge that can extend the MOEA and guide the evolutionary search behavior, i.e., in form of elitist chromosome and dependency aware operators, to produce better solutions, .

FEMOSAA exploits MOEA which particularly fits for cases where the relative weights between objectives are unknown or it is too difficult to quantify them. In this perspective, the benefit of FEMOSAA is that it does not require one to specify weights, which could be labor intensive. Indeed, we acknowledge that there are scenarios where the relative importance between conflicting objectives is explicitly known, and their weights can be precisely specified. In those cases, the concept of MOEA and knee selection employed by FEMOSAA might be less sensible. However, it is possible for FEMOSAA to work with single-objective evolutionary algorithms using aggregation of the objectives while deactivating the knee selection, in which case the SAS can still benefit from the power of the elitist chromosome representation and the dependency aware operators.

Another point which is worth mentioning is that MOEA does not guarantee optimal solutions; however, it is very efficient in producing good approximation to complex and non-linear problems that would be otherwise unsolvable by exact optimization. Thus, we would not recommend to use FEMOSAA on SAS that are simple, small in the search space and can be handled by exact search that leads to an optimal solution.

\subsubsection{FEMOSAA Running Time}
Indeed, in contrast to FEMOSAA, the classic rule and policy based decision making approach is very fast when adapting a SAS. However, its effectiveness of adaptation quality relies on several important factors, including: 
\begin{enumerate}[leftmargin=0.5cm]

\item The full knowledge of every possible condition that the SAS may encounter.
\item Some theoretical assumptions on the SAS and the environment that underpins the rules and policies. 
\item The manual reasoning of the optimal adaptation decision under a given condition (i.e., the mapping between a condition and an adaptation decision). 
\end{enumerate}

For SAS works in highly dynamic and uncertain environments (e.g., in cloud computing and software defined networking), rule and policy based approaches would fail due to the fact that they heavily rely on human knowledge and there are emergent conditions that have not been accounted for, given the requirements of points (1) and (2). Further, SAS often has a large variability, i.e., a large number of alternatives as the two SAS we studied. This makes point (3) in the rule and policy based approaches unrealistic; in fact, an exact search optimization would also fail under the problem with a large search space, as such a problem itself is intractable. In contrast, search-based software engineering techniques, particularly evolutionary algorithms, offer a promising way to our SAS optimization problem. This is because being a metaheuristic, evolutionary algorithms are dynamic in nature that are able to perform optimization without in-depth knowledge and assumptions of the problem in hand (e.g., the property of the SAS and the environment). In addition, the notion of natural evolution and population permits its ability to find approximately optimal solutions even for intractable problems.

As we illustrated in Section~\ref{sec:exp-overhead}, FEMOSAA achieves a runtime overhead of seconds under the two SAS studied. This may cause delay in adaptation under extreme scenarios where the transition needs to be completed in microseconds. However, for other cases where the requirements of transition time can be relaxed, the better quality of adaptation that is generated by the underlying evolutionary algorithm has made the cost of its runtime overhead negligible, especially considering that the SAS optimization problem would otherwise be unsolvable with rule and policy based approaches. In fact, as we have shown throughout Section~\ref{sec:exp}, the proposed elitist chromosome representation and dependency aware operators in FEMOSAA have enabled the underlying evolutionary algorithm to reach a better adaptation quality with even smaller runtime overhead.

\subsubsection{Threats to Validity}
Some threats to validity of FEMOSAA can be discussed as the following:

\textbf{Threats to construct validity} is concerned with whether the used metrics can indeed reflect what we intend to measure. In this work, our experiments have selected wider range of quality attributes, i.e., response time, energy consumption, throughput and cost. Those quality attributes and their metrics are the most commonly assessed quality aspects for SAS from the literature~\cite{mingay2007green}\cite{wada2012e3}\cite{fusion}\cite{plato}. Further, we assessed the aggregated results of different quality objectives using HV and ED, which are widely applied metrics to measure the quality of solutions for multi-objective optimization problems~\cite{Wang:2016}\cite{icse-nier}. Threats to construct validity could be also related to the stochastic nature of the considered MOEAs in experiments which can influence the measurements, especially for those that do not account for dependency and knee in the optimization. Indeed, to draw a meaningful conclusion of the measurements for stochastic search-based optimization, repeated runs are necessary as suggested in~\cite{arcuri2011practical}. We have mitigated such a bias by following the design introduced in~\cite{arcuri2011practical}, including: conducting 102 optimization runs\footnote{The number of runs is an implied result of the client emulator's setup for \emph{RUBiS-SAS} and \emph{SOA-SAS}} for FEMOSAA and the others; exploiting a statistical test to verify the significance of comparisons and reporting the effect size for confirming the measured results are not occurred due to chance. Further, in the \emph{RUBiS-SAS} case, FEMOSAA has been evaluated and measured following the realistic FIFA98 workload trace.


\textbf{Threats to internal validity} is related to the values of parameters for the MOEAs. The setup in this work has been carefully tailored such that it produces good trade-off between the quality of optimization and the overhead. However, these values might vary depending on the characteristics in the context, e.g., the given feature model, the types of SAS and the environmental conditions, which itself could be a topic for future research. Threats to internal validity can be also arised from the accidental bugs in the experimental implementations, which is always possible when writing any kind of software, especially for SAS which is naturally more complex than others. However, we have tried to mitigate this unavoidable phenomenon by (i) exploiting well-established open-source framework whenever possible (e.g., jMetal~\cite{durillo2011jmetal}); (ii) following the exact guidance given in some of the compared work; and (iii) debugging through formal software testing procedure. 

\textbf{Threats to external validity} is linked to the benchmark and scenario that are used in the experiments. To improve generalization of the experimental evaluations, we have also evaluated FEMOSAA with three widely applied but distinct MOEAs under two running SAS and different workload patterns, which can diversify the runtime behaviors of the SAS. In particular, one of the SAS (i.e., \emph{RUBiS-SAS}) contains a stack of real-world software, which help to emulate more realistic scenario of the running SAS. Though our experimentation on the cases approximate real and industrial scale, it is difficult to claim complete generality; such claim would require much large number of independent domain-specific cases and needs to be performed by independent future adopters. In future work, we plan to evaluate FEMOSAA in other extreme context, e.g., in mobile environment where computational resource is rather limited. However, it is worth noting that the efforts for building the experiments and deploying the running SAS are non-trivial, which could involve potential expenses.



\section{Related Work}
 \label{sec:rw}
Search-based optimization has been widely applied for SAS, either constitutes as a general framework or being tailored to a specific domain, e.g., service systems or cloud systems. In this section, we provide an overview of the most notable and relevant research in this area while examining them in the light of FEMOSAA.

 \subsection{Evolutionary Optimization for SAS}
A common way of optimizing multiple objectives is to simply aggregate different objectives (e.g., weighted sum) such that they can be resolved in a single objective function. Hence, there exists some work that leverages single-objective evolutionary algorithms for optimizing SAS based on an objective aggregation. PLATO~\cite{plato} and VALKYRIE~\cite{valkyrie} are two examples. However, as opposed to the automatic transposition approach in FEMOSAA, they have relied on manual encoding of the SAS into chromosome representation and they do not consider dependency between features. Further, it is well-known that the relative weights are difficult to be tailored by the engineers; and a single aggregation could restrict the search, causing limitation when searching for good solutions spread over the search space.

Exploiting MOEAs to handle the trade-off between conflicting objectives is an emerging trend for optimizing SAS at runtime. Among others, DUSE~\cite{duse} uses NSGA-II to produce a set of non-dominated solutions for SAS. Similarly, NSGA-II and other EAs, as well as other meta-heuristics algorithms, have also been applied in~\cite{Frey:2013}\cite{nsgaii-cloud1}\cite{nsgaii-cloud2}\cite{tsc-chen-2015}\cite{chen-gecco}, which focus on general SAS and the specific SAS for cloud and service systems. Nevertheless, their encoding of the SAS has been manual and no dependency between features are considered. Consequently, a significant number of function evaluations would be wasted to search for invalid solutions, which as we have shown, can degrade the quality of solutions found. Furthermore, they rely on the non-dominated set, from which any solution can be used for adaptation, which can entail imbalanced trade-off. On contrary, FEMOSAA finds a knee solution that is generally preferable. Moreover, FEMOSAA relies on automatic approach where the elitist features are identified and the dependency is extracted to guide MOEA. 

The closely related and most recent work is probably the one produced by Pascual et al.~\cite{Pascual2015392}, where they proposed a framework for self-adaptive mobile system, which uses the feature model with MOEA while considering feature dependencies in the optimization. However, FEMOSAA is different from their work in various aspects:

\begin{itemize}[leftmargin=0.5cm]

\item Pascual et al.'s work relies on a binary representation of all features and only categorical features are considered; while FEMOSAA encodes elitist features into the chromosome only to form a polyadic representation which reduces the number of genes, and thereby considerably shrinking the search space.

\item Instead of modifying the reproduction operators, the work by Pascual et al. still allow the operators to explore and generate invalid solutions but fixes those solutions using a random repair strategy, i.e., each gene, which violates the categorical dependency, is fixed. However, such a fix is not guided by the dependency chain, thus there is no guarantee that the fixed gene would not cause additional violations (if there is a chain of dependency), and thereby it cannot ensure that a valid solution is always produced. Indeed, in their work, the process is repeated if the previous fix has not resolved all the violations, and the repair stops when a maximum number of repeats has been reached. On the contrary, FEMOSAA extracts dependencies and directly injects them to both the mutation and crossover operators, which are explicitly guided by the dependency chains and the related value trees; this fundamentally prevents invalid solutions from being explored.

\item The nature of binary representation in Pascual et al.'s work implies that it is difficult for them to handle numeric dependencies, which are covered by FEMOSAA.

\item FEMOSAA considers knee selection for adaptation while Pascual et al.'s work selects any non-dominated solution for adaptation, which could be highly imbalanced.

\end{itemize}


 \subsection{Evolutionary Optimization in Software Product Line Engineering}
 
The feature model is widely used in Software Product Line (SPL) for modeling variability, which is similar to our use of feature model on SAS. In particular, MOEA has been applied to SPL, see~\cite{sayyad2013scalable}\cite{sayyad2013value}\cite{sip} for example. However, unlike SPL, where the objectives are highly concerned with designing software products that expose features' richness, diversity and their known defects, \emph{etc}, our focus in SAS is on optimizing non-functional quality attributes, seeking adaptation decisions that can better respond to dynamics and uncertainty in the environment with limited or no human intervention. In addition, SPL assume design time while the SAS mainly focus on runtime.

Although given the differences mentioned above, some SPL work exhibits resemblance to FEMOSAA: they both aim to transpose the design of the feature model into the context of MOEA. Therefore, here we compare FEMOSAA with those SPL preventative approaches, where we specifically look at how the feature model is transposed into MOEA and their strategies to handle feature dependencies. This consideration is particularly essential to explain why it is insufficient to directly apply the transposition used in the existing studies of the SPL domain, and what are the improvements we have made in this aspect. 

Among others, Sayyad et al.~\cite{sayyad2013scalable}\cite{sayyad2013value} exploit NSGA-II and IBEA for finding optimal design of product line using binary encoding of the features into chromosome representation. In contrast, FEMOSAA does not rely on the lengthy binary representation and it encodes a polyadic chromosome using elitists feature, which as we have shown, leading to better optimization results and running time. Hierons et al.~\cite{sip} proposed an extended MOEA encoding method for the feature model, in which the basic encoding is still binary. Similar to our motivation, Hierons et al.~\cite{sip} seek simplification by eliminating features which are the root of a $\mathcal{OR}$ group as their variability can be represented by their children, e.g., \emph{Cache Mode} in Figure~\ref{fig:fm}. However, FEMOSAA goes one step ahead by discarding the binary chromosome representation; this is achieved through selecting the elitists features that cannot be removed without affecting the variability while minimizing the length of encoding, which does not only eliminating any features whose variability can be represented by their children (not restricted to the features that are root of $\mathcal{OR}$ group as in~\cite{sip}), e.g., \emph{Cache} and \emph{Cache Mode} in Figure~\ref{fig:fm}, but also eliminating those whose variability can be represented by their parent, e.g., \emph{CPU}'s children in Figure~\ref{fig:fm}. Unlike their work, we do not simply remove all mandatory features; we retain those with a $\mathcal{XOR}$ group of children as they would help us to considerably simplify the chromosome representation even more (e.g., \emph{CPU} in Figure~\ref{fig:fm}). Further, they have ignored numeric features which are common in SAS. The numeric features, if not specifically handled, would lead to high computational overhead under binary encoding.


Another fact that distinguishes FEMOSAA and existing SPL work on the transposition process is the way in which the dependencies are handled. Hierons et al.~\cite{sip} and Sayyad et al.~\cite{sayyad2013scalable}\cite{sayyad2013value} simply formulate dependency compliance as an additional objective to be optimized. Their formulation stems from the fact that it is too complex to explicitly handle dependency during the evolution due to the inherited difficulty of the binary encoding. The basic assumption is that, with a large number of generations, the MOEA would eventually discovery many valid solutions. While this schema may be sensible for design time optimization problems, it is ill-suited for SAS where the optimizations occur at runtime, because the extra dimension of dependency objective would impose too much additional difficulty of the problem and, at the same time, still fail to guarantee valid solutions. With the binary chromosome representation, Henard et al.~\cite{henard2015combining} aim to overcome this by combing MOEA with an off-the-shelf CSP solver, in which the solvers act similarly to our dependency aware operators. However, those solvers are general and have not been specifically tailored to the needs of problem, which are often computationally expensive and they would not guarantee valid solutions either. FEMOSAA, on the other hand, extracts the dependencies with respect to the elitist chromosome representation at design time, and those dependencies are injected into the operators of MOEA to fundamentally prevent the exploration of invalid solutions during runtime evolution. Further, FEMOSAA handles complex dependencies related to the numeric features.

 \subsection{Non-evolutionary Optimization for SAS}
Beside evolutionary algorithms, exact algorithms are also utilized for optimizing SAS~\cite{fusion}\cite{fuzzy}\cite{moses}, as they guarantee to find the optimal solution and can easily work with single objective. Here, we aim to consider the most noticeable non-evolutionary optimization approaches for SAS which do not tie to the special characteristics of a specific application domain. FUSION~\cite{fusion} is a general framework that optimizes SAS using exact algorithm. At design time, it also applies the feature model to represent the design of a SAS. However, unlike FEMOSAA, it does not consider numeric features and they formulate the problem as Integer Optimization problem using binary encoding of all the features. Eshafani et al.~\cite{fuzzy} also follow similar idea of FUSION but they additionally use Fuzzy logic to constitute the objective functions. MOSES~\cite{moses} is a framework that is designed for self-adaptive service systems, where the optimization is formulated as Linear Programming problem that is convex and can be solved exactly. 

One fundamental issue with the non-evolutionary and exact algorithm is that they fail to scale when the search space is large, which is often the case for modern SAS. Furthermore, exact approaches tend to be highly sensitive to the nature of the problem, e.g., whether they are convex or concave. This could impose extra difficulty because the analysis of problem nature for SAS is very difficult, if not impossible, due to the dynamic and uncertain nature of the context. Additionally, exploiting exact algorithms often need to work on an objective aggregation, which limits their applicability and capability.

In contrast, FEMOSAA relies on MOEA, which is a type of stochastic evolutionary optimization algorithms that is specifically designed to handle multi-objective optimization problems. It is known that MOEA can efficiently find optimal (or near-optimal) solutions for problems with large search space and is capable to reveal fine-grained trade-off surface without the need of objective aggregation. In addition, MOEA is problem agnostic and thereby it is less sensitive to the nature of a given problem.

  \subsection{Other Approaches for Self-Adaptation at Runtime}
Advanced control theory has also been used for SAS decision making because of their low latency. Among others, Filieri et al.~\cite{filieri2015automated} propose a multi-objective controller where each objective would have independent sensors and actuators; the reasoning relies on aggregation of objectives, however. Simplex~\cite{shevtsov2016keep} is also another recent control theoretic method for SAS, where simplex optimization algorithm is used in conjunction with the updates of controller's gains.
  
Reinforcement Learning (RL)~\cite{bu2013coordinated} is another thread that regards the SAS optimization problem as a learning problem. However, in RL, there is no explicit optimization process due to the absence of clear model, and therefore, the adaptation decision is often tailored in a trial-and-error manner which could impose an expensive exploration phase.

Without an explicit search behavior, both control theoretic and learning based approaches lack the ability to perform exploration without affecting the SAS. In addition, they do not consider dependency constraints and they are difficult to be adopted for effectively reasoning about trade-offs at runtime. In contrast, FEMOSAA exploits MOEA and the feature model, aiming to explicitly handle multi-objective optimization while considering dependencies.


\section{Conclusion}
 \label{sec:con}
This paper presents FEMOSAA, a novel framework that systematically and automatically synergizes the feature model of SAS and a MOEA, to optimize the SAS at runtime.  At design time, FEMOSAA finds the elitist features, including categorical and numeric ones, to create a polyadic chromosome representation; it then extracts the dependencies between the genes, which are then used to extend the underlying MOEA for runtime optimization. The feature model serves as the engineers' domain knowledge that can reduce the search space (fundamentally and virtually) and guide the search, henceforth increasing the chance for finding better solutions. Further, FEMOSAA finds the knee solutions that achieve a balanced trade-off. By extensively comparing FEMOSAA with its variants and the state-of-the-art frameworks on two complex real-world SAS, using three widely applied MOEAs and under two workload patterns for optimizing various conflicting objectives, we show that FEMOSAA produces statistically better and more balanced results for trade-off with reasonable overhead. In particular, the most notable observations of FEMOSAA are that:

\begin{itemize}[leftmargin=0.5cm]

\item  Applying the elitist chromosome representation to encode the problem into MOEA helps to produce better quality and smaller runtime overhead for optimizing SAS, but such improvement tends to become marginal when the number of effected solution is small, e.g., the workload is heavy.

\item The dependency aware operators can properly guide the search, finding solutions with better convergence and diversity, leading to better quality of SAS optimization.  However, applying dependency aware operators without ensuring the balance of the selected adaptation solution might obscure its effectiveness.

\item The knee selection helps to find more balanced solution for adaptation. However, applying knee selection without ensuring the quality of searched valid solution can obscure its effectiveness.

\item In contrast to the state-of-the-arts framework from the literature, FEMOSAA, with all the three studied MOEAs, produces statistically and practically better quality for optimizing SAS at runtime. 


\item Overall, FEMOSAA has very competitive runtime overhead in contrast to the state-of-the-art frameworks. Further, the extra efforts spent in dependency aware operators and knee selection are negligible; sometimes, they can even slightly speedup the running time of MOEAs.

\end{itemize}

Our work impacts and advances the synergy between software engineering for SAS and evolutionary computation, combining the strengths from both fields. Particularly, with FEMOSAA, software engineers can exploit MOEAs to tackle SAS optimization without prior extensive expertise of the MOEA. On the other hand, automatic transposition of the feature model into MOEA's context can improve MOEA, and make the domain knowledge systematic and comprehensible for MOEA researchers, who in turn can design more effective algorithms for SAS. In contrast to many SBSE work, our deeper synergy takes one step further by automatically and dynamically extracting the domain information of SAS to extend the internal structure of MOEA. In future work, we plan to apply FEMOSAA in other domains of SAS and extend it for managing more conflicting objectives.




\balance

\bibliographystyle{acm}
\bibliography{references}

\end{document}

%% file: tikz/knee-example.tex
\begin{tikzpicture}
    \begin{axis}[
      width=5cm,
    height=5cm,
        xlabel=Objective 1,
        ylabel=Objective 2,
       legend style={at={(0.3,0.8),font=\fontsize{7}{8}\selectfont},
      anchor=north west,legend columns=1}]

\addplot[only marks,mark=o] plot coordinates {
(0.0,1.0)
(0.01,0.9)
(0.02,0.8)
(0.06,0.7)
(0.12,0.6)
(0.2,0.5)
(0.3,0.4)
(0.42,0.3)
(0.56,0.2)
(0.72,0.1)
(0.9,0.0)

};

\addplot[only marks,mark=*,black] plot coordinates {
(0.3,0.4)
(0.42,0.3)
};

\addlegendentry{pareto-optimal}
\addlegendentry{knee}
    \end{axis}
    \end{tikzpicture}

%% file: tikz/knee.tex
\begin{tikzpicture}
\begin{axis}[%
width=5cm,
    height=5cm,
    xlabel=Energy Consumption (watt),
    ylabel=Response Time (ms),
     legend style={at={(0.5,0.9),font=\fontsize{7}{8}\selectfont},
      anchor=north west,legend columns=1}]
\addplot[only marks,mark=*,black] plot coordinates {
(2.8,31)
    };

\addplot[only marks,mark=otimes*,style={fill=lightgray}] plot coordinates {
(2.6,35.4)
(4.2,29.5)
};

\addplot[only marks,mark=*,style={fill=lightgray}] plot coordinates {
(2.7,33.7)
(3.5,30.2)
(3.2,30.9)
    };

    \draw (axis cs:2.6,35.4) -- (axis cs:4.2,29.5);
    \draw [dashed] (axis cs:2.8,31) -- (axis cs:3.36,32.6);
   \addlegendentry{knee}
\addlegendentry{extreme}
\end{axis}
\end{tikzpicture}

%% file: tikz/workload.tex
\begin{tikzpicture}
    \begin{axis}[
      width=15cm,
    height=6cm,
    xmin=0,
    xmax=24,
        xlabel=Days,
        ylabel=Users Count]
    \addplot[smooth,mark=o,blue] plot coordinates {
   (0,47)
(1,121)
(2,167)
(3,419)
(4,436)
(5,407)
(6,196)
(7,175)
(8,483)
(9,339)
(10,315)
(11,299)
(12,280)
(13,144)
(14,143)
(15,305)
(16,402)
(17,405)
(18,421)
(19,573)
(20,178)
(21,159)
(22,518)
(23,610)
(24,236)
    };
   
    \end{axis}
    \end{tikzpicture}

%% file: tikz/soa-env.tex
\begin{tikzpicture}
    \begin{axis}[
      width=15cm,
    height=6cm,
    xmin=0,
    xmax=102,
        xlabel=Time point,
        ylabel=Relative Standard Deviation]
    \addplot[smooth,mark=o,blue] plot coordinates {
  (1,0.5598781683545676)
(2,0.5577668885764526)
(3,0.5560550466922252)
(4,0.5809903481804344)
(5,0.5557609833783113)
(6,0.5631040546927574)
(7,0.5671721812717974)
(8,0.5766828924709453)
(9,0.5754014716777731)
(10,0.5692605758674869)
(11,0.5704293017401422)
(12,0.5684793300168369)
(13,0.5699545355521076)
(14,0.5709104102942282)
(15,0.5779262774729406)
(16,0.5819186376436061)
(17,0.5831649836057103)
(18,0.5850040750657443)
(19,0.5882151340824664)
(20,0.6055284484486501)
(21,0.59822638865834)
(22,0.6266092641264538)
(23,0.6014554601526401)
(24,0.5852442142164849)
(25,0.593872306386467)
(26,0.6221926457450775)
(27,0.5972373768865806)
(28,0.5892310884289209)
(29,0.5891902887612805)
(30,0.5876254688957394)
(31,0.6062978870887651)
(32,0.5872744997642643)
(33,0.6074643819687763)
(34,0.5916490306891786)
(35,0.6022175078177188)
(36,0.5929133813546841)
(37,0.5954089758663231)
(38,0.5935352473245473)
(39,0.6175658520543852)
(40,0.5903672231887879)
(41,0.5948379331877979)
(42,0.6067661240518817)
(43,0.614347087762929)
(44,0.6130868237344695)
(45,0.6071120704454128)
(46,0.6146365786009473)
(47,0.6118323045204574)
(48,0.6109308191818994)
(49,0.6152851968400803)
(50,0.6396099587106816)
(51,0.643473325015516)
(52,0.6380659462414335)
(53,0.6432235091700665)
(54,0.6464737050439574)
(55,0.6364237905437392)
(56,0.6299163530862999)
(57,0.6574113701667911)
(58,0.6323743237406728)
(59,0.6164543870470915)
(60,0.6183014560768237)
(61,0.6461649056588066)
(62,0.6216241863115287)
(63,0.613709800348128)
(64,0.642942462908493)
(65,0.6413019400956)
(66,0.6591597857990978)
(67,0.6409199125240554)
(68,0.654635884945472)
(69,0.6385065589059731)
(70,0.6379693264332971)
(71,0.6286908164073939)
(72,0.631185420171705)
(73,0.6448295005960204)
(74,0.6681871669396009)
(75,0.6395190041886842)
(76,0.6417454711883992)
(77,0.6580075075101048)
(78,0.684244676693945)
(79,0.6831404348960315)
(80,0.6775542607090856)
(81,0.6844957084116058)
(82,0.6633908750118558)
(83,0.6708962312196841)
(84,0.6624243530555167)
(85,0.665672811618681)
(86,0.6844924990704713)
(87,0.658227844100732)
(88,0.685237380964268)
(89,0.68845407037546)
(90,0.6752052045221766)
(91,0.6785059074148766)
(92,0.6954651514864651)
(93,0.6679710845652012)
(94,0.6555291491680046)
(95,0.6551930770487524)
(96,0.6659558337755469)
(97,0.6407473857344709)
(98,0.6331674575829247)
(99,0.6625893707251804)
(100,0.6324681297904391)
(101,0.657699365974985)
(102,0.6323272929178849)
    };
   
    \end{axis}
    \end{tikzpicture}

%% file: tikz/moead-rw-quality.tex
\begin{tikzpicture}
    \begin{axis}[ 
        grid = both,
        minor xtick={2,2.5,...,6},
        minor ytick={0,50,...,500},
        grid style={dashed},
      width=7cm,
    height=7cm,
    xmax=6.3,
    ymax=500,
    xmin=2,
    ymin=0,
          ylabel=GM of Response Time (ms),
        xlabel=GM of Energy Consumption (watt),
          y label style={align=center,at={(0.05,0.5)}},
       legend style={at={(0.02,0.98),font=\fontsize{9}{6}\selectfont},
      anchor=north west,legend columns=1}]

\addplot[only marks,mark=star,mark size=4pt] plot coordinates {
(3.13,38.39)
};

\addplot[only marks,mark=diamond,mark size=4pt] plot coordinates {
(3.60,115.59)
};

\addplot[only marks,mark=square,mark size=4pt] plot coordinates {
(3.26,103.99)
};

\addplot[only marks,mark=triangle,mark size=4pt] plot coordinates {
(3.80,83.62)
};

\addplot[only marks,mark=o,mark size=4pt] plot coordinates {
(4.91,168.47)
};

\addlegendentry{FEMOSAA (3.13,38.39)}
\addlegendentry{FEMOSAA-K (3.60,115.59)}
\addlegendentry{FEMOSAA-D (3.26,103.99)}
\addlegendentry{FEMOSAA-N (3.80,83.62)}
\addlegendentry{FEMOSAA-0/1 (4.91,168.47)}
    \end{axis}
    \end{tikzpicture}

%% file: tikz/moead-rw-sta.tex
\begin{tikzpicture}
\begin{axis}[
    ybar=0.05cm,
    bar width=0.25cm,
 enlarge x limits=0.3,
 enlarge y limits=0.05,
 ymax=1.1,
    legend style={at={(0.65,0.98),font=\fontsize{8.5}{6}\selectfont},
      anchor=north,legend columns=1},
      nodes near coords,
      every node near coord/.append style={rotate=90, anchor=west},
      xlabel={HV=Hypervolume, ED=Euclidean Distance},
          y label style={at={(0.15,0.5)}},
    symbolic x coords={Valid Solutions,HV,ED},
    xtick=data,
    width=7cm,
    height=6.9cm
    ]
    
    \addplot[fill=black!100!white] coordinates {(Valid Solutions,1) (HV,1) (ED,0)};
\addplot[fill=black!60!white] coordinates {(Valid Solutions,0.13) (HV,0.30) (ED,0.33)};
\addplot[fill=black!40!white] coordinates {(Valid Solutions,1) (HV,0.46) (ED,0.26)};
\addplot[fill=black!20!white] coordinates {(Valid Solutions,0.09) (HV,0.41) (ED,0.26)};
\addplot[fill=white!100!white] coordinates {(Valid Solutions,0) (HV,0) (ED,0.71)};

\legend{FEMOSAA,FEMOSAA-K,FEMOSAA-D,FEMOSAA-N,FEMOSAA-0/1}

\end{axis}
\end{tikzpicture}

%% file: tikz/nsgaii-rw-quality.tex
\begin{tikzpicture}
    \begin{axis}[ 
        grid = both,
          minor xtick={2,2.5,...,6},
        minor ytick={0,50,...,500},
        grid style={dashed},
      width=7cm,
    height=7cm,
    xmax=6.3,
    ymax=500,
    xmin=2,
    ymin=0,
          ylabel=GM of Response Time (ms),
        xlabel=GM of Energy Consumption (watt),
          y label style={align=center,at={(0.05,0.5)}},
       legend style={at={(0.02,0.98),font=\fontsize{9}{6}\selectfont},
      anchor=north west,legend columns=1}]

\addplot[only marks,mark=star,mark size=4pt] plot coordinates {
(3.85,61.04)
};

\addplot[only marks,mark=diamond,mark size=4pt] plot coordinates {
(4.67,108.64)
};

\addplot[only marks,mark=square,mark size=4pt] plot coordinates {
(4.61,114.93)
};

\addplot[only marks,mark=triangle,mark size=4pt] plot coordinates {
(4.81,118.57)
};

\addplot[only marks,mark=o,mark size=4pt] plot coordinates {
(6.08,380.73)
};

\addlegendentry{FEMOSAA (3.85,61.04)}
\addlegendentry{FEMOSAA-K (4.67,108.64)}
\addlegendentry{FEMOSAA-D (4.61,114.93)}
\addlegendentry{FEMOSAA-N (4.81,118.57)}
\addlegendentry{FEMOSAA-0/1 (6.08,380.73)}
    \end{axis}
    \end{tikzpicture}

%% file: tikz/nsgaii-rw-sta.tex
\begin{tikzpicture}
\begin{axis}[
    ybar=0.05cm,
    bar width=0.25cm,
 enlarge x limits=0.3,
 enlarge y limits=0.05,
 ymax=1.1,
    legend style={at={(0.65,0.98),font=\fontsize{8.5}{6}\selectfont},
      anchor=north,legend columns=1},
      nodes near coords,
      every node near coord/.append style={rotate=90, anchor=west},
      xlabel={HV=Hypervolume, ED=Euclidean Distance},
          y label style={at={(0.15,0.5)}},
    symbolic x coords={Valid Solutions,HV,ED},
    xtick=data,
    width=7cm,
    height=6.9cm
    ]
    
    \addplot[fill=black!100!white] coordinates {(Valid Solutions,1) (HV,1) (ED,0)};
\addplot[fill=black!60!white] coordinates {(Valid Solutions,0.12) (HV,0.54) (ED,0.20)};
\addplot[fill=black!40!white] coordinates {(Valid Solutions,1) (HV,0.55) (ED,0.19)};
\addplot[fill=black!20!white] coordinates {(Valid Solutions,0.1) (HV,0.47) (ED,0.23)};
\addplot[fill=white!100!white] coordinates {(Valid Solutions,0) (HV,0) (ED,0.71)};

\legend{FEMOSAA,FEMOSAA-K,FEMOSAA-D,FEMOSAA-N,FEMOSAA-0/1}

\end{axis}
\end{tikzpicture}

%% file: tikz/ibea-rw-quality.tex
\begin{tikzpicture}
    \begin{axis}[ 
        grid = both,
          minor xtick={2,2.5,...,6},
        minor ytick={0,50,...,500},
        grid style={dashed},
      width=7cm,
    height=7cm,
    xmax=6.3,
    ymax=500,
    xmin=2,
    ymin=0,
          ylabel=GM of Response Time (ms),
        xlabel=GM of Energy Consumption (watt),
          y label style={align=center,at={(0.05,0.5)}},
       legend style={at={(0.02,0.98),font=\fontsize{9}{6}\selectfont},
      anchor=north west,legend columns=1}]

\addplot[only marks,mark=star,mark size=4pt] plot coordinates {
(3.35,78.02)
};

\addplot[only marks,mark=diamond,mark size=4pt] plot coordinates {
(4.64,102.27)
};

\addplot[only marks,mark=square,mark size=4pt] plot coordinates {
(4.38,147.00)
};

\addplot[only marks,mark=triangle,mark size=4pt] plot coordinates {
(4.65,92.72)
};

\addplot[only marks,mark=o,mark size=4pt] plot coordinates {
(5.41,285.64)
};

\addlegendentry{FEMOSAA (3.35,78.02)}
\addlegendentry{FEMOSAA-K (4.64,102.27)}
\addlegendentry{FEMOSAA-D (4.38,147.00)}
\addlegendentry{FEMOSAA-N (4.65,92.72)}
\addlegendentry{FEMOSAA-0/1 (5.41,285.64)}
    \end{axis}
    \end{tikzpicture}

%% file: tikz/ibea-rw-sta.tex
\begin{tikzpicture}
\begin{axis}[
    ybar=0.05cm,
    bar width=0.25cm,
 enlarge x limits=0.3,
 enlarge y limits=0.05,
 ymax=1.1,
    legend style={at={(0.65,0.98),font=\fontsize{8.5}{6}\selectfont},
      anchor=north,legend columns=1},
      nodes near coords,
      every node near coord/.append style={rotate=90, anchor=west},
      xlabel={HV=Hypervolume, ED=Euclidean Distance},
          y label style={at={(0.15,0.5)}},
    symbolic x coords={Valid Solutions,HV,ED},
    xtick=data,
    width=7cm,
    height=6.9cm
    ]
    
    \addplot[fill=black!100!white] coordinates {(Valid Solutions,1) (HV,1) (ED,0)};
\addplot[fill=black!60!white] coordinates {(Valid Solutions,0.16) (HV,0.33) (ED,0.32)};
\addplot[fill=black!40!white] coordinates {(Valid Solutions,1) (HV,0.33) (ED,0.30)};
\addplot[fill=black!20!white] coordinates {(Valid Solutions,0.14) (HV,0.34) (ED,0.32)};
\addplot[fill=white!100!white] coordinates {(Valid Solutions,0) (HV,0) (ED,0.71)};

\legend{FEMOSAA,FEMOSAA-K,FEMOSAA-D,FEMOSAA-N,FEMOSAA-0/1}

\end{axis}
\end{tikzpicture}

%% file: tikz/moead-r-quality.tex
\begin{tikzpicture}
    \begin{axis}[ 
        grid = both,
         minor xtick={2,2.5,...,6},
        minor ytick={0,50,...,500},
        grid style={dashed},
      width=7cm,
    height=7cm,
    xmax=6.3,
    ymax=500,
    xmin=2,
    ymin=0,
          ylabel=GM of Response Time (ms),
        xlabel=GM of Energy Consumption (watt),
          y label style={align=center,at={(0.05,0.5)}},
       legend style={at={(0.02,0.98),font=\fontsize{9}{6}\selectfont},
      anchor=north west,legend columns=1}]

\addplot[only marks,mark=star,mark size=4pt] plot coordinates {
(2.60,30.28)
};

\addplot[only marks,mark=diamond,mark size=4pt] plot coordinates {
(3.34,80.04)
};

\addplot[only marks,mark=square,mark size=4pt] plot coordinates {
(2.81,85.84)
};

\addplot[only marks,mark=triangle,mark size=4pt] plot coordinates {
(3.07,99.52)
};

\addplot[only marks,mark=o,mark size=4pt] plot coordinates {
(3.38,262.38)
};

\addlegendentry{FEMOSAA (2.60,30.28)}
\addlegendentry{FEMOSAA-K (3.34,80.04)}
\addlegendentry{FEMOSAA-D (2.81,85.84)}
\addlegendentry{FEMOSAA-N (3.07,99.52)}
\addlegendentry{FEMOSAA-0/1 (3.38,262.38)}
    \end{axis}
    \end{tikzpicture}

%% file: tikz/moead-r-sta.tex
\begin{tikzpicture}
\begin{axis}[
    ybar=0.05cm,
    bar width=0.25cm,
 enlarge x limits=0.3,
 enlarge y limits=0.05,
 ymax=1.1,
    legend style={at={(0.65,0.98),font=\fontsize{8.5}{6}\selectfont},
      anchor=north,legend columns=1},
      nodes near coords,
      every node near coord/.append style={rotate=90, anchor=west},
      xlabel={HV=Hypervolume, ED=Euclidean Distance},
          y label style={at={(0.15,0.5)}},
    symbolic x coords={Valid Solutions,HV,ED},
    xtick=data,
    width=7cm,
    height=6.9cm
    ]
    
    \addplot[fill=black!100!white] coordinates {(Valid Solutions,1) (HV,1) (ED,0)};
\addplot[fill=black!60!white] coordinates {(Valid Solutions,0.09) (HV,0.04) (ED,0.48)};
\addplot[fill=black!40!white] coordinates {(Valid Solutions,1) (HV,0.57) (ED,0.17)};
\addplot[fill=black!20!white] coordinates {(Valid Solutions,0.15) (HV,0.28) (ED,0.33)};
\addplot[fill=white!100!white] coordinates {(Valid Solutions,0) (HV,0) (ED,0.71)};

\legend{FEMOSAA,FEMOSAA-K,FEMOSAA-D,FEMOSAA-N,FEMOSAA-0/1}

\end{axis}
\end{tikzpicture}

%% file: tikz/nsgaii-r-quality.tex
\begin{tikzpicture}
    \begin{axis}[ 
        grid = both,
         minor xtick={2,2.5,...,6},
        minor ytick={0,50,...,500},
        grid style={dashed},
      width=7cm,
    height=7cm,
    xmax=6.3,
    ymax=500,
    xmin=2,
    ymin=0,
          ylabel=GM of Response Time (ms),
        xlabel=GM of Energy Consumption (watt),
          y label style={align=center,at={(0.05,0.5)}},
       legend style={at={(0.02,0.98),font=\fontsize{9}{6}\selectfont},
      anchor=north west,legend columns=1}]

\addplot[only marks,mark=star,mark size=4pt] plot coordinates {
(3.01,37.19)
};

\addplot[only marks,mark=diamond,mark size=4pt] plot coordinates {
(3.08,94.00)
};

\addplot[only marks,mark=square,mark size=4pt] plot coordinates {
(3.05,89.56)
};

\addplot[only marks,mark=triangle,mark size=4pt] plot coordinates {
(3.04,138.00)
};

\addplot[only marks,mark=o,mark size=4pt] plot coordinates {
(4.47,263.77)
};

\addlegendentry{FEMOSAA (3.01,37.19)}
\addlegendentry{FEMOSAA-K (3.08,94.00)}
\addlegendentry{FEMOSAA-D (3.05,89.56)}
\addlegendentry{FEMOSAA-N (3.04,138.00)}
\addlegendentry{FEMOSAA-0/1 (4.47,263.77)}
    \end{axis}
    \end{tikzpicture}

%% file: tikz/nsgaii-r-sta.tex
\begin{tikzpicture}
\begin{axis}[
    ybar=0.05cm,
    bar width=0.25cm,
 enlarge x limits=0.3,
 enlarge y limits=0.05,
 ymax=1.1,
    legend style={at={(0.75,0.98),font=\fontsize{8.5}{6}\selectfont},
      anchor=north,legend columns=1},
      nodes near coords,
      every node near coord/.append style={rotate=90, anchor=west},
      xlabel={HV=Hypervolume, ED=Euclidean Distance},
          y label style={at={(0.15,0.5)}},
    symbolic x coords={Valid Solutions,HV,ED},
    xtick=data,
    width=7cm,
    height=6.9cm
    ]
    
    \addplot[fill=black!100!white] coordinates {(Valid Solutions,1) (HV,1) (ED,0)};
\addplot[fill=black!60!white] coordinates {(Valid Solutions,0.13) (HV,0.71) (ED,0.13)};
\addplot[fill=black!40!white] coordinates {(Valid Solutions,1) (HV,0.75) (ED,0.12)};
\addplot[fill=black!20!white] coordinates {(Valid Solutions,0.14) (HV,0.54) (ED,0.22)};
\addplot[fill=white!100!white] coordinates {(Valid Solutions,0) (HV,0) (ED,0.71)};

\legend{FEMOSAA,FEMOSAA-K,FEMOSAA-D,FEMOSAA-N,FEMOSAA-0/1}

\end{axis}
\end{tikzpicture}

%% file: tikz/ibea-r-quality.tex
\begin{tikzpicture}
    \begin{axis}[ 
        grid = both,
         minor xtick={2,2.5,...,6},
        minor ytick={0,50,...,500},
        grid style={dashed},
      width=7cm,
    height=7cm,
    xmax=6.3,
    ymax=500,
    xmin=2,
    ymin=0,
          ylabel=GM of Response Time (ms),
        xlabel=GM of Energy Consumption (watt),
          y label style={align=center,at={(0.05,0.5)}},
       legend style={at={(0.02,0.98),font=\fontsize{9}{6}\selectfont},
      anchor=north west,legend columns=1}]

\addplot[only marks,mark=star,mark size=4pt] plot coordinates {
(2.34,55.85)
};

\addplot[only marks,mark=diamond,mark size=4pt] plot coordinates {
(2.93,78.86)
};

\addplot[only marks,mark=square,mark size=4pt] plot coordinates {
(2.76,56.30)
};

\addplot[only marks,mark=triangle,mark size=4pt] plot coordinates {
(2.94,57.93)
};

\addplot[only marks,mark=o,mark size=4pt] plot coordinates {
(3.39,118.13)
};

\addlegendentry{FEMOSAA (2.34,55.85)}
\addlegendentry{FEMOSAA-K (2.93,78.86)}
\addlegendentry{FEMOSAA-D (2.76,56.30)}
\addlegendentry{FEMOSAA-N (2.94,57.93)}
\addlegendentry{FEMOSAA-0/1 (3.39,118.13)}
    \end{axis}
    \end{tikzpicture}

%% file: tikz/ibea-r-sta.tex
\begin{tikzpicture}
\begin{axis}[
    ybar=0.05cm,
    bar width=0.25cm,
 enlarge x limits=0.3,
 enlarge y limits=0.05,
 ymax=1.1,
    legend style={at={(0.68,0.98),font=\fontsize{8.5}{6}\selectfont},
      anchor=north,legend columns=1},
      nodes near coords,
      every node near coord/.append style={rotate=90, anchor=west},
      xlabel={HV=Hypervolume, ED=Euclidean Distance},
          y label style={at={(0.15,0.5)}},
    symbolic x coords={Valid Solutions,HV,ED},
    xtick=data,
    width=7cm,
    height=6.9cm
    ]
    
    \addplot[fill=black!100!white] coordinates {(Valid Solutions,1) (HV,1) (ED,0)};
\addplot[fill=black!60!white] coordinates {(Valid Solutions,0.11) (HV,0.28) (ED,0.34)};
\addplot[fill=black!40!white] coordinates {(Valid Solutions,1) (HV,0.59) (ED,0.20)};
\addplot[fill=black!20!white] coordinates {(Valid Solutions,0.14) (HV,0.41) (ED,0.29)};
\addplot[fill=white!100!white] coordinates {(Valid Solutions,0) (HV,0) (ED,0.71)};

\legend{FEMOSAA,FEMOSAA-K,FEMOSAA-D,FEMOSAA-N,FEMOSAA-0/1}

\end{axis}
\end{tikzpicture}

%% file: tikz/moead-sas-quality.tex
\begin{tikzpicture}
    \begin{axis}[ 
        grid = both,
        minor xtick={10,20,...,130},
        minor ytick={0,5,...,80},
        grid style={dashed},
      width=7cm,
    height=7cm,
    xmax=130,
    ymax=80,
        ymin=10,
    xmin=30,
          ylabel=GM of Throughput (request/second),
        xlabel=GM of Cost (\$),
          y label style={align=center,at={(0.05,0.5)}},
       legend style={at={(0.02,0.99),font=\fontsize{9}{6}\selectfont},
      anchor=north west,legend columns=1}]

\addplot[only marks,mark=star,mark size=4pt] plot coordinates {
(38.39,35.72)
};

\addplot[only marks,mark=diamond,mark size=4pt] plot coordinates {
(66.28,37.18)
};

\addplot[only marks,mark=square,mark size=4pt] plot coordinates {
(43.88,27.66)
};

\addplot[only marks,mark=triangle,mark size=4pt] plot coordinates {
(72.98,35.41)
};

\addplot[only marks,mark=o,mark size=4pt] plot coordinates {
(111.60,21.66)
};

\addlegendentry{FEMOSAA (38.39,35.72)}
\addlegendentry{FEMOSAA-K (66.28,37.18)}
\addlegendentry{FEMOSAA-D (43.88,27.66)}
\addlegendentry{FEMOSAA-N (72.98,35.41)}
\addlegendentry{FEMOSAA-0/1 (111.60,21.66)}
    \end{axis}
    \end{tikzpicture}

%% file: tikz/moead-sas-sta.tex
\begin{tikzpicture}
\begin{axis}[
    ybar=0.07cm,
    bar width=0.2cm,
 enlarge x limits=0.25,
 enlarge y limits=0.05,
 ymax=1.1,
    legend style={cells={align=left},at={(0.72,0.98),font=\fontsize{8.5}{6}\selectfont},
      anchor=north,legend columns=1},
      nodes near coords,
      every node near coord/.append style={rotate=90, anchor=west},
      xlabel={HV=Hypervolume, ED=Euclidean Distance},
          y label style={at={(0.15,0.5)}},
    symbolic x coords={Valid Solutions,HV,ED},
    xtick=data,
    width=7cm,
    height=6.9cm
    ]
    
    \addplot[fill=black!100!white] coordinates {(Valid Solutions,1) (HV,0.94) (ED,0.03)};
\addplot[fill=black!60!white] coordinates {(Valid Solutions,0.83) (HV,0.62) (ED,0.19)};
\addplot[fill=black!40!white] coordinates {(Valid Solutions,1) (HV,0.48) (ED,0.24)};
\addplot[fill=black!20!white] coordinates {(Valid Solutions,0.84) (HV,0.49) (ED,0.24)};
\addplot[fill=white!100!white] coordinates {(Valid Solutions,0) (HV,0) (ED,0.71)};

\legend{FEMOSAA,FEMOSAA-K,FEMOSAA-D,FEMOSAA-N,FEMOSAA-0/1}

\end{axis}
\end{tikzpicture}

%% file: tikz/nsgaii-sas-quality.tex
\begin{tikzpicture}
    \begin{axis}[ 
        grid = both,
        minor xtick={10,20,...,130},
        minor ytick={0,5,...,80},
        grid style={dashed},
      width=7cm,
    height=7cm,
    xmax=130,
    ymax=80,
        ymin=10,
    xmin=30,
          ylabel=GM of Throughput (request/second),
        xlabel=GM of Cost (\$),
          y label style={align=center,at={(0.05,0.5)}},
       legend style={at={(0.02,0.99),font=\fontsize{9}{6}\selectfont},
      anchor=north west,legend columns=1}]

\addplot[only marks,mark=star,mark size=4pt] plot coordinates {
(56.86,39.87)
};

\addplot[only marks,mark=diamond,mark size=4pt] plot coordinates {
(95.11,39.37)
};

\addplot[only marks,mark=square,mark size=4pt] plot coordinates {
(73.55,44.82)
};

\addplot[only marks,mark=triangle,mark size=4pt] plot coordinates {
(102.39,37.47)
};

\addplot[only marks,mark=o,mark size=4pt] plot coordinates {
(125.32,25.69)
};

\addlegendentry{FEMOSAA (56.86,39.87)}
\addlegendentry{FEMOSAA-K (95.11,39.37)}
\addlegendentry{FEMOSAA-D (73.55,44.82)}
\addlegendentry{FEMOSAA-N (102.39,37.47)}
\addlegendentry{FEMOSAA-0/1 (125.32,25.69)}
    \end{axis}
    \end{tikzpicture}

%% file: tikz/nsgaii-sas-sta.tex
\begin{tikzpicture}
\begin{axis}[
    ybar=0.07cm,
    bar width=0.2cm,
 enlarge x limits=0.25,
 enlarge y limits=0.05,
 ymax=1.1,
    legend style={cells={align=left},at={(0.72,0.98),font=\fontsize{8.5}{6}\selectfont},
      anchor=north,legend columns=1},
      nodes near coords,
      every node near coord/.append style={rotate=90, anchor=west},
      xlabel={HV=Hypervolume, ED=Euclidean Distance},
          y label style={at={(0.15,0.5)}},
    symbolic x coords={Valid Solutions,HV,ED},
    xtick=data,
    width=7cm,
    height=6.9cm
    ]
    
      \addplot[fill=black!100!white] coordinates {(Valid Solutions,1) (HV,0.83) (ED,0.08)};
\addplot[fill=black!60!white] coordinates {(Valid Solutions,0.46) (HV,0.36) (ED,0.29)};
\addplot[fill=black!40!white] coordinates {(Valid Solutions,1) (HV,0.76) (ED,0.12)};
\addplot[fill=black!20!white] coordinates {(Valid Solutions,0.50) (HV,0.25) (ED,0.36)};
\addplot[fill=white!100!white] coordinates {(Valid Solutions,0) (HV,0) (ED,0.71)};

\legend{FEMOSAA,FEMOSAA-K,FEMOSAA-D,FEMOSAA-N,FEMOSAA-0/1}

\end{axis}
\end{tikzpicture}

%% file: tikz/ibea-sas-quality.tex
\begin{tikzpicture}
    \begin{axis}[ 
        grid = both,
        minor xtick={10,20,...,130},
        minor ytick={0,5,...,80},
        grid style={dashed},
      width=7cm,
    height=7cm,
    xmax=130,
    ymax=80,
    ymin=10,
    xmin=30,
          ylabel=GM of Throughput (request/second),
        xlabel=GM of Cost (\$),
          y label style={align=center,at={(0.05,0.5)}},
       legend style={at={(0.02,0.99),font=\fontsize{9}{6}\selectfont},
      anchor=north west,legend columns=1}]

\addplot[only marks,mark=star,mark size=4pt] plot coordinates {
(65.07,44.83)
};

\addplot[only marks,mark=diamond,mark size=4pt] plot coordinates {
(99.67,39.31)
};

\addplot[only marks,mark=square,mark size=4pt] plot coordinates {
(77.39,44.82)
};

\addplot[only marks,mark=triangle,mark size=4pt] plot coordinates {
(104.93,36.18)
};

\addplot[only marks,mark=o,mark size=4pt] plot coordinates {
(119.17,23.91)
};

\addlegendentry{FEMOSAA (65.07,44.83)}
\addlegendentry{FEMOSAA-K (99.67,39.31)}
\addlegendentry{FEMOSAA-D (77.39,44.82)}
\addlegendentry{FEMOSAA-N (104.93,36.18)}
\addlegendentry{FEMOSAA-0/1 (119.17,23.91)}
    \end{axis}
    \end{tikzpicture}

%% file: tikz/ibea-sas-sta.tex
\begin{tikzpicture}
\begin{axis}[
    ybar=0.07cm,
    bar width=0.2cm,
 enlarge x limits=0.25,
 enlarge y limits=0.05,
 ymax=1.1,
    legend style={cells={align=left},at={(0.72,0.98),font=\fontsize{8.5}{6}\selectfont},
      anchor=north,legend columns=1},
      nodes near coords,
      every node near coord/.append style={rotate=90, anchor=west},
      xlabel={HV=Hypervolume, ED=Euclidean Distance},
          y label style={at={(0.15,0.5)}},
    symbolic x coords={Valid Solutions,HV,ED},
    xtick=data,
    width=7cm,
    height=6.9cm
    ]
    
       \addplot[fill=black!100!white] coordinates {(Valid Solutions,1) (HV,1) (ED,0)};
\addplot[fill=black!60!white] coordinates {(Valid Solutions,0.33) (HV,0.30) (ED,0.33)};
\addplot[fill=black!40!white] coordinates {(Valid Solutions,1) (HV,0.77) (ED,0.11)};
\addplot[fill=black!20!white] coordinates {(Valid Solutions,0.39) (HV,0.19) (ED,0.39)};
\addplot[fill=white!100!white] coordinates {(Valid Solutions,0) (HV,0) (ED,0.71)};

\legend{FEMOSAA,FEMOSAA-K,FEMOSAA-D,FEMOSAA-N,FEMOSAA-0/1}

\end{axis}
\end{tikzpicture}

%% file: tikz/moead-femosaa.tex
\begin{tikzpicture}
    \begin{axis}[
      width=5cm,
    height=5cm,
    xmin=8,
    xmax=18,
    ymax=100,
    ytick = {0,20,...,100},
        xlabel=Energy Consumption (watt),
        ylabel=Response Time (ms),
       legend style={at={(0.3,0.8),font=\fontsize{7}{8}\selectfont},
      anchor=north west,legend columns=1}]
    \addplot[only marks,mark=o,black,y filter/.code={\pgfmathparse{\pgfmathresult*1000}\pgfmathresult}] plot coordinates {
(17.931876622945012,0.0011829171041075467)
(13.63175954204351,0.004475529340068504)
(12.972632760288313,0.00796455359060753)
(12.769868856597657,0.017365779627769248)
(12.256453095774459,0.016414715460830953)
(11.644630582916541,0.022160862129241246)
(11.498234740325886,0.022667543940958653)
(11.46213455601667,0.023984031041454013)
(11.46213455601667,0.023984031041454013)
(11.41807408986925,0.02241312464199037)
(11.433972056033559,0.023952129651808943)
(11.332611201384484,0.023947333210749093)
(11.41807408986925,0.02241312464199037)
(9.87938135562867,0.02356909115865221)
(11.433972056033559,0.023952129651808943)
(9.523165032636186,0.023617691529340398)
(11.332611201384484,0.023947333210749093)
(11.332611201384484,0.023947333210749093)
(11.398899716555427,0.02351239062097954)
(11.41807408986925,0.02241312464199037)
(11.433972056033559,0.023952129651808943)
(11.462730728429555,0.0924283630542223)
(11.297765066600828,0.09067786245731649)
(11.297765066600828,0.09067786245731649)
(11.427288421233008,0.09170095186571839)
(9.30700326289995,0.0233413006077743)
(11.297765066600828,0.09067786245731649)
(11.297765066600828,0.09067786245731649)
(11.297765066600828,0.09067786245731649)
(11.426662559189843,0.022608159134350218)
(11.363032518829323,0.09093094009321448)
(11.426662559189843,0.022608159134350218)
(9.360578598677687,0.02390891325912678)
(9.728217638806857,0.024328353051233355)
(9.93950064087281,0.02396072587553685)
(9.93950064087281,0.02396072587553685)
(11.371478241898272,0.08501146223604987)
(9.183337074355332,0.02385761545164938)
(9.93950064087281,0.02396072587553685)
(10.082515757222527,0.023775581817826442)
(10.103831743199569,0.02349052669195633)
(11.384023880521958,0.08935224852717835)
(10.23448563704266,0.023652069530361385)
(10.293798759288482,0.02309586655102245)
(10.302154817085409,0.0881778499313509)
(10.302154817085409,0.0881778499313509)
(10.415725056857175,0.023276253131666258)
(10.415725056857175,0.023276253131666258)
(10.437473615392914,0.08965874202903196)
(10.56708984991287,0.09307548785060407)
(10.56708984991287,0.09307548785060407)
(10.588336249176507,0.023479809085787777)
(10.67382378664665,0.09326505051010042)
(10.675881259987541,0.02315749467639126)
(10.704493379526763,0.023875660307159024)
(10.721598687576645,0.023324884288100357)
(10.751491733122018,0.0234413842006285)
(10.751491733122018,0.0234413842006285)
(10.751491733122018,0.0234413842006285)
(10.759940762493095,0.023608480649665767)
(10.872598818536442,0.02388101579265315)
(10.886369775415462,0.023215737814019116)
(10.928039418502056,0.02314719648725937)
(10.928039418502056,0.02314719648725937)
(10.928039418502056,0.02314719648725937)
(10.9371408063017,0.09075558766691688)
(10.943582158292145,0.09109326215209132)
(11.389871485456167,0.09210417918791679)
(10.943582158292145,0.09109326215209132)
(10.943582158292145,0.09109326215209132)
(10.943582158292145,0.09109326215209132)
(10.943582158292145,0.09109326215209132)
(10.992367083134893,0.022800109549491823)
(11.005859933458947,0.09169642177351292)
(11.005859933458947,0.09169642177351292)
(11.009419778864606,0.09271555772774001)
(11.009419778864606,0.09271555772774001)
(11.009419778864606,0.09271555772774001)
(11.015893130047647,0.023448896044472665)
(11.015893130047647,0.023448896044472665)
(11.389871485456167,0.09210417918791679)
(11.389871485456167,0.09210417918791679)
(11.015893130047647,0.023448896044472665)
(11.058927720369477,0.023619377783339116)
(11.073105512924325,0.09200623730340991)
(11.078469062037794,0.02328517145734009)
(11.0871007953184,0.02311723054393282)
(11.09433944061691,0.02372563696769083)
(11.109055281464059,0.08682668805599798)
(11.18803418689874,0.02230094706854791)
(11.258582032405304,0.023278872885307108)
(11.258582032405304,0.023278872885307108)
(11.408456819210429,0.09143488123518255)
(11.284582415309673,0.02399197851055423)
(11.284582415309673,0.02399197851055423)
(11.284582415309673,0.02399197851055423)
(11.297765066600828,0.09067786245731649)
(8.891192686875936,0.023821976556205767)
(8.891192686875936,0.023821976556205767)
(11.426662559189843,0.022608159134350218)
    };

\addplot[only marks,mark=*,style={fill=lightgray},y filter/.code={\pgfmathparse{\pgfmathresult*1000}\pgfmathresult}] plot coordinates {
(17.931876622945012,0.0011829171041075467)
(13.63175954204351,0.004475529340068504)
(12.972632760288313,0.00796455359060753)
(12.256453095774459,0.016414715460830953)
(11.644630582916541,0.022160862129241246)
(9.30700326289995,0.0233413006077743)
(10.293798759288482,0.02309586655102245)
(10.992367083134893,0.022800109549491823)
(11.18803418689874,0.02230094706854791)
(8.891192686875936,0.023821976556205767)
(8.891192686875936,0.023821976556205767)
};

\addplot[only marks,mark=*,black,y filter/.code={\pgfmathparse{\pgfmathresult*1000}\pgfmathresult}] plot coordinates {
(13.63175954204351,0.004475529340068504)
};

\addlegendentry{searched}
\addlegendentry{non-dominated}
\addlegendentry{selected}
    \end{axis}
    \end{tikzpicture}

%% file: tikz/nsgaii-femosaa.tex
\begin{tikzpicture}
    \begin{axis}[
      width=5cm,
    height=5cm,
    xmin=10,
   xmax=34,
   xtick = {0,5,...,34},
    ymax=100,
    ytick = {0,20,...,100},
        xlabel=Energy Consumption (watt),
        ylabel=Response Time (ms),
       legend style={at={(0.3,0.8),font=\fontsize{7}{8}\selectfont},
      anchor=north west,legend columns=1}]

       \addplot[only marks,mark=o,black,y filter/.code={\pgfmathparse{\pgfmathresult*1000}\pgfmathresult}] plot coordinates {
(17.372493450688935,0.004422571948190981)
(18.071309436195296,0.001805454862138917)
(16.279808097126036,0.004647632140089698)
(24.462817148310425,0.0011423904338751301)
(15.93760847858691,0.02022337798048229)
(16.279808097126036,0.004647632140089698)
(18.976685815679744,0.001163048059999863)
(11.366428042536064,0.023648101061404934)
(18.071309436195296,0.001805454862138917)
(24.462817148310425,0.0011423904338751301)
(17.416763918367618,0.004269835655522228)
(16.279808097126036,0.004647632140089698)
(13.945133470269623,0.022832938136879477)
(15.5798640198741,0.02230526531079237)
(17.372493450688935,0.004422571948190981)
(22.48519806407228,0.0011532507845739902)
(21.125959942708906,0.0012079092297380195)
(19.98933466782349,0.0012632843865184886)
(19.244040347165743,0.0013679526235958425)
(19.465671777377803,0.0013015358600460111)
(21.125959942708906,0.0012079092297380195)
(19.98933466782349,0.0012632843865184886)
(12.862691285840999,0.05889266277831044)
(12.862691285840999,0.05889266277831044)
(18.909297673215693,0.003974711468986728)
(16.864994434489326,0.007670297437034025)
(14.016098140840116,0.02648508087034061)
(15.757416160581094,0.02306651169636004)
(15.757416160581094,0.02306651169636004)
(18.723091584324695,0.005729544835904846)
(18.52651733229062,0.006795816685536942)
(23.646447108406406,0.0012002401284537374)
(33.27683488975811,0.001193976164681425)
(33.27683488975811,0.001193976164681425)
(33.27683488975811,0.001193976164681425)
(19.26658507913229,0.0014798418013079859)
(19.755256070176667,0.0013216809390526852)
(21.547388430482968,0.001208057274985488)
(21.547388430482968,0.001208057274985488)
(21.547388430482968,0.001208057274985488)
(21.547388430482968,0.001208057274985488)
(18.950860636998684,0.004553635155653153)
(17.54150457377524,0.02081129545020806)
(18.294611688243496,0.012714619109606393)
(17.54150457377524,0.02081129545020806)
(18.477819383611923,0.008319009161497654)
(15.055807938959145,0.08497832474880869)
(16.139762151388886,0.06137712927683821)
(16.4111658558761,0.023082181127469566)
(16.020943577759468,0.06329015966992269)
(16.4111658558761,0.023082181127469566)
(16.20948511690071,0.057262143906508994)
(16.4111658558761,0.023082181127469566)
(16.139762151388886,0.06137712927683821)
(18.681947416695103,0.00823110439609158)
(18.941972674355743,0.007922498595439231)
(18.681947416695103,0.00823110439609158)
(18.681947416695103,0.00823110439609158)
(19.52793817995209,0.0016291545545481197)
(19.357888149718466,0.003183864590066542)
(19.357888149718466,0.003183864590066542)
(21.29520496794882,0.0013597545896491955)
(20.71403787256997,0.0014085046447566774)
(21.29520496794882,0.0013597545896491955)
(19.843961049260447,0.001451224275921129)
(24.15132034324375,0.001211090144614156)
(24.343440844969095,0.0012101843555450747)
(24.343440844969095,0.0012101843555450747)
(24.343440844969095,0.0012101843555450747)
(21.571511429488286,0.0012557599414646215)
(18.959639486268326,0.0051634507701757044)
(18.60868360971868,0.012566677278839391)
(18.50413186059216,0.019933825187169203)
(15.830277670392565,0.09215958439966088)
(16.47689381630581,0.04735883154928411)
(17.484057015218315,0.04175851775276523)
(17.66143387259208,0.02738069883810672)
(17.97311091615649,0.023662134733369866)
(17.911020820254222,0.025282581558842526)
(17.97311091615649,0.023662134733369866)
(17.484057015218315,0.04175851775276523)
(17.484057015218315,0.04175851775276523)
(16.450613867557653,0.06520490730592363)
(16.163078519758297,0.06839086645918295)
(18.74670987495632,0.00843263836001612)
(18.74670987495632,0.00843263836001612)
(18.74670987495632,0.00843263836001612)
(18.74670987495632,0.00843263836001612)
(25.242328569830253,0.0012627457512040976)
(16.393666866234334,0.06923867527101082)
(18.126393898592315,0.04917479567927835)
(18.686096591191053,0.013110664158568841)
(17.396861279793004,0.04985810576046221)
(16.6555476758076,0.05923568659539646)
(18.83287268745267,0.011798830557929687)
(19.37399152969457,0.003921180818529594)
(20.829698209026184,0.0024929571962142186)
(19.62774103779438,0.003126475467922007)
(16.499429567407507,0.06491420790333453)
(24.198728493143136,0.0013148893768562112)
    };

\addplot[only marks,mark=*,style={fill=lightgray},y filter/.code={\pgfmathparse{\pgfmathresult*1000}\pgfmathresult}] plot coordinates {
(17.372493450688935,0.004422571948190981)
(18.071309436195296,0.001805454862138917)
(16.279808097126036,0.004647632140089698)
(24.462817148310425,0.0011423904338751301)
(15.93760847858691,0.02022337798048229)
(16.279808097126036,0.004647632140089698)
(18.976685815679744,0.001163048059999863)
(11.366428042536064,0.023648101061404934)
(18.071309436195296,0.001805454862138917)
(24.462817148310425,0.0011423904338751301)
(17.416763918367618,0.004269835655522228)
(16.279808097126036,0.004647632140089698)
(13.945133470269623,0.022832938136879477)
(15.5798640198741,0.02230526531079237)
(17.372493450688935,0.004422571948190981)
(22.48519806407228,0.0011532507845739902)
};

\addplot[only marks,mark=*,black,y filter/.code={\pgfmathparse{\pgfmathresult*1000}\pgfmathresult}] plot coordinates {
(16.279808097126036,0.004647632140089698)
};

\addlegendentry{searched}
\addlegendentry{non-dominated}
\addlegendentry{selected}
    \end{axis}
    \end{tikzpicture}

%% file: tikz/ibea-femosaa.tex
\begin{tikzpicture}
    \begin{axis}[
      width=5cm,
    height=5cm,
    xmin=14,
    xmax=31,
    ymax=100,
    ytick = {0,20,...,100},
           xlabel=Energy Consumption (watt),
        ylabel=Response Time (ms),
       legend style={at={(0.3,0.8),font=\fontsize{7}{8}\selectfont},
      anchor=north west,legend columns=1}]
    \addplot[only marks,mark=o,black,y filter/.code={\pgfmathparse{\pgfmathresult*1000}\pgfmathresult}] plot coordinates {
    (19.360896147387557,0.08737486857867136)
(25.136136690486797,0.0015324643010688533)
(22.943490945664074,0.014424756501896792)
(21.267207796510565,0.0048760869353107975)
(17.516972834222834,0.08575229331977102)
(19.259184349984146,0.0051529335681539555)
(21.601298114607697,0.03129725336797268)
(22.962083309003418,0.0077036502455636304)
(21.555108343611305,0.007026188192325644)
(19.107566780706218,0.0823353918006124)
(20.771838148824816,0.004430675748631188)
(19.951359843086788,0.007173164142748367)
(19.951359843086788,0.007173164142748367)
(19.9071751905473,0.09319400839494932)
(19.96611082638177,0.09100134492103011)
(17.54672312879326,0.09419859967711153)
(24.71470820271273,0.001531770375724282)
(23.176423574879003,0.0014128568794836637)
(21.72453488195737,0.022487665772583165)
(24.463389760676737,0.005453252165663061)
(22.631570367507376,0.010991457106576792)
(18.26564589199052,0.007125924662234762)
(18.108567559031048,0.02344262795823191)
(21.555108343611305,0.007026188192325644)
(19.754202292554943,0.0022472199177991574)
(18.121318483217816,0.09254380876509041)
(21.25180322394129,0.016234763023581046)
(18.228551301710823,0.07880636864066717)
(16.98661853566216,0.024750979791662457)
(24.024174817382185,0.0017509252055114618)
(26.17344894441133,0.002431749502528427)
(18.03311623748992,0.03117470712050235)
(19.92353472485334,0.0013983870944280189)
(15.316943598185437,0.049056679854928904)
(20.143480344812126,0.007103044161180441)
(23.176423574879003,0.0014128568794836637)
(17.509506027445287,0.009039846680167229)
(21.629215806447334,0.007811782411504316)
(15.641146944846954,0.04863578941697101)
(22.392624274232286,0.014689252529805099)
(24.376778024678078,0.004705473229333783)
(17.042015928696657,0.08676490935618883)
(19.847553386518012,0.005011577704908628)
(20.029342839116126,0.06788306234420775)
(19.951359843086788,0.007173164142748367)
(21.971327901430165,0.021462232039353525)
(26.17344894441133,0.002431749502528427)
(18.228551301710823,0.07880636864066717)
(19.916510855915945,0.09331761505168544)
(22.962083309003418,0.0077036502455636304)
(21.971327901430165,0.021462232039353525)
(19.847553386518012,0.005011577704908628)
(18.040610828435494,0.09179005567379349)
(19.88372564718713,0.0027762118953749103)
(19.7956196333213,0.06918733685417353)
(15.641146944846954,0.04863578941697101)
(21.601298114607697,0.03129725336797268)
(24.376778024678078,0.004705473229333783)
(21.69429372622916,0.011766357786361266)
(21.20778731867326,0.007798739854544855)
(24.04845631195295,0.006115447518396884)
(21.555108343611305,0.007026188192325644)
(17.516972834222834,0.08575229331977102)
(15.118538501111924,0.0516416660553639)
(26.09507916186094,0.0014600499823694602)
(24.243719111352384,0.005791999459009367)
(21.18476872274982,0.007529926040954602)
(24.601170125056022,0.005649072658565917)
(19.951359843086788,0.007173164142748367)
(19.58149384541429,0.004846824429756039)
(30.026302962911608,0.0015901990408506703)
(21.20778731867326,0.007798739854544855)
(25.136136690486797,0.0015324643010688533)
(21.69429372622916,0.011766357786361266)
(19.535357571214327,0.0050859379972724664)
(21.684631698243493,0.010001082446274047)
(19.951359843086788,0.007173164142748367)
(14.829192858274034,0.051431875273032904)
(21.20778731867326,0.007798739854544855)
(17.516972834222834,0.08575229331977102)
(21.230904996949786,0.007143495930978287)
(23.176423574879003,0.0014128568794836637)
(24.548593554123748,0.0049451740624815235)
(21.41517984350074,0.024374278914227086)
(17.111220339996603,0.08676035370803804)
(23.089842566849775,0.0015403239532799137)
(24.04845631195295,0.006115447518396884)
(22.813967591031886,0.010123758202150186)
(21.370090379567635,0.01196759255007488)
(24.601170125056022,0.005649072658565917)
(19.9071751905473,0.09319400839494932)
(26.17344894441133,0.002431749502528427)
(25.316044098397555,0.0035355577663914667)
(15.250621346048101,0.05147312162245588)
(20.14315186645358,0.08680587076199428)
(26.17344894441133,0.002431749502528427)
(23.176423574879003,0.0014128568794836637)
(21.69429372622916,0.011766357786361266)
(21.684631698243493,0.010001082446274047)
(21.684631698243493,0.010001082446274047)
    };

\addplot[only marks,mark=*,style={fill=lightgray},y filter/.code={\pgfmathparse{\pgfmathresult*1000}\pgfmathresult}] plot coordinates {
(19.259184349984146,0.0051529335681539555)
(18.26564589199052,0.007125924662234762)
(19.754202292554943,0.0022472199177991574)
(16.98661853566216,0.024750979791662457)
(19.92353472485334,0.0013983870944280189)
(15.316943598185437,0.049056679854928904)
(17.509506027445287,0.009039846680167229)
(15.641146944846954,0.04863578941697101)
(15.641146944846954,0.04863578941697101)
(19.58149384541429,0.004846824429756039)
(19.535357571214327,0.0050859379972724664)
(14.829192858274034,0.051431875273032904)
};

\addplot[only marks,mark=*,black,y filter/.code={\pgfmathparse{\pgfmathresult*1000}\pgfmathresult}] plot coordinates {
(17.509506027445287,0.009039846680167229)
};

\addlegendentry{searched}
\addlegendentry{non-dominated}
\addlegendentry{selected}
    \end{axis}
    \end{tikzpicture}

%% file: tikz/moead-femosaa-nothing.tex
\begin{tikzpicture}
    \begin{axis}[
      width=5cm,
    height=5cm,
    xmin=9,
   xmax=11,
    ymax=100,
     ytick = {0,20,...,100},
          xlabel=Energy Consumption (watt),
        ylabel=Response Time (ms),
           legend style={at={(0.3,0.8),font=\fontsize{7}{8}\selectfont},
      anchor=north west,legend columns=1},]

\addplot[only marks,mark=o,black,y filter/.code={\pgfmathparse{\pgfmathresult*1000}\pgfmathresult}] plot coordinates {
(10.693871718578626,0.02410800335318633)
(10.341717036967332,0.023826244679709466)
(10.524284181504477,0.024035171795825528)
(10.641717036967332,0.023826244679709466)
(9.495637401350919,0.02384382414472927)
(10.524284181504477,0.024035171795825528)
(10.524284181504477,0.024035171795825528)
(9.696737879605918,0.0928221571445652)
(10.422869711794936,0.09630245093419848)
(9.960849307055732,0.024092435172170884)
(10.230256636309555,0.09952203539042198)
(10.305904995217547,0.024013696468477893)
};

\addplot[only marks,mark=*,style={fill=lightgray},y filter/.code={\pgfmathparse{\pgfmathresult*1000}\pgfmathresult}] plot coordinates {
(10.641717036967332,0.023826244679709466)
(10.441717036967332,0.023836244679709466)
(9.495637401350919,0.02384382414472927)
};

\addplot[only marks,mark=*,black,y filter/.code={\pgfmathparse{\pgfmathresult*1000}\pgfmathresult}] plot coordinates {
(10.641717036967332,0.023826244679709466)
};

\addlegendentry{searched}
\addlegendentry{non-dominated}
\addlegendentry{selected}
    \end{axis}
    \end{tikzpicture}

%% file: tikz/nsgaii-femosaa-nothing.tex
\begin{tikzpicture}
    \begin{axis}[
      width=5cm,
    height=5cm,
    xmin=12,
    xmax=18,
    ymax=150,
        xlabel=Energy Consumption (watt),
        ylabel=Response Time (ms),
       legend style={at={(0.3,0.8),font=\fontsize{7}{8}\selectfont},
      anchor=north west,legend columns=1}]

\addplot[only marks,mark=o,black,y filter/.code={\pgfmathparse{\pgfmathresult*1000}\pgfmathresult}] plot coordinates {
(12.89415232540159,0.14449458115858374)
(17.922460139175513,0.08421478182638166)
};

\addplot[only marks,mark=*,style={fill=lightgray},y filter/.code={\pgfmathparse{\pgfmathresult*1000}\pgfmathresult}] plot coordinates {
(12.89415232540159,0.14449458115858374)
(17.922460139175513,0.08421478182638166)
};

\addplot[only marks,mark=*,black,y filter/.code={\pgfmathparse{\pgfmathresult*1000}\pgfmathresult}] plot coordinates {
(17.922460139175513,0.08421478182638166)
};

\addlegendentry{searched}
\addlegendentry{non-dominated}
\addlegendentry{selected}
    \end{axis}
    \end{tikzpicture}

%% file: tikz/ibea-femosaa-nothing.tex
\begin{tikzpicture}
    \begin{axis}[
      width=5cm,
    height=5cm,
    xmin=17,
    xmax=20,
    ymax=100,
       xlabel=Energy Consumption (watt),
        ylabel=Response Time (ms),
       legend style={at={(0.1,0.5),font=\fontsize{7}{8}\selectfont},
      anchor=north west,legend columns=1}]
    \addplot[only marks,mark=o,black,y filter/.code={\pgfmathparse{\pgfmathresult*1000}\pgfmathresult}] plot coordinates {
(18.92167342746446,0.08667462725496185)
(19.9135395048189,0.07571236453926276)
(18.418653726029028,0.08901330412071987)
(17.793514361967677,0.07350022890300362)
(19.205000032353492,0.09062207545757472)
(19.90065680083801,0.07359029066096905)
(19.154141577542315,0.08716046462326033)
(19.154141577542315,0.08716046462326033)
    };

\addplot[only marks,mark=*,style={fill=lightgray},y filter/.code={\pgfmathparse{\pgfmathresult*1000}\pgfmathresult}] plot coordinates {
(17.793514361967677,0.07350022890300362)
};

\addplot[only marks,mark=*,black,y filter/.code={\pgfmathparse{\pgfmathresult*1000}\pgfmathresult}] plot coordinates {
(17.793514361967677,0.07350022890300362)
};

\addlegendentry{searched}
\addlegendentry{non-dominated}
\addlegendentry{selected}
    \end{axis}
    \end{tikzpicture}

%% file: tikz/all-rw-quality.tex
\begin{tikzpicture}
    \begin{axis}[ 
        grid = both,
        minor xtick={3,4,...,16},
        minor ytick={0,50,...,650},
        grid style={dashed},
      width=7cm,
    height=7cm,
    xmax=16.5,
    ymax=650,
          ylabel=GM of Response Time (ms),
        xlabel=GM of Energy Consumption (watt),
          y label style={align=center,at={(0.05,0.5)}},
       legend style={at={(0.02,0.98),font=\fontsize{9}{6}\selectfont},
      anchor=north west,legend columns=1}]

\addplot[only marks,mark=star,mark size=4pt] plot coordinates {
(3.13,38.39)
};

\addplot[only marks,mark=diamond,mark size=4pt] plot coordinates {
(3.85,61.04)
};

\addplot[only marks,mark=square,mark size=4pt] plot coordinates {
(3.35,78.02)
};

\addplot[only marks,mark=triangle,mark size=4pt] plot coordinates {
(4.81,118.57)
};

\addplot[only marks,mark=o,mark size=4pt] plot coordinates {
(4.88,157.48)
};

\addplot[only marks,mark=otimes,mark size=4pt] plot coordinates {
(15.81,336.79)
};

\addlegendentry{F*MOEA/D-STM (3.13,38.39)}
\addlegendentry{F*NSGA-II (3.85,61.04)}
\addlegendentry{F*IBEA (3.35,78.02)}
\addlegendentry{DUSE (4.81,118.57)}
\addlegendentry{PLATO (4.88,157.48)}
\addlegendentry{FUSION (15.81,336.79)}
    \end{axis}
    \end{tikzpicture}

%% file: tikz/all-rw-sta.tex
\begin{tikzpicture}
\begin{axis}[
    ybar=0.07cm,
    bar width=0.2cm,
 enlarge x limits=0.25,
 enlarge y limits=0.05,
 ymax=1.1,
    legend style={cells={align=left},at={(0.77,0.98),font=\fontsize{8.5}{6}\selectfont},
      anchor=north,legend columns=1},
      nodes near coords,
      every node near coord/.append style={rotate=90, anchor=west},
      xlabel={HV=Hypervolume, ED=Euclidean Distance},
          y label style={at={(0.15,0.5)}},
    symbolic x coords={Valid Solutions,HV,ED},
    xtick=data,
    width=7cm,
    height=6.9cm
    ]
    
    \addplot[fill=black!100!white] coordinates {(Valid Solutions,1) (HV,1) (ED,0)};
    \addplot[fill=black!80!white] coordinates {(Valid Solutions,1) (HV,0.87) (ED,0.05)};
\addplot[fill=black!60!white] coordinates {(Valid Solutions,1) (HV,0.85) (ED,0.07)};
\addplot[fill=black!40!white] coordinates {(Valid Solutions,0.1) (HV,0.63) (ED,0.15)};
\addplot[fill=black!20!white] coordinates {(Valid Solutions,0.08) (HV,0.52) (ED,0.21)};
\addplot[fill=white!100!white] coordinates {(Valid Solutions,0.65) (HV,0) (ED,0.71)};

\legend{F*MOEA/\\D-STM,F*NSGA-II,F*IBEA,DUSE,PLATO,FUSION}

\end{axis}
\end{tikzpicture}

%% file: tikz/all-r-quality.tex
\begin{tikzpicture}
    \begin{axis}[ 
        grid = both,
        minor xtick={3,4,...,16},
        minor ytick={0,50,...,650},
        grid style={dashed},
      width=7cm,
    height=7cm,
    xmax=16.5,
    ymax=650,
          ylabel=GM of Response Time (ms),
        xlabel=GM of Energy Consumption (watt),
          y label style={align=center,at={(0.05,0.5)}},
       legend style={at={(0.02,0.98),font=\fontsize{9}{6}\selectfont},
      anchor=north west,legend columns=1}]

\addplot[only marks,mark=star,mark size=4pt] plot coordinates {
(2.60,39.28)
};

\addplot[only marks,mark=diamond,mark size=4pt] plot coordinates {
(3.01,37.19)
};

\addplot[only marks,mark=square,mark size=4pt] plot coordinates {
(2.34,55.85)
};

\addplot[only marks,mark=triangle,mark size=4pt] plot coordinates {
(3.04,138.00)
};

\addplot[only marks,mark=o,mark size=4pt] plot coordinates {
(3.46,174.59)
};

\addplot[only marks,mark=otimes,mark size=4pt] plot coordinates {
(13.21,278.60)
};

\addlegendentry{F*MOEA/D-STM (2.60,39.28)}
\addlegendentry{F*NSGA-II (3.01,37.19)}
\addlegendentry{F*IBEA (2.34,55.85)}
\addlegendentry{DUSE (3.04,138.00)}
\addlegendentry{PLATO (3.46,174.59)}
\addlegendentry{FUSION (13.21,278.60)}
    \end{axis}
    \end{tikzpicture}

%% file: tikz/all-r-sta.tex
\begin{tikzpicture}
\begin{axis}[
    ybar=0.07cm,
    bar width=0.2cm,
 enlarge x limits=0.25,
 enlarge y limits=0.05,
 ymax=1.1,
    legend style={cells={align=left},at={(0.77,0.98),font=\fontsize{8.5}{6}\selectfont},
      anchor=north,legend columns=1},
      nodes near coords,
      every node near coord/.append style={rotate=90, anchor=west},
      xlabel={HV=Hypervolume, ED=Euclidean Distance},
          y label style={at={(0.15,0.5)}},
    symbolic x coords={Valid Solutions,HV,ED},
    xtick=data,
    width=7cm,
    height=6.9cm
    ]
    
    \addplot[fill=black!100!white] coordinates {(Valid Solutions,1) (HV,0.97) (ED,0.01)};
    \addplot[fill=black!80!white] coordinates {(Valid Solutions,1) (HV,0.94) (ED,0.03)};
\addplot[fill=black!60!white] coordinates {(Valid Solutions,1) (HV,0.92) (ED,0.04)};
\addplot[fill=black!40!white] coordinates {(Valid Solutions,0.14) (HV,0.54) (ED,0.21)};
\addplot[fill=black!20!white] coordinates {(Valid Solutions,0.13) (HV,0.39) (ED,0.29)};
\addplot[fill=white!100!white] coordinates {(Valid Solutions,0.63) (HV,0) (ED,0.71)};

\legend{F*MOEA/\\D-STM,F*NSGA-II,F*IBEA,DUSE,PLATO,FUSION}

\end{axis}
\end{tikzpicture}

%% file: tikz/all-sas-quality.tex
\begin{tikzpicture}
    \begin{axis}[ 
        grid = both,
        minor xtick={15,20,...,105},
        minor ytick={0,5,...,100},
        grid style={dashed},
      width=7cm,
    height=7cm,
    xmax=105,
    ymax=100,
          ylabel=GM of Throughput (request/second),
        xlabel=GM of Cost (\$),
          y label style={align=center,at={(0.05,0.5)}},
       legend style={at={(0.02,0.99),font=\fontsize{9}{6}\selectfont},
      anchor=north west,legend columns=1}]

\addplot[only marks,mark=star,mark size=4pt] plot coordinates {
(38.39,35.72)
};

\addplot[only marks,mark=diamond,mark size=4pt] plot coordinates {
(56.86,39.87)
};

\addplot[only marks,mark=square,mark size=4pt] plot coordinates {
(65.08,44.83)
};

\addplot[only marks,mark=triangle,mark size=4pt] plot coordinates {
(102.39,37.47)
};

\addplot[only marks,mark=o,mark size=4pt] plot coordinates {
(90.82,27.90)
};

\addplot[only marks,mark=otimes,mark size=4pt] plot coordinates {
(25.80,5.90)
};

\addlegendentry{F*MOEA/D-STM (38.39,35.72)}
\addlegendentry{F*NSGA-II (56.86,39.87)}
\addlegendentry{F*IBEA (65.08,44.83)}
\addlegendentry{DUSE (102.39,37.47)}
\addlegendentry{PLATO (90.82,27.90)}
\addlegendentry{FUSION (25.80,5.90)}
    \end{axis}
    \end{tikzpicture}

%% file: tikz/all-sas-sta.tex
\begin{tikzpicture}
\begin{axis}[
    ybar=0.07cm,
    bar width=0.2cm,
 enlarge x limits=0.25,
 enlarge y limits=0.05,
 ymax=1.1,
    legend style={cells={align=left},at={(0.67,0.98),font=\fontsize{8.5}{6}\selectfont},
      anchor=north,legend columns=1},
      nodes near coords,
      every node near coord/.append style={rotate=90, anchor=west},
      xlabel={HV=Hypervolume, ED=Euclidean Distance},
          y label style={at={(0.15,0.5)}},
    symbolic x coords={Valid Solutions,HV,ED},
    xtick=data,
    width=7cm,
    height=6.9cm
    ]
    
    \addplot[fill=black!100!white] coordinates {(Valid Solutions,1) (HV,0.80) (ED,0.08)};
    \addplot[fill=black!80!white] coordinates {(Valid Solutions,1) (HV,0.58) (ED,0.20)};
\addplot[fill=black!60!white] coordinates {(Valid Solutions,1) (HV,0.49) (ED,0.26)};
\addplot[fill=black!40!white] coordinates {(Valid Solutions,0.57) (HV,0) (ED,0.50)};
\addplot[fill=black!20!white] coordinates {(Valid Solutions,0.24) (HV,0.14) (ED,0.43)};
\addplot[fill=white!100!white] coordinates {(Valid Solutions,0.03) (HV,0) (ED,0.50)};

\legend{F*MOEA/\\D-STM,F*NSGA-II,F*IBEA,DUSE,PLATO,FUSION}

\end{axis}
\end{tikzpicture}